\pdfoutput=1
\documentclass[a4paper, 11pt]{article}
\usepackage{jheppub, amsmath, amssymb, wasysym, slashed, hyperref, color, multirow}
\usepackage[T1]{fontenc}
\usepackage[table]{xcolor}
\definecolor{nicered}{rgb}{.7,.1,.1}
\definecolor{nicegreen}{rgb}{.1,.5,.1}
\definecolor{darkblue}{rgb}{0,0,.5}
\definecolor{el}{rgb}{.9, .8, .7}
\definecolor{mu}{rgb}{.8, .7, .8}
\hypersetup{colorlinks, citecolor=nicegreen,linkcolor=nicered, urlcolor=darkblue}

\author[a]{Miha Nemev\v sek}
\affiliation[a]{Jo\v zef Stefan Institute, Jamova 39, Ljubljana, Slovenia}
\emailAdd{miha.nemevsek@ijs.si}

\author[b,c]{Fabrizio Nesti}
\affiliation[b]{Dipartimento di Fisica, Theoretical section, Universit\`a di Trieste, Strada Costiera 11, I-34151 Trieste, Italy}
\affiliation[c]{Ru\dj er Bo\v{s}kovi\'c Institute, Bijeni\v{c}ka cesta 54, 10000, Zagreb, Croatia}
\emailAdd{fabrizio.nesti@irb.hr}

\author[d,e,f]{Juan Carlos Vasquez}
\affiliation[d]{ Centro Cient\'ifico Tecnol\'ogico de Valpara\'iso-CCTVal, Universidad T\'ecnica Federico Santa Mar\'ia, Valpara\'iso, Chile}
\affiliation[e]{ICTP, Strada Costiera, 11 - 34151 Trieste, Italy}
\affiliation[f]{SISSA/INFN,via Bonomea, 265 - 34136 Trieste, Italy}
\emailAdd{jcvasque@sissa.it}

\title{\huge Majorana Higgses at colliders}
                                            
{\large \abstract{Collider signals of heavy Majorana neutrino mass origin are studied in the minimal Left-Right symmetric model, where their mass is generated spontaneously together with the breaking of lepton number. The right-handed triplet Higgs boson $\Delta$, responsible for such breaking, can be copiously produced at the LHC through the Higgs portal in the gluon fusion and less so in gauge mediated channels. At $\Delta$ masses below the opening of the $VV$ decay channel, the two observable modes are pair-production of heavy neutrinos via the triplet gluon fusion $gg \to \Delta \to NN$ and pair production of triplets from the Higgs $h \to \Delta \Delta \to 4N$ decay. The latter features tri- and quad same-sign lepton final states that break lepton number by four units and have no significant background. In both cases up to four displaced vertices may be present and their displacement may serve as a discriminating variable. The backgrounds at the LHC, including the jet fake rate, are estimated and the resulting sensitivity to the Left-Right breaking scale extends well beyond 10 TeV. In addition, sub-dominant radiative modes are surveyed: the $\gamma \gamma$, $Z \gamma$ and lepton flavour violating ones. Finally, prospects for $\Delta$ signals at future $e^+ e^-$ colliders are presented.}}

\arxivnumber{1612.06840}

\begin{document}
\maketitle
\flushbottom

%
%
\section{Introduction} \label{SecIntro}

An attractive feature of the Standard model (SM) is the economy of the Higgs mechanism~\cite{Higgs:1964pj} that simultaneously provides spontaneous origin of gauge boson and charged fermion masses~\cite{Weinberg:1967tq} such that
\begin{equation} \label{eqSMforigin}
  \Gamma_{h \to f \overline f} \propto m_f^2.
\end{equation}
This picture has now been successfully confirmed by the LHC, apart from the first two generations of fermions~\cite{Khachatryan:2016vau}. The success of the SM may suggest that perhaps the masses of all fundamental particles, including neutrinos, are protected by a gauge symmetry and get generated through spontaneous breaking.

Despite the success of the SM, the description of weak interactions is glaringly asymmetric and neutrinos are massless. Left-Right (LR) symmetric theories~\cite{LROrigin} remove both deficiencies. Parity is broken spontaneously~\cite{LRSpont} and the minimal LR symmetric model (LRSM)~\cite{MinkowskiMS79} generates the Majorana mass~\cite{Majorana:1937vz} for the right-handed (RH) neutrino $N$ through spontaneous breaking of $SU(2)_R$. Furthermore, the LRSM naturally explains the lightness of left-handed neutrinos via the celebrated see-saw mechanism~\cite{MinkowskiMS79, seesaw,Schechter:1980gr}
\begin{equation} \label{eqSeesaw}
  M_\nu = - M_D^T \, M_N^{-1} \, M_D^{\,}.
\end{equation}
The neutrino Dirac mass $M_D$ is generated by the SM-like Higgs and in direct analogy, a ``Majorana'' Higgs $\Delta$ provides the Majorana mass $M_N$ from its condensate. Therefore, as with the prediction of the SM in~\eqref{eqSMforigin}, one would like to observe 
\begin{equation} \label{eqLRNorigin}
  \Gamma_{\Delta \to NN} \propto m_N^2,
\end{equation}
as a proof of the spontaneous generation of heavy neutrino mass. This would be a significant step towards the complete determination of neutrino mass origin. 

While the see-saw in~\eqref{eqSeesaw} is clearly appealing because light neutrino mass matrix $M_\nu$ becomes suppressed in the presence of heavy neutrinos, it also complicates the assessment of the nature of mass origin. One may conceivably use colliders to determine the $M_N$, either through the 'golden' heavy neutrino channel as proposed by Keung and Senjanovi\'c (KS)~\cite{Keung:1983uu} that is being actively searched for at the LHC~\cite{Nemevsek:2011hz, KSatLHC, Mitra:2016kov}, or via the ``Majorana'' Higgs decays in~\eqref{eqLRNorigin} discussed here. An observation of either one would clearly signal lepton number violation (LNV) at colliders and reveal the Majorana nature of heavy neutrinos. Together with the information from neutrinoless double beta ($0\nu2\beta$) decay~\cite{Mohapatra:1980yp, Tello:2010am} it may thus be possible to reconstruct $M_N$~\cite{Das:2012ii, Vasquez:2014mxa}.

In contrast to the SM, where the Dirac mass of charged fermions is uniquely determined, the $M_D$ in the seesaw~\eqref{eqSeesaw} cannot be unambiguously computed~\cite{Casas:2001sr}, due to the quadratic nature of the seesaw matrix equation, even if $M_N$ and $M_{\nu}$ were known from colliders and neutrino oscillations, respectively. In the LRSM this becomes possible thanks to restoration of parity that removes the ambiguity in the Dirac mass matrix. Thus, $M_D$ is predicted~\cite{Nemevsek:2012iq} and testable in sub-dominant decays of $N$~\cite{Nemevsek:2012iq}, searches in $W$~\cite{Datta:1993nm, Izaguirre:2015pga}, $h$~\cite{BhupalDev:2012zg}, the electron EDM~\cite{Nieves:1986uk, Nemevsek:2012iq} and $0\nu2\beta$~\cite{Nemevsek:2012iq, 0nu2bMix}. 

In the context of the LRSM, other low energy processes in the quark flavor sector played an important role in the past~\cite{Beall:1981ze, Ecker:1985vv, Mohapatra:1983ae, Zhang:2007fn, Maiezza:2010ic, Bertolini:2012pu} by setting a lower limit on the LR scale. Recent updates of $K$ and $B$ oscillations~\cite{Bertolini:2014sua}, together with CP-odd $\varepsilon, \varepsilon'$~\cite{Bertolini:2012pu} and the neutron EDM~\cite{Maiezza:2014ala}, converged on a lower limit of $M_{W_{R}} \gtrsim 3 \text{ TeV}$, barring the issue of strong CP~\cite{Maiezza:2014ala}.

Flavor changing processes have a significant impact on the Higgs sector of the minimal model, in particular on the flavor-changing scalar of the bi-doublet that needs to be heavy~\cite{Senjanovic:1979cta}, beyond the reach of the LHC but potentially accessible to a 100 TeV machine~\cite{Dev:2016dja}. The required large mass may cause issues with perturbativity and unitarity that results in a lower bound on the mass of $W_R$ and some of the triplet scalar components in the LRSM~\cite{Maiezza:2016bzp}. However, the neutral component of the RH triplet is not affected and since it is a SM singlet, its mass can safely be below the TeV scale. This is precisely the region of interest for determination of spontaneous origin of $M_N$, which is the subject of this work.

The RH triplet Higgs as the source of spontaneous mass origin for heavy neutrinos within the minimal LRSM was proposed in~\cite{MinkowskiMS79}. Phenomenological collider studies of the Higgs sector in the LRSM was sketched in~\cite{Gunion:1986im}, where decays to heavy neutrinos were pointed out for both the RH triplet and the SM Higgs. Higgs decay to heavy sterile neutrinos was mentioned also in~\cite{Pilaftsis:1991ug} and analyzed in~\cite{Graesser:2007yj} with effective operators. Recently, a more detailed collider study of the LNV decays of the SM Higgs to $NN$ within the LRSM was done in~\cite{Maiezza:2015lza}. Here, we extend the analysis to the phenomenology of the RH triplet at colliders: \S\ref{SecModel} gives a short review of LRSM features, the decay and production channels of the Higgses of interest are computed in~\S\ref{SecDecay} and \S\ref{SecProduction}. The \S\ref{SecLHC} discusses signals and backgrounds at the LHC while \S\ref{SecEpEm} gives an outlook on $e^+ e^-$ machines. We conclude in \S\ref{SecConclude} and leave details the discussion on triple Higgs vertices, loop functions and jet fakes for appendices.

%
%
\section{The minimal Left-Right model} \label{SecModel}

The minimal LR symmetric model (LRSM)~\cite{MinkowskiMS79, Mohapatra:1980yp} is based on the gauge group $SU(2)_L\times SU(2)_R\times U(1)_{B-L}$, with an additional discrete symmetry that may be generalized parity $\mathcal{P}$ or charge conjugation $\mathcal{C}$. The fermions belong to LR doublets of quarks $Q_{L,R}^T = (u, d) _{L, R}$ and leptons $L_{L,R}^T = (\nu, \ell)_{L, R}$, while the Higgs sector consists of a bi-doublet $\phi(2,2,0)$ and two LR symmetric triplets $\Delta_L = (3,1,2)$ and $\Delta_R = (1,3,2)$. The latter is the RH triplet and its fields are denoted as
\begin{equation} \Delta_R \equiv \begin{pmatrix}
    \Delta_R^+/\sqrt{2} & \Delta_R^{++} \\
    \Delta_R^0 + v_R & -\Delta_R^+/\sqrt{2} \end{pmatrix}.
\end{equation}
The $v_R$ vev is predominantly responsible for the breaking of $SU(2)_R$, thus the real part of $\Delta_R^0$ is {\em the} Higgs of the LRSM and its couplings to gauge bosons and RH neutrinos determine their masses. In the minimal model with parity broken at low scales $g_R \simeq g_L \equiv g$ and
\begin{equation} \label{eqMassWRZLR}
  M_{W_R} = g \, v_R , \quad M_{Z_{LR}} \simeq \sqrt{3} \, M_{W_R}.
\end{equation}
The Yukawa Majorana terms that couple leptons to triplets give masses to heavy Majorana neutrinos $N$~\cite{Mohapatra:1980yp}
\begin{equation} \label{eqMajoranaYukawa}
  \mathcal L_{Y_\Delta} = Y_\Delta L_R^T C \Delta_R L_R + \text{h.c.,  such that }
   M_N = 2 v_R Y_\Delta = V_R^{\,} m_N V_R^T,
\end{equation}
where $m_N$ is diagonal and $V_R$ is the RH analog of the PMNS matrix. It determines the flavor structure of leptonic $SU(2)_R$ gauge interactions and Yukawa couplings with the triplets, relevant for lepton flavor violating (LFV) decays. 

The second step of breaking the $SU(2)_L$ is completed by the vev of the bi-doublet
\begin{equation}
  \langle \phi \rangle = \begin{pmatrix} v_1 & 0 \\ 0 & v_2 \, e^{i \alpha} \end{pmatrix} \text{ and }
  \phi = \begin{pmatrix} \phi_1^0 + v_1 & \phi_2^+ \\ \phi_1^- & \phi_2^0 + v_2 \, e^{i \alpha} \end{pmatrix} ,
\end{equation}
and gives the fermions their Dirac masses~\cite{LROrigin, LRSpont,Mohapatra:1980yp, Maiezza:2010ic, Nemevsek:2012iq}. The neutral field components of $\phi$, in particular the SM-like Higgs, can mix with $\Delta_0$, which is the real part of $\Delta_R^0$. The mixings with $\Delta_L$ are suppressed by the small $\langle \Delta_L \rangle \propto v^2/v_R$, and the mixing with the heavy Flavor-Changing (FC) scalar $H$ is phenomenologically constrained by flavor physics~\cite{Maiezza:2016bzp}. Therefore, it is sensible to reduce the mass matrix to the $2 \times 2$ case involving only the SM-like Higgs $h$ and the triplet-like $\Delta$.
\begin{align}
  \begin{pmatrix} h \\ \Delta \end{pmatrix} = \begin{pmatrix} c_\theta & s_\theta \\ -s_\theta & c_\theta \end{pmatrix} 
  \begin{pmatrix} 
    h_0 \\ \Delta_0 \end{pmatrix},
\end{align}
where $h_0 = \text{re } \phi_1^0 \, \left(\frac{v_1}{v}\right) + \text{re } \phi_2^0 \, \left(\frac{v_2}{v}\right)$, $s_\theta = \sin \theta$, $c_\theta = \cos \theta$ is the mixing angle. One should keep in
mind the existing constraints on the allowed mixing with the SM Higgs. To a good approximation,
$\Delta$ behaves as a SM singlet, therefore the studies in~\cite{singletHiggs} apply.  The allowed
mixing angle depends on $m_\Delta$ and one has typically $s_\theta < 0.2$--$0.4$ for the mass range
under consideration in this work, while future prospects are discussed in~\cite{Buttazzo:2015bka}.

For the sake of illustration, we give the expression of the LRSM potential parameters (see
\cite{Maiezza:2016bzp} and references therein for the definition of the LRSM potential), in terms of
the $h$ and $\Delta$ masses and mixing $\theta$ in the limit of vanishing mixing with the heavy FC Higgses
\begin{gather} \label{eqmhDth}
  \lambda_1 = \frac{m_h^2 c_\theta^2 + m_\Delta^2 s_\theta^2}{4\,v^2} + s_{\beta}^2 \left(2 \lambda_2 + \lambda_3 \right),
  \quad \quad 
  \rho_1 = \frac{m_\Delta^2 c_\theta^2 + m_h^2 s_\theta^2}{4\,v_R^2} ,
  \\
  \alpha_1 = \frac{s_{2\theta} c_\theta (m_\Delta^2-m_h^2)}{4\,v\, v_R} + 
  s_{\beta}^2 \left( \frac{m_H^2 - 4 v^2 c_{\beta}^2 \left(2 \lambda_2 + \lambda_3 \right)}{v_R^2} \right).
\end{gather}
where $t_{\beta/2}=\tan(\beta/2)\equiv v_2/v_1$. Because in general $\Delta$ receives its mass from $v_R$, the relevant couplings $\rho_1$ and $\alpha_1$ turn out to be small in the range of masses relevant for this work, i.e. $m_\Delta \lesssim 200 \text{ GeV}$; see Appendix~\ref{SecAppTripleH} for complete details.

In this particular limit the $h-\Delta$ tri-linear couplings are unambiguously determined by the masses of $h$ and $\Delta$, the mixing angle $s_\theta$ and the LR scale. Such a formulation is especially convenient for phenomenological studies because there is no need to worry about inter-dependencies of LRSM potential parameters responsible for the production and decay rates. The expressions are collected in the Appendix~\ref{SecAppTripleH} including the formulae for small non-vanishing $\Delta-H$ mixing. 

%
%
\section{Decay modes of Higgses} \label{SecDecay}

\begin{figure} \centering
  \includegraphics[height=6.1 cm]{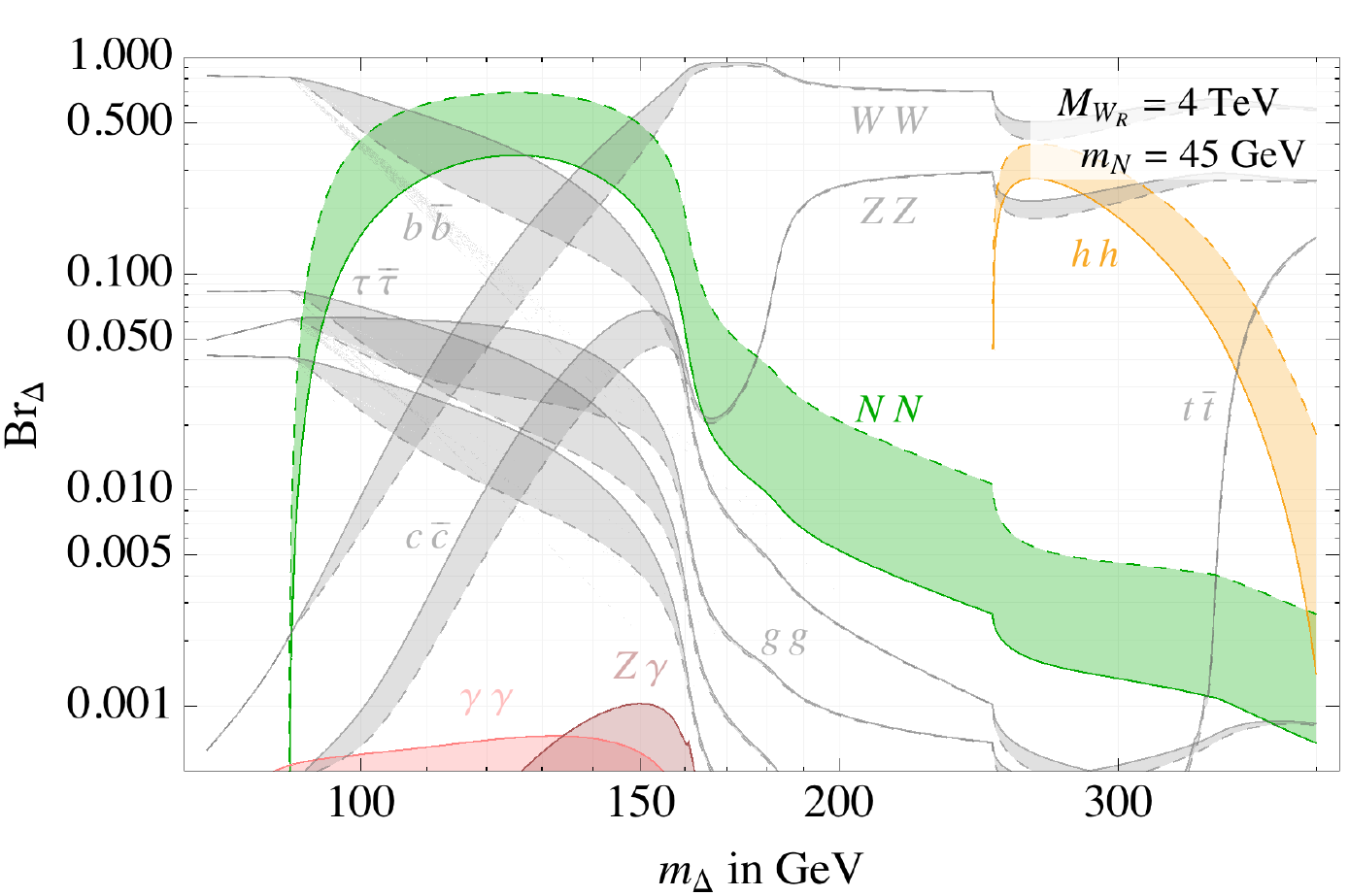} \hfill
  \raisebox{1mm}{\includegraphics[height=5.6cm]{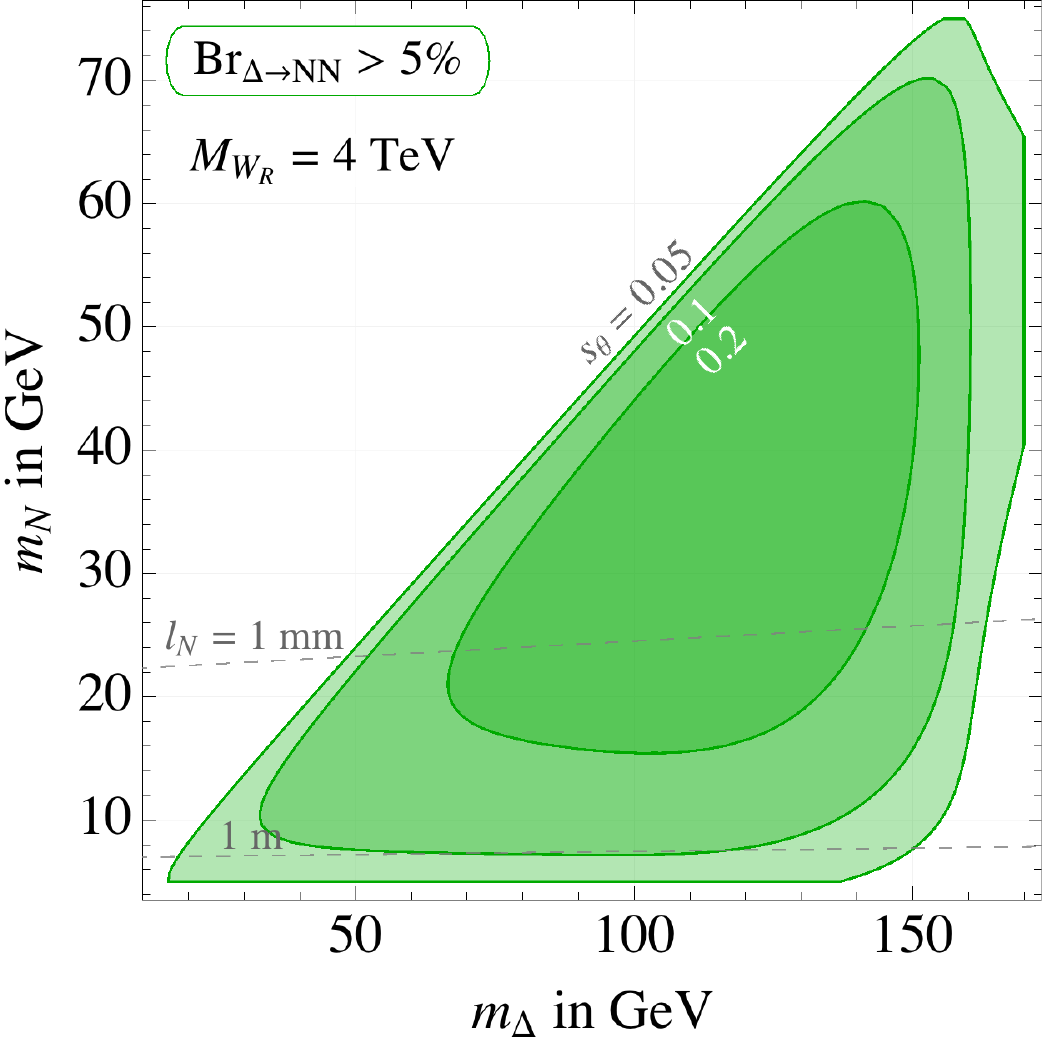}}
  \caption{Left: Branching ratios of $\Delta$ to a pair of Majorana neutrinos (green) with $m_N = 45 \text{ GeV}$, pairs of Higgses (red) and other SM particles in grey, $\gamma\gamma$ in pink and $Z \gamma$ in purple. The shaded areas cover the variation of $s_\theta$ from $5 \%$ in dashed lines to $10 \%$ in solid ones.  Right: Regions of appreciable branching ratio to $N$ ($\text{Br}_{\Delta \to NN} >5 \, \%$) for $s_\theta = 5, 10 \text{ and } 20 \%$; in both cases $M_{W_R}$ is fixed at $4 \text{ TeV}$, $m_H = 17 \text{ TeV}$ and $r_{++}=0.3$ (see Appendix). The dashed lines show the lifetime of $N$ in the lab frame, taking into account the mean boost of $N$ at $\sqrt s = 13 \text{ TeV}$.} 
  \label{figBrD}
\end{figure}

The $h$ and $\Delta$ decay modes to fermionic and bosonic final states are described in the following sub-sections. The branching ratios of $\Delta$ are collected in Fig.~\ref{figBrD}.

%
\subsection{Decays to fermions} \label{SubSecFermions}

The tree-level coupling of $\Delta$ to Majorana neutrinos come from the Yukawa term in~\eqref{eqMajoranaYukawa}. Since $\Delta_R$ is responsible for spontaneous breaking, its coupling to a pair of $N$s is flavor diagonal in the mass basis and proportional to $\text{diag}(m_{N_i})$ (henceforth we drop the family index for clarity). This is the essence of the heavy neutrino Higgs mechanism one would like to test. Specifically, it leads to decay rates
\begin{align} \label{eqGamDhNN}
  \Gamma_{\Delta \to N N} &= c_\theta^2 \frac{\alpha_w}{8} m_\Delta \left(\frac{m_N}{M_{W_R}} \right)^2 
  \beta_{\Delta N}^{3/2}, & 
  \Gamma_{h \to N N} &= s_\theta^2 \frac{\alpha_w}{8} m_h \left(\frac{m_N}{M_{W_R}} \right)^2 
  \beta_{h N}^{3/2},
\end{align}
where a factor of $2$ in the amplitude is due to the Majorana nature of $N$ and $1/2$ for the same final state particles, with $\beta_{i N} = 1-(2 m_N/m_i)^2$ and $\alpha_w = g^2/(4\pi)$. Apart from the $NN$ channel, the two body decay rates
\begin{equation}
  \Gamma_{\Delta \to f \overline f} = s_\theta^2 \, \Gamma_{h \to f \overline f} \left(m_h \to m_\Delta, s_\theta = 0 \right)\,
\end{equation}
to SM fermions ($f$) open up when the Higgs mixing is present. For $m_\Delta < 10 \text{ GeV}$, these lead to displaced vertices of pairs of SM fermions~\cite{Clarke:2013aya}.

In order to probe and ultimately determine the origin of heavy neutrino mass, the $\text{Br}_{\Delta \to NN}$ should be appreciable. Its behaviour can be understood from Fig.~\ref{figBrD}. As long as $2 \, m_N < m_\Delta$ and $\Delta$ is below the $VV$ threshold, the $NN$ final state dominates in proportion to $c_\theta$. The region on the right panel of Fig.~\ref{figBrD} therefore defines the parameter space of interest for collider studies performed in \S \ref{SecLHC}.

%
\paragraph{Lepton flavor violating decays.} At one loop, the heavy neutrinos and doubly charged scalars\footnote{Here we focused on the dominant mode mediated by $\Delta_R^{++}$; the LFV-$\gamma\gamma$ interplay and the conclusions are the same if $W_R$ loops are also taken into account.} mediate lepton flavor violating (LFV) decays
\begin{align} \label{eqGamDLFV}
  \Gamma_{\Delta \to \ell_i^{\pm} \ell_j^{\mp}} &= \frac{m_\Delta}{8 \pi} \left( \frac{\alpha_w}{16 \pi} \right)^2 
    \left| \frac{M_N^{\,} M_N^\dagger}{M_{W_R}^2} \right|^2_{ij} \left| \tilde f(\beta_\Delta)\right|^2
    \left( \frac{m_i v_{\Delta \Delta_R^{++} \Delta_R^{--}}}{m_{\Delta_R^{++}}^2} \right)^2,
\end{align}
with $\beta_\Delta = (2 m_{\Delta_R^{++}}/m_\Delta)^2$ and the triple Higgs coupling $v_{\Delta \Delta_R^{++} \Delta_R^{--}}$ shown in~\eqref{eqVDDRpp}, while the loop function $\tilde f$ can be found in~\eqref{eqFtLFV} of Appendix~\ref{SecAppLoops}. It turns out that the $v_{h \Delta_R^{++} \Delta_R^{--}}$ in~\eqref{eqVhDRpp} is determined by the same parameters as $v_{\Delta \Delta_R^{++} \Delta_R^{--}}$ but is also further constrained by the $h \to \gamma \gamma$ data that sets the lower bound on $m_{\Delta_R^{++}} \gtrsim 1 \text{ TeV}$~\cite{Maiezza:2016bzp}.

The flavor structure of $\Delta$, $N$ and leptonic flavor violating decays below is determined by the heavy neutrino Majorana mass matrix $M_N$. It sets the $\Delta_R^{++}$ (and $W_R$) amplitudes for $\Delta$ decay and also governs low energy processes. In particular, the $\ell \to 3 \ell'$~\cite{Cirigliano:2004mv, Tello:2010am, Tello:2012qda} and $\ell \to \ell' \gamma$ rates are
\begin{align}
  \Gamma_{\ell_i \to \ell_j \gamma} &\simeq \frac{\alpha_w^2}{36 \pi} \, \frac{m_i^5}{m_{\Delta_R^{++}}^4}
    \frac{\alpha}{16 \pi} \left| \frac{M_N^{\,} M_N^\dagger}{M_{W_R}^2} \right|_{ij}^2,
  &
  \Gamma_{\ell_i \to 3 \ell_j} &\simeq \frac{\alpha_w^2}{36 \pi} \, \frac{m_i^5}{m_{\Delta_R^{++}}^4}
    \frac{3}{64} \left| \frac{M^{\,}_{Nij} M^*_{N jj}}{M_{W_R}^2} \right|^2.
\end{align}
It turns out that due to the chirality flipping and current constraints on radiative $\ell_i \to \ell_j \gamma$ decays, the LFV final states of $h$ are largely suppressed even for the $\tau$ channel when $N$ is heavy. In the case of $\Delta$ decays, the LFV modes are rather subdominant to other radiative channels such as $\gamma \gamma$ and $\gamma Z$ and especially the SM final states in the presence of $s_\theta$.

%
\subsection{Decays to bosons}

Depending on its mass and mixing, $\Delta$ can decay to various bosonic pairs. For example, when $m_\Delta > 250 \text{ GeV}$, the $\Delta \to hh$ channel opens up and its rate is set unambiguously by $s_\theta$. At the same time the SM gauge boson channels open up
\begin{equation}
  \Gamma_{\Delta \to V V} = s_\theta^2 \, \Gamma_{h \to V V} \left(m_h \to m_\Delta, s_\theta = 0 \right),
\end{equation}
with $V = W$ or $Z$ on and off-shell.\footnote{An additional amplitude may also be present when the LR gauge boson mixing is non-zero.} These rates grow with the Higgs mass due to the longitudinal would-be-Goldstones and tend to dominate the $\text{Br}_\Delta$ above the 160 GeV threshold when $s_\theta > 1 \%$, as seen from the left panel of Fig.~\ref{figBrD}. LHC prospects for signals of such final states were studied by~\cite{Banerjee:2015hoa} in a related $B-L$ model (see also \cite{Kang:2015uoc} for the $Z' \to NN$ channel).

Finally, $\Delta$ can decay to four SM fermions via two off-shell RH gauge bosons, or $V^*V^*$ through LR gauge boson mixing. These turn out to be slower than the radiative $\gamma \gamma$ and $Z \gamma$ rates computed below.

\begin{figure} \centering
  \includegraphics[height=6.1 cm]{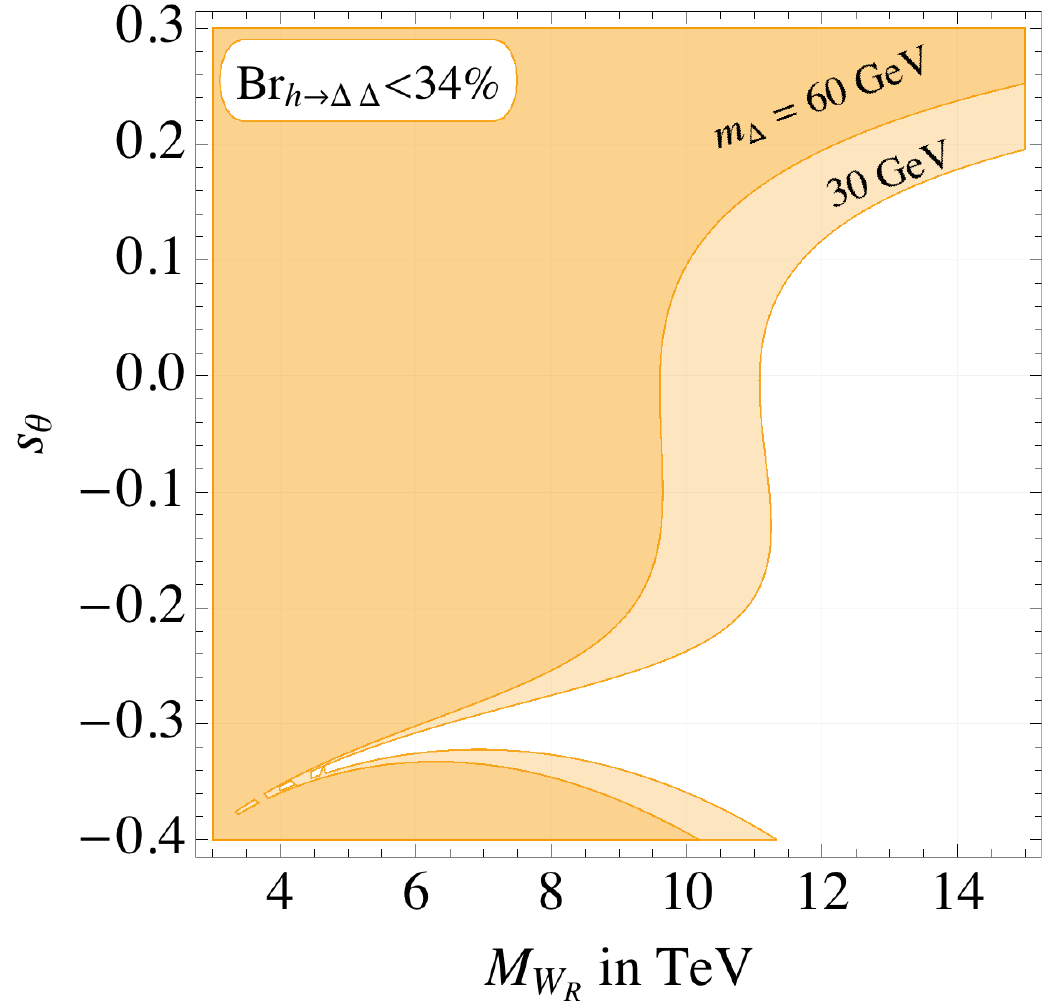} \hfill
  \includegraphics[height=6.1 cm]{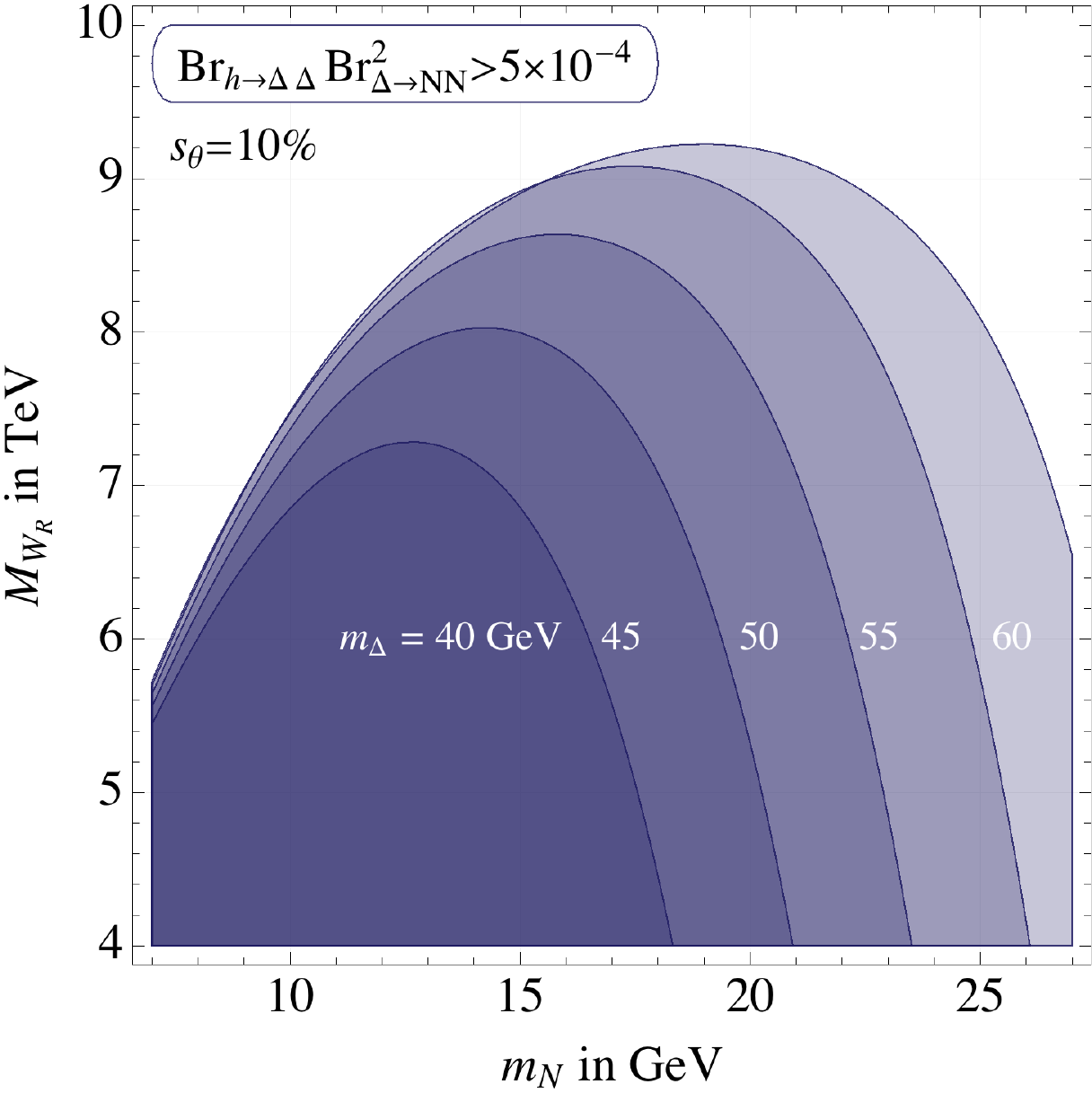}
  \caption{Left: Regions of $h\to \Delta \Delta$ allowed by the current Higgs data~\cite{deFlorian:2016spz} with $m_H = 17 \text{ TeV}$ and $r_{++}=0.3$. Right: Regions of larger than $5 \times 10^{-4}$ branching ratio of Higgs decaying to four Majorana neutrinos. Contours correspond to different $m_\Delta$ and $s_\theta = 10 \%$.}
  \label{figBrhDD}
\end{figure}

%
\paragraph{Decays to scalars.} If there are scalars below $m_\Delta$, two body decays open up, e.g. $\Delta \to hh$. At the same time, if $\Delta$ happens to be below half the Higgs mass, the $h \to \Delta \Delta$ mode appears. Decay rates for both cases are
\begin{align} \label{eqGamhDSS}
  \Gamma_{h \to S \overline S} &= \frac{1}{16 \pi \left(1 + \delta_{S \overline S} \right)} \frac{1}{m_h} v_{hSS}^2 \beta_{hS}^{1/2}, &
  \Gamma_{\Delta \to S \overline S} &= \frac{1}{16 \pi \left(1 + \delta_{S \overline S} \right)} \frac{1}{m_\Delta} v_{\Delta SS}^2 \beta_{\Delta S}^{1/2}, 
\end{align}
where $\delta_{S \overline S} = 0 (1)$ for charged (neutral) particles and $\beta_{iS} = 1- \left(2 m_S/m_i\right)^2$, $i = h,\Delta$. The Higgs tri-linear couplings $v_i$ entering in Eq.~\eqref{eqGamhDSS} are discussed at some length in the Appendix~\ref{SecAppTripleH}. 

The quantitative effect of Higgs mixing on scalar decay modes is understood from Figs.~\ref{figBrD} and~\ref{figBrhDD}. Above the electroweak scale for $m_\Delta \gtrsim 300 \text{ GeV}$, the $\Delta \to hh$ channel opens up rapidly (Fig.~\ref{figBrD}, left). Conversely, in the low mass region when $m_\Delta < m_h/2$, the $h \to \Delta \Delta$ rate becomes sizeable (Fig.~\ref{figBrhDD}, left).  Since the $\Delta$ proceeds to decay to pairs of $N$s, one ends up with quite an exotic Higgs decay to four Majorana neutrinos. The relevant quantity that sets the event rate is the $\text{Br}^{\,}_{h \to \Delta\Delta} \text{Br}_{\Delta \to NN}^2$, shown on the right of Fig.~\ref{figBrhDD}. 

From the left plot of Fig.~\ref{figBrhDD} it is also clear that the pair production of $\Delta$ is more favourable in the case of positive $\theta$, while for negative $\theta$ the $v_{h\Delta\Delta}$ is mildly suppressed, at low scale $M_{W_R}$, as discussed in the appendix after Eq.~\ref{vhDDlin}.

%
\paragraph{The di-photon and $Z \gamma$ channels.} Decay rates for these two radiative processes are
\begin{align} \label{eqDGagaZga}
  \Gamma_{\Delta \to \gamma \gamma} &= \frac{m_\Delta^3}{64 \pi} \left( \frac{\alpha}{4 \pi} \right)^2 \left| F_\Delta \right|^2, 
  &
  \Gamma_{\Delta \to Z \gamma} &= \frac{m_\Delta^3}{32 \pi c_w^2} \left( \frac{\alpha}{4 \pi} \right)
    \left( \frac{\alpha_w}{4 \pi} \right) \left| G_\Delta \right|^2 \beta_{\Delta Z}^3,
\end{align}
where $\beta_{\Delta Z} =  1- M_Z^2/m_\Delta^2$, $N_f =3(1)$ for quarks (leptons), $Q$ is the electromagnetic charge and $\hat v = T_{3L}/2 - Q s_w^2$. The loop functions are
\begin{align} \label{EqFDh}
  F_\Delta &= c_\theta F_{W_R} + \sum_S v_{\Delta S S}^{\,} Q_S^2 F_S^{\,}, 
    - s_\theta F_W - s_\theta  \sum_f N_f^{\,} Q_f^2 F_f^{\,},
  \\ \label{EqGDh}
  G_\Delta &= c_\theta G_{W_R} + \sum_S  v_{\Delta S S}  \, Q_S \hat v_S \, G_S
    - s_\theta G_W - s_\theta \sum_f N_f Q_f \hat v_f G_f,
\end{align}
with expressions for $F$ and $G$ given in Eqs.~\eqref{eqLoopFf}-\eqref{eqLoopGf} of Appendix~\ref{SecAppLoops} and tri-linear couplings $v_{\Delta S S}$ in Appendix~\ref{SecAppTripleH}.

The dominant contributions to these rates come from the doubly charged scalar exchange and the mixing with the SM Higgs. This mixing brings in a fairly large quartic ($\alpha_3$) that is required in low scale LRSM by the tree-level FCNC $H$ exchange, see e.g.~\cite{Maiezza:2016bzp}. Despite this enhancement, the radiative rates are rather small when $s_\theta \gtrsim 10^{-3}$ and their $\text{Br}_{\Delta \to \gamma \gamma, Z \gamma} \lesssim 1 \permil$ are sub-dominant, as seen on Fig.~\ref{figBrD}. Below the $Z$ mass, the di-photon channel dominates but as soon as kinematically allowed, the $Z \gamma$ takes over.

%
%
\section{Production at the LHC} \label{SecProduction}

Let us turn to $\Delta$ production at hadronic colliders in the Higgs portal and gauge-mediated channels. It can be fairly easily produced at the LHC, once the mixing with the SM Higgs is present. The three available production modes are single, associated and pair production. 

%
\medskip
\noindent{\bf Single} production of $\Delta$ occurs through the Higgs mixing. In addition to the associated $W \Delta, Z \Delta$, heavy quark $QQ \Delta$ and the VBF $qq \Delta$ channels, the dominant mode is the gluon fusion one, known to $\text{N}^3\text{LO}$~\cite{Anastasiou:2016hlm}. Even though the cross-section decreases steeply with $m_\Delta$, it provides an appreciable production rate with $\Delta$ mass well in the $\mathcal O(100) \text{ GeV}$ range.
 
\begin{figure}
  \centering
  \includegraphics[height=.32\columnwidth]{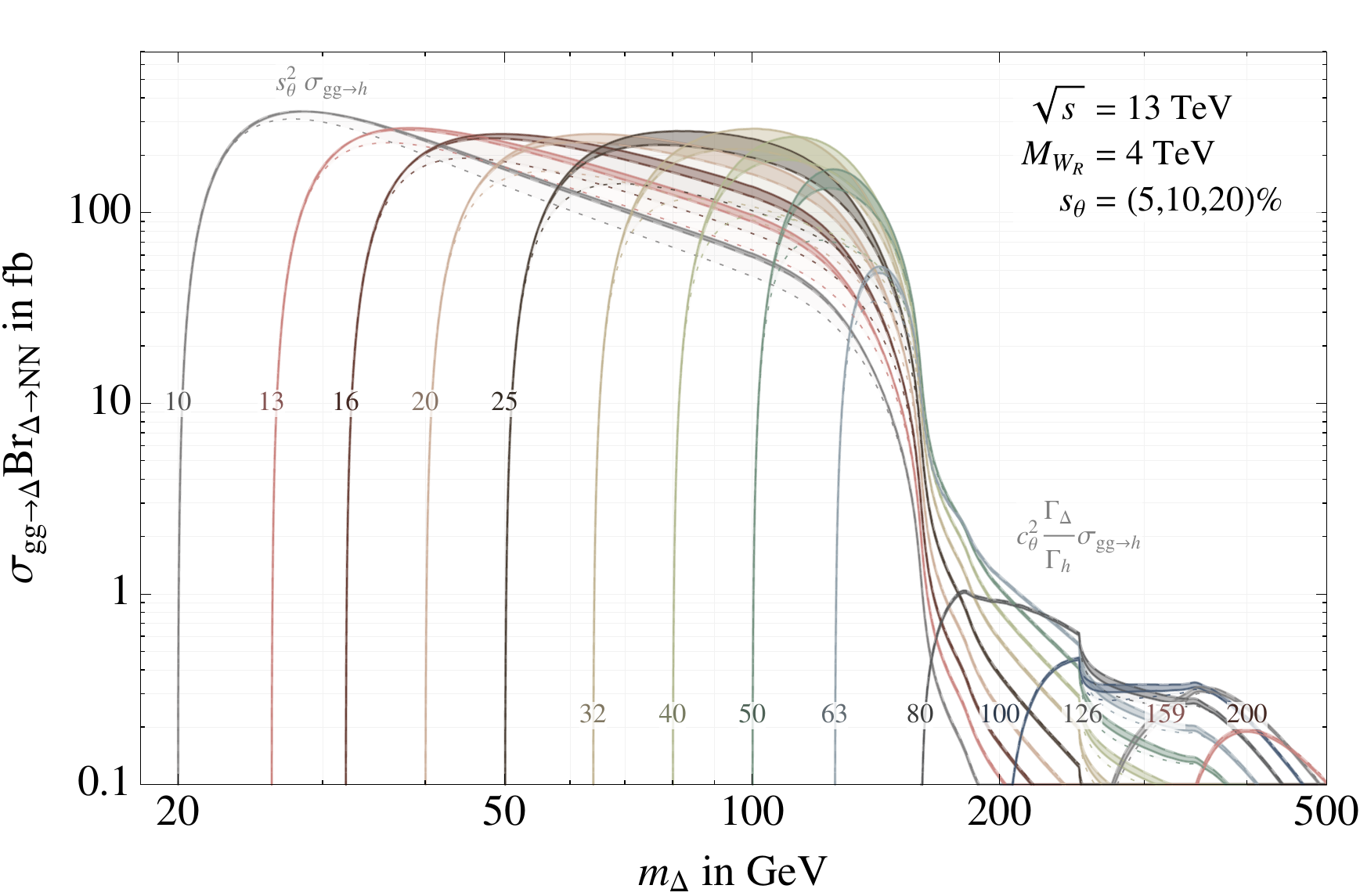} \hfill
  \includegraphics[height=.32\columnwidth]{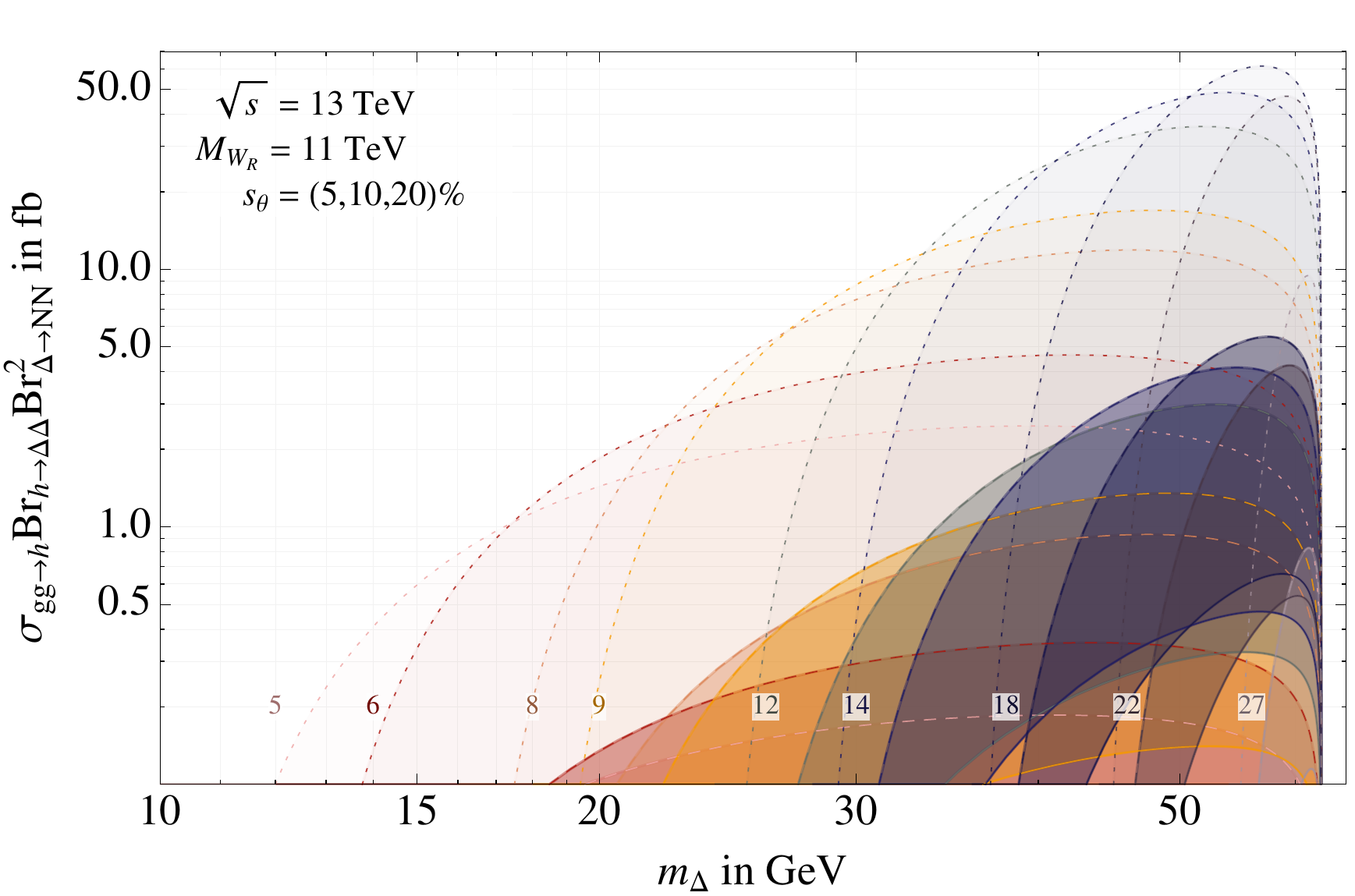}
  \caption{Left: Production cross-section of heavy Majorana neutrino pairs via $\Delta$ gluon fusion. Colors correspond to different masses of $N$, ranging from $10$ to $200 \text{ GeV}$. Right: Production of four $N$s in $gg \to h \to \Delta \Delta$ for different masses of $N$ down to $5 \text{ GeV}$. In both plots, $\sqrt s = 13 \text{ TeV}, m_H = 17 \text{ TeV}, r_{++} = 0.3$ and for each $m_N$ the values of $s_\theta$ considered are 5, 10 and 20\% with respective dotted, dashed and solid line.}
  \label{figSigBrhgg3TeV}
\end{figure}

The $NN$ event rate is determined by the $\text{Br}_{\Delta \to NN}$ from Fig.~\ref{figBrD} and allows for copious production of heavy Majorana neutrinos. While the $\text{Br}_{\Delta \to NN}$ increases with $s_\theta$, the production is reduced and there is an interplay between the two. Below the $VV$ threshold, the Higgs width is relatively small, such that $\text{Br}_{\Delta \to NN} = \mathcal O(1)$ and the event rate scales with $s_\theta^2$. Once the gauge boson threshold is crossed, the Higgs width increases substantially and suppresses the $\text{Br}_\Delta$
\begin{equation} \sigma_{gg \to \Delta \to NN}  \simeq \sigma_{gg \to h}(m_\Delta) \times
    \begin{cases} 
      s_\theta^2, \, &m_\Delta \lesssim 2 M_W, \\
      c_\theta^2 \, \frac{\Gamma(\Delta \to NN)}{\Gamma_h}, \, &m_\Delta \gtrsim 2 M_W.
    \end{cases}
\end{equation}
This behaviour is seen in Fig.~\ref{figSigBrhgg3TeV}, where the opening of SM channels leads to a sharp reduction of the $NN$ production rate at about 160 GeV. Above this threshold the dominant channels are $VV$, $t \overline t$ and $hh$, as seen from Fig.~\ref{figBrD}. For $m_\Delta \gtrsim 160 \text{ GeV}$, the rate is nearly insensitive to $s_\theta$ and decreases only for small mixing $s_\theta \lesssim 0.03$.

It is also clear from Fig.~\ref{figSigBrhgg3TeV} that below 160 GeV the production of $N$ pairs is on $\mathcal O(10-100) \text{ fb}$, even for relatively small mixing. This is a promising prospect for the LHC and prompts the collider study in \S \ref{SecLHC}.

\begin{figure} \centering
  \includegraphics[height=.315\columnwidth]{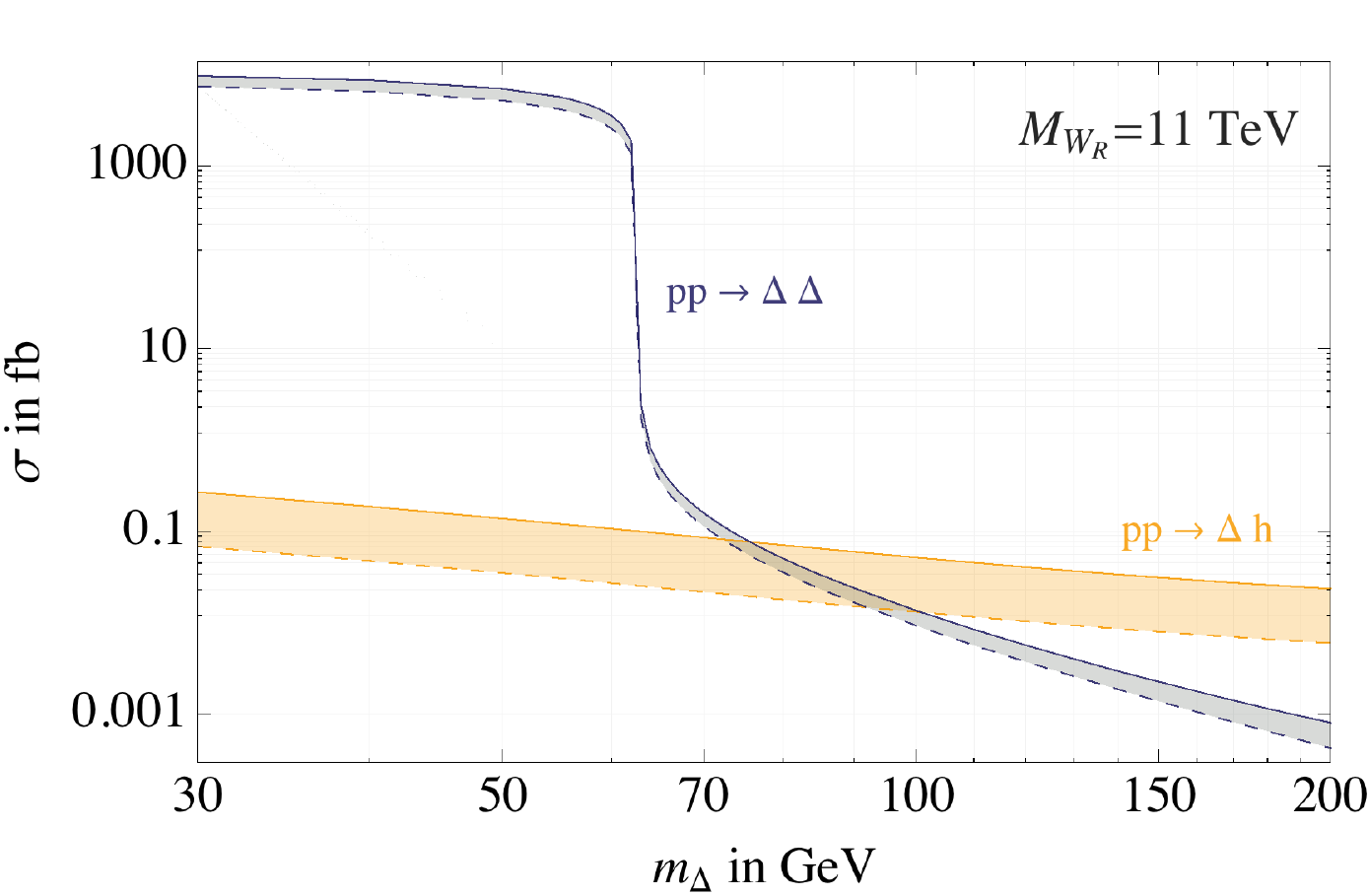}\hfill
  \includegraphics[height=.315\columnwidth]{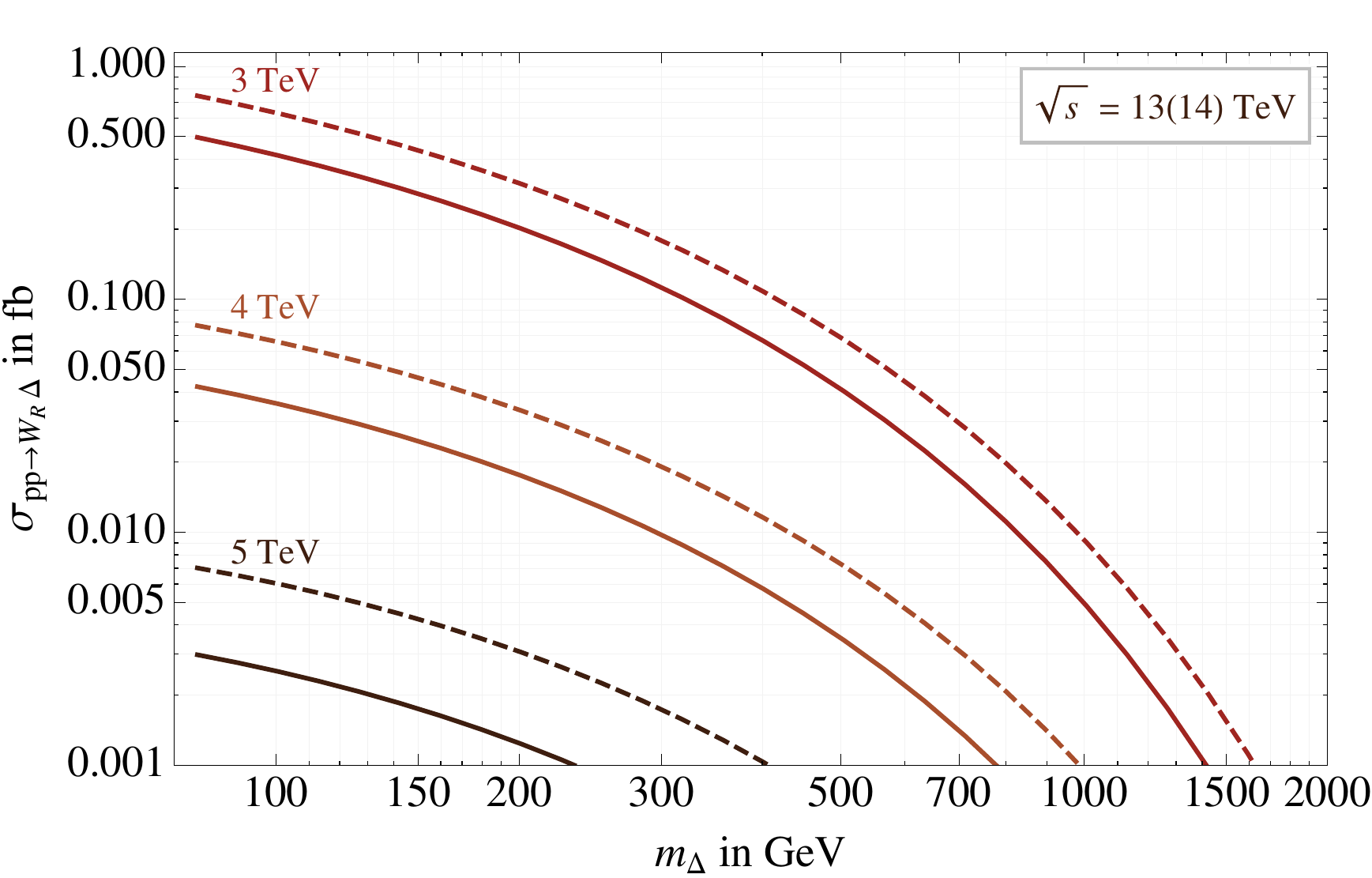}
  \caption{Left: The pair (associated) production of $\Delta$ in blue (yellow) through the Higgs portal at 13 TeV. Shading describes the variation of $s_\theta$ from $5 \%$ in dashed to $10 \%$ in solid lines, $M_{W_R} = 11 \text{ TeV}$. Right: Cross-section for Drell-Yan production of $W_R \Delta$ for $M_{W_R} = 3,4,5 \text{ TeV}$ and $\sqrt s = 13 (14) \text{ TeV}$ in solid (dashed).}
  \label{figDProd}
\end{figure}

%
\medskip
\noindent
{\bf Associated {\rm and} pair} production of $\Delta h$ and $\Delta$ pairs proceed through the off-shell Higgs~\cite{Hessler:2014ssa, Dev:2016dja} ($\Delta^*$ is smaller due to~\eqref{eqVhDD} and~\eqref{eqVDDD}) with parton-level cross-sections
\begin{equation} \label{eqSigAssocPair}
  \hat \sigma_{gg \to \Delta S} \simeq \frac{c_\theta^2}{64 \pi (1+\delta_{\Delta S})} \, \hat s
  \left(\frac{\alpha_s}{4 \pi}\right)^2 \frac{v_{hS\Delta}^2}{
  (\hat s - m_h^2)^2 + \hat s \Gamma_h^2} \left| F_b + F_t \right|^2 \sqrt{\beta_{\hat s \Delta S}} ,
\end{equation}
where $\beta_{\hat s \Delta S} = 1 - 2 \left( m_S^2 + m_\Delta^2 \right)/\hat s + \left(m_S^2 - m_\Delta^2 \right)^2/\hat s^2$ and $F_f(4 m_f^2/\hat s)$ is defined in Eq.~\eqref{eqLoopFf}. Using Eqs.~\eqref{eqVhhh}-\eqref{eqVDDD}, the associated and pair production rates of $\Delta$, (and $\Delta_R^{++}$, $\Delta_L$) are obtained by integrating over the PDFs and shown on the left panel of Fig.~\ref{figDProd}. Pair production of $\Delta$ is dominant below the Higgs decay threshold, while the associated production stays below 1 fb and we do not dwell on it any further.

Pair production of $\Delta$ in Higgs decays gives a striking prospect for the LHC since their subsequent decays to $NN$ would result in a final state with four heavy Majorana neutrinos.\footnote{The associated $h \Delta \to \Delta\Delta\Delta$ and SM Higgs pair production of $hh \to \Delta\Delta\Delta\Delta$ may lead to even more spectacular signals with $6N$ and $8N$ in the final state, with up to 8 displaced vertices. The estimated rates are $\lesssim 10^{-4} \text{ fb}$, therefore we do not study these in more detail.} The plot on the right of Fig.~\ref{figSigBrhgg3TeV} shows that below the $m_h/2$ threshold, a sizeable cross-section can be expected for this exotic final state.

%
\paragraph{Gauge production.} For $m_\Delta \ll M_{W_R}$, there exist a rather small gauge mediated production from $W_R$ (and $Z_{LR}$ that is heavier in the minimal model). The kinematics of this channel is quite different from the gluon fusion channel and may be triggered more efficiently. The main channels are through the Drell-Yan process in the $s$-channel and vector boson fusion of $W_R$.\footnote{These are also the production channels for $\Delta_R^{++}$, studied in~\cite{Maalampi:2002vx}, in which case an additional combinatorial factor of two is present in the amplitude.} The production cross-section for an on-shell $W_R$ and $\Delta$ is
\begin{align}
  \hat \sigma_{q_i q_j \to W_R \Delta}\left(\hat s \right) &= \frac{\pi}{8} \alpha_w^2 \left|V_{i j}\right|^2
  \frac{\sqrt \lambda (\lambda/\hat s^2 + 12 M_{W_R}^2/ \hat s)}{( \hat s - M_{W_R}^2 )^2 + \Gamma_{W_R}^2 M_{W_R}^2},
  \\
    \sigma_{pp \to W_R \Delta} &= \frac{1}{9} \sum_{ij} \int_{\tau_{\min}}^1 d \tau \int_\tau^1 d x \frac{1}{\tau x} 
   f_{q_j} \left(x, \mu_F \right) f_{q_i} \left(\tau/x, \mu_F \right) \, \hat \sigma_{q_i q_j \to W_R \Delta}(\tau s),
\end{align}
where $\lambda = m_\Delta^4 - 2 m_\Delta^2 (\hat s + M_{W_R}^2 ) + (\hat s - M_{W_R}^2 )^2$ and  $\tau_{\min} = ( m_\Delta^2 + M_{W_R}^2)/s$, the sum over all (anti)quarks is taken with $f_q$ the usual partonic functions and $\mu_F^2 = m_\Delta^2+M_{W_R}^2$. As seen in Fig.~\ref{figDProd}, the cross-section is $\lesssim 0.5 \text{ fb}$, and is increasingly suppressed by parton distributions at higher $M_{W_R}$.  Even though this cross section is small, it is interesting that $\Delta$ strahlung by $W_R$ gives rise to final states with $\Delta L=2$ when $W_R$ decays to jets, but also to $\Delta L=4$ when $W_R$ decays to $\ell N$, as in the KS process. 

Finally, it is worth commenting that new vertices such as $W_L W_R \Delta$ appear in the presence of LR gauge boson mixing. This gives rise to new channels of production such as $W_R \to \Delta W_L$.  Because such mixing is small ($\lesssim 10^{-4}$) also such processes are negligible for the present analysis. Only in the limit of vanishing Higgs mixing $s_\theta\to 0$ they can start to play a role, e.g.\ in the $\Delta$ decay. 
 
%
%
\section{Signals at the LHC} \label{SecLHC}

\begin{figure}
  \centering \hspace{.5cm}
  \includegraphics[width=.4\columnwidth]{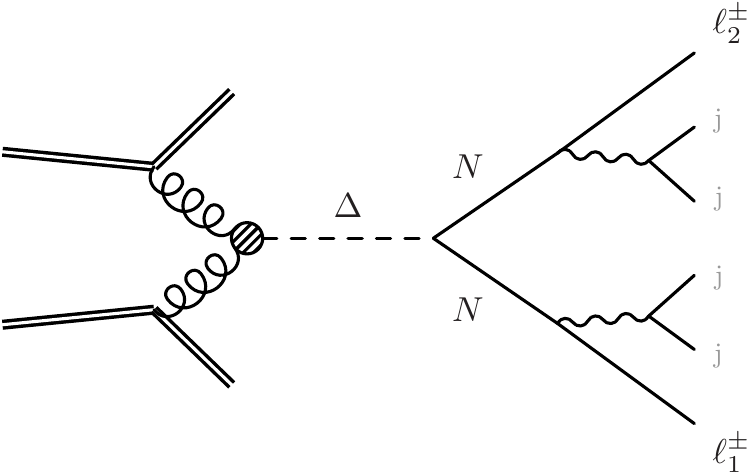} \hfill
  \includegraphics[width=.4\columnwidth]{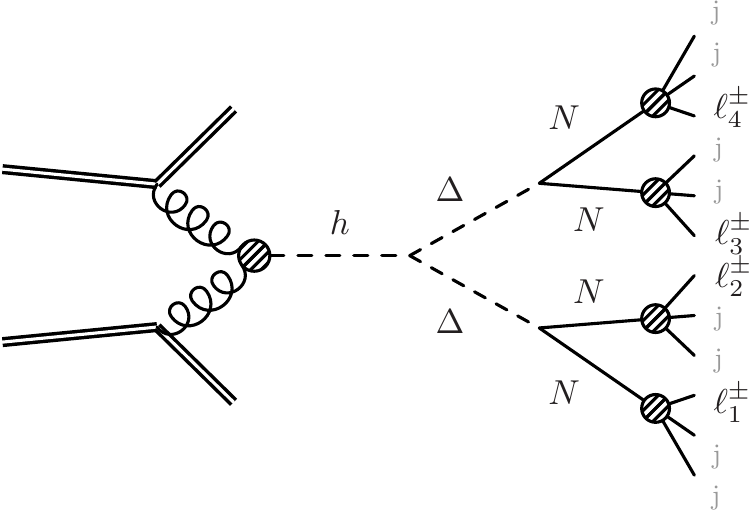} \hspace{.5cm}
  \caption{Left: Feynman diagrams for pair production of $N$ through the $\Delta$ resonance that leads to two same-sign leptons and a $\Delta L = 0,2$ signal. Right: Pair-production of $\Delta$ via an exotic Higgs decay with four leptons in final states with $\Delta L = 0,2,4$.}
  \label{figFeynDiag}
\end{figure}

As discussed in the section above, the two most promising $\Delta$-mediated processes to look for $N$ at the LHC, are the single gluon fusion and $\Delta$ pair production from $h$ decay with respective diagrams shown on Fig.~\ref{figFeynDiag}. This section contains a study of efficiencies, relevant backgrounds, signal characteristics and estimated sensitivities for the two cases of interest.

For the signal generation, an updated extension~\cite{LRMSMixModel} of the FeynRules~\cite{Alloul:2013bka} implementation of the LRSM~\cite{Roitgrund:2014zka} was used. The events were then generated at LO using MadGraph 5~\cite{Alwall:2014hca}, hadronized with Pythia 6~\cite{Sjostrand:2006za}\footnote{Pythia 8 gives statistically the same result.} and passed to Delphes 3~\cite{deFavereau:2013fsa} for detector simulation. The gluon fusion production cross-section was rescaled to the NNLO+NNLL QCD and NLO EW value recommended in~\cite{HiggsXSecWG}. The backgrounds were simulated with MadGraph 5 at LO and rescaled with a common constant $K$-factor of 1.5 for $t \overline t$~\cite{TTbarXSec}, $WZ$~\cite{Melia:2011tj} and $Zh$~\cite{HiggsXSecWG} and a factor of 1.34 for $ZZ$, $Wh$ and the other $VVV, VVh$ processes.

To estimate the detector response, the default Delphes 3 ATLAS card was modified with electrons~\cite{ATLAS:2016iqc} and muons~\cite{Aad:2016jkr} treated separately and with the experimental values on triggering thresholds~\cite{ATLASTrig} taken into account.
\begin{description}
  \item[Electrons.] For electrons, we apply the measured efficiencies from~\cite{ATLAS:2016iqc} with no electrons below $p_T < 6 \text{ GeV}$. We take into account the mono(di)-electron triggers with $24(12) \text{ GeV}$ and define tight(loose) isolation with $p_T^{\text{varcone20}} < 0.06 (0.15)$~\cite{ATLAS:2016iqc}. Because of these requirements, the electron signal is reduced with respect to the muon case, especially in the low $m_\Delta$ regime. 
  \item[Muons.] In case of muons, the efficiencies are taken from~\cite{Aad:2016jkr} with zero efficiency for $p_T$ below 5 GeV. In the major part of parameter space the single muon trigger ($p_{T \mu_1} > 20 \text{ GeV}$) is more efficient, while for $m_\Delta \lesssim 80 \text{ GeV}$ the di-muon trigger with $p_{T \mu_2} > 10 \text{ GeV}$ turns out to be better. The overall selection efficiency of triggering on the signal goes from $10 \, \%$ for low masses to $\sim 80 \, \%$ at $m_\Delta \simeq 160 \text{ GeV}, m_N \simeq 75 \text{ GeV}$. Again, the isolation criteria are $p_T^{\text{varcone30}} < 0.06(0.15)$ for tight (loose) electrons.
\end{description}
Jets were clustered using the anti-$k_T$ jet algorithm with $\Delta R = 0.4$ and $p_{Tj min} = 20\,\text{GeV}$. Cuts, event counting and sensitivity estimates were performed with the help of MadAnalysis~5~\cite{Conte:2012fm}.

%
\subsection{$pp \to \Delta \to N N$} \label{subsecDto2N}

Because of the Majorana character of $N$, each on-shell heavy neutrino decays through the off-shell $W_R$ to a charged lepton, or an anti-lepton with equal probability, and two jets. Thus in half or the events one gets LNV signals and benefits from the low SM backgrounds.

As discussed in the sub-section~\ref{SubSecFermions}, the leptonic mixing in $V_R$ may be non-zero which would lead to LFV final states, while being in agreement with low energy constraints from Eq.~\eqref{eqGamDLFV}. For the sake of simplicity, we perform our analysis without flavor mixing, i.e. by setting $V_R$ to be diagonal. Furthermore, since the region of interest for $m_N$ here is below $80 \text{ GeV}$ and $V_R^q \simeq V_{ckm}$ in LRSM with restored LR parity~\cite{Zhang:2007fn, Maiezza:2010ic, Senjanovic:2014pva}, one does not expect $b$ jets in the final state. 

As a result, the single production of $\Delta$ leads to the final state with two same or opposite-sign leptons together with 4 non-$b$ jets, as shown in Fig.~\ref{figFeynDiag} left. This event topology is the same as the LNV decay of the 125 GeV Higgs $h \to N N$ studied in~\cite{Maiezza:2015lza}.

\begin{table} 
\centering
\begin{tabular}{| r | c c c c c c c c c c |}
 \hline 
  			& \cellcolor{el} $t \overline t$ & \cellcolor{el} $t \overline t h$ &  \cellcolor{el}$t \overline t Z$ &
			\cellcolor{el} $t \overline t W$ & \cellcolor{el} $WZ$ & \cellcolor{el} $Wh$ & \cellcolor{el} $ZZ$ & 
			\cellcolor{el} $Zh$ & \cellcolor{el} $W W jj$ &  \cellcolor{el} \text{Jet Fake}
  \\ \hline
$e e + n_{\color{black}j}^{\color{nicered}*}$ 
				&  806 & 4 & 5 & 26 & 1241 & 87 & 147 & 16 & 1.5 &2651 
  \\
$\slashed E_T$	 	&  313 & 0.5  & 0.7 & 3 & 400 & 21 & 129 & 7 & 0.2 & 782 
  \\
$p_T$			&  112 & 0.2 & 0.1 & 0.7 & 174 & 8.4 & 63 & 4 & 0.05 & 284
  \\
$m_T$ 			&  60 & 0.1  & 0.04 & 0.3 & 80 & 4 & 56 & 2 & 0.03 & 106
  \\
$m^{\text{inv}}$		&  35	 & 0.03 & 0.03 & 0.2 & 25 & 2 & 36 & 2 & 0 & 80
\\
$l_{Te} $  			& 0  & 0 & 0  &  0  & 0.7 & 0.1 &  0.9 & 0.05 &  0.001& 2

  \\ \hline \hline 
  			& \cellcolor{mu} $t \overline t$ & \cellcolor{mu} $t \overline t h$ &  \cellcolor{mu}$t \overline t Z$ &
			\cellcolor{mu} $t \overline t W$ & \cellcolor{mu} $WZ$ & \cellcolor{mu} $Wh$ & \cellcolor{mu} $ZZ$ & 
			\cellcolor{mu} $Zh$ & \cellcolor{mu} $W W jj$ &  \cellcolor{mu} \text{Jet Fake}
  \\ \hline
$\mu \mu + n_{\color{black}j}^{\color{nicered}*}$ 
				& 670 & 4  & 6 & 32 & 750 & 133 & 68 & 16 & 2 & 1676
  \\
$\slashed E_T$	 	& 130 & 0.5 & 0.9 & 3.5 & 200 & 32 & 33 & 6 & 0.3 & 391
  \\
$p_T$			& 57 & 0.2  & 0.2 & 1 & 95 & 17 & 16 & 3 & 0.1 & 152
  \\
$m_T$ 			& 32 	& 0.1  & 0.1 & 0.5 & 51 & 9 & 12 & 2 & 0.05 & 49
  \\
$m^{\text{inv}}$		& 17	& 0.04  & 0.04 & 0.2 & 23 & 5 & 8 & 1 & 0.01 & 40
  \\
$l_{T\mu}$  		& 0  & 0 & 0  &  0  & 1.4 & 0.4 &  1 & 0.15 &  0.005 & 3 
\\
 \hline
\end{tabular}
\caption{SM background processes at 13 TeV and $100 \text{ fb}^{-1}$ for the same-sign tight di-leptons and $n_j = 1-3$ final state, see text for the cut descriptions. $^{\color{nicered}*}$The same-sign lepton selection includes $p_T$ trigger for di-muons at 10 GeV and di-electrons at 12 GeV.}
\label{tabBckgDL2}
\end{table}

\paragraph{Basic selections and backgrounds.} For the signal event selection we demand exactly two same-sign, same-flavor leptons and $n_j = 1, 2$ or $3$ jets. We do not attempt to extract the signal in the opposite-sign case due to the overwhelming SM background, although for large $N$ displacements the inclusion of this channel may be feasible.

In both $e$ and $\mu$ channels the same cuts as in~\cite{Maiezza:2015lza} are imposed to further enhance the sensitivity. Their impact is seen in Table~\ref{tabBckgDL2}, here we describe them in a bit more detail. As seen from Fig.~\ref{figFeynDiag}, there is no missing energy at the parton level, therefore we demand $\slashed{E}_T < 30\, \text{GeV}$, which is the ``$\slashed{E}_T$'' cut in Table~\ref{tabBckgDL2}. The ``$p_T$'' cut refers to the transverse momentum of the leading-$p_T$ charged lepton: that lepton is still fairly soft, therefore we impose $p_T(\ell_1) < 55 \text{ GeV}$. In addition, we impose a ``$m_T$'' cut, by requiring the transverse mass $m^T_{\ell \slashed p_T} < 30\,\text{GeV}$ (where $m^T_{\ell \slashed p_T}$ is the transverse invariant mass of the charged lepton momentum with the missing momentum). We also impose invariant masses $m_{\ell \ell} < 80\,\text{GeV},$ $m_{\ell \slashed p_T} < 60\,\text{GeV}$, collectively called the ``$m^{\text{inv}}$'' cut. This selection efficiency turns out to be around a few percent or less.\footnote{The $W \Delta$ and $Z \Delta$ selection efficiency is slightly better, but the cross-sections are considerably lower. In the end, these two channels contribute around $3-5 \%$ to the gluon fusion one with the same event selection procedure.} 

The major suppression factor is due to the soft momenta of final state particles $p_T \simeq m_\Delta/6$. It becomes evident in the low $m_\Delta$ region and is more pronounced for electrons due to higher $p_T$ thresholds.

\begin{figure} \centering
  \includegraphics[width=.325\columnwidth]{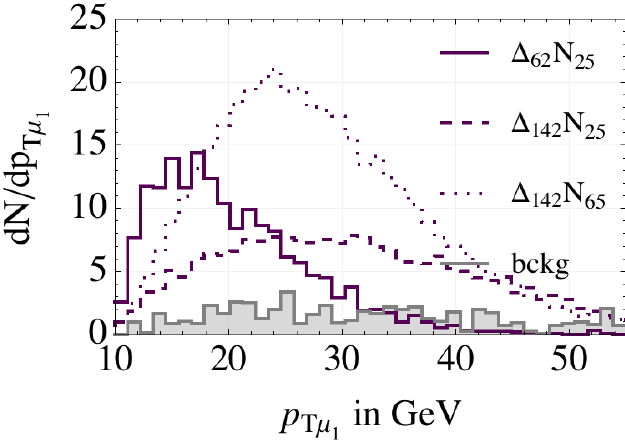} \hfill
  \includegraphics[width=.332\columnwidth]{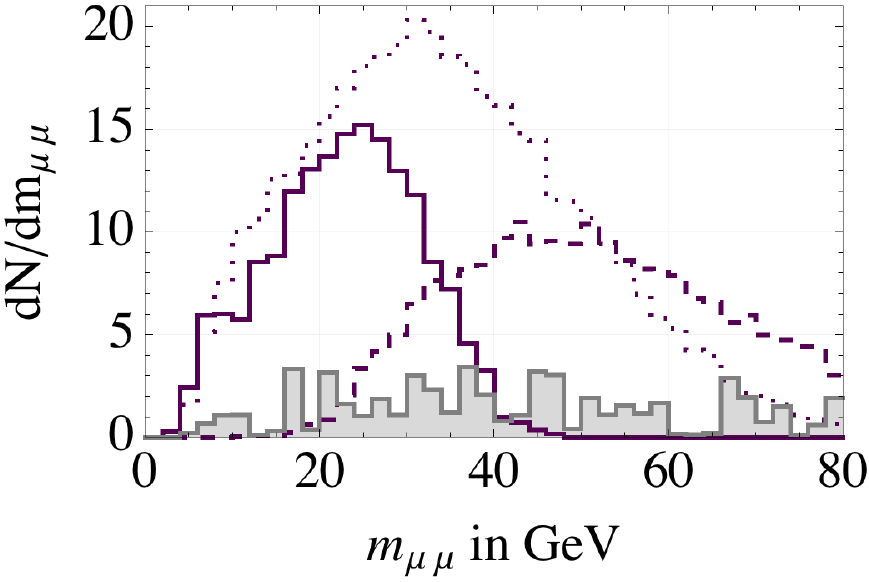} \hfill 
  \includegraphics[width=.325\columnwidth]{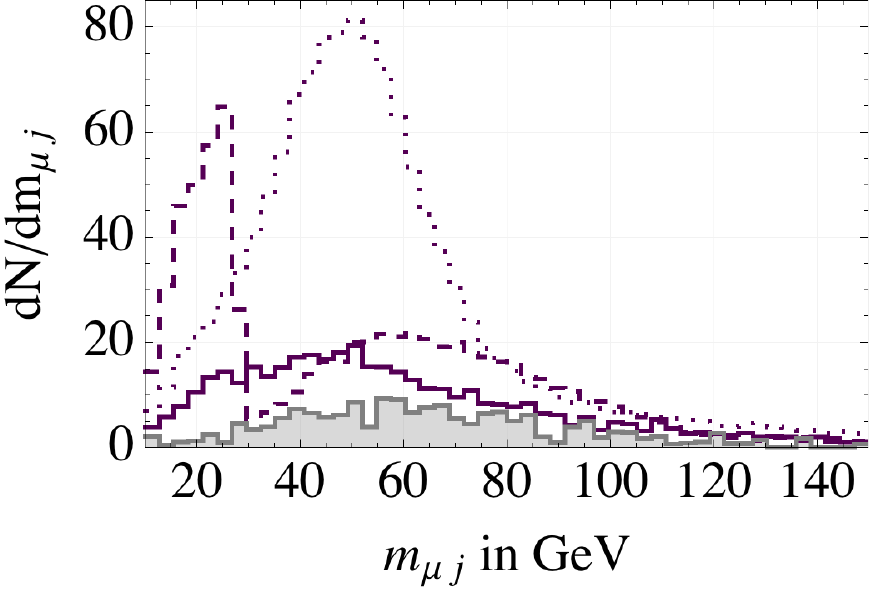}
  
  \vspace{1mm}
  \includegraphics[width=.333\columnwidth]{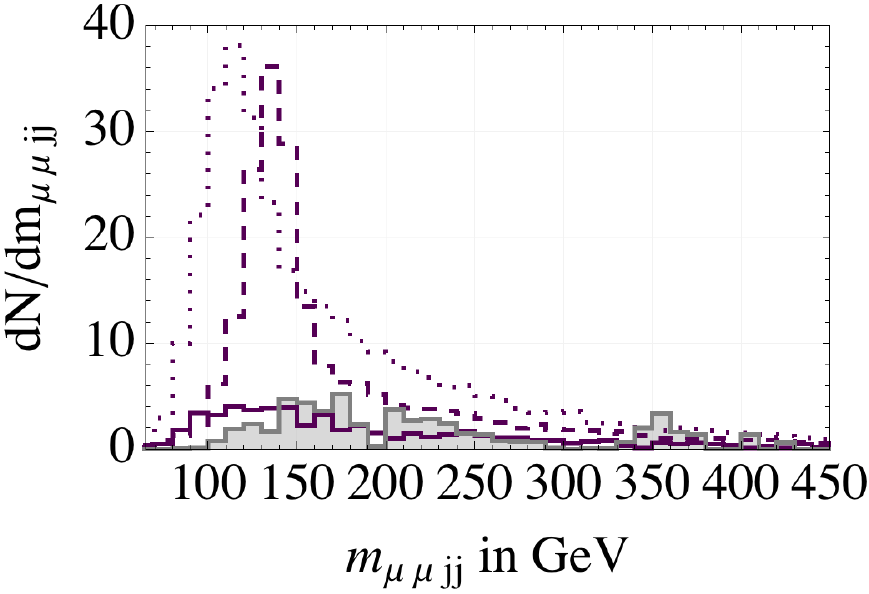} \hfill
  \includegraphics[width=.319\columnwidth]{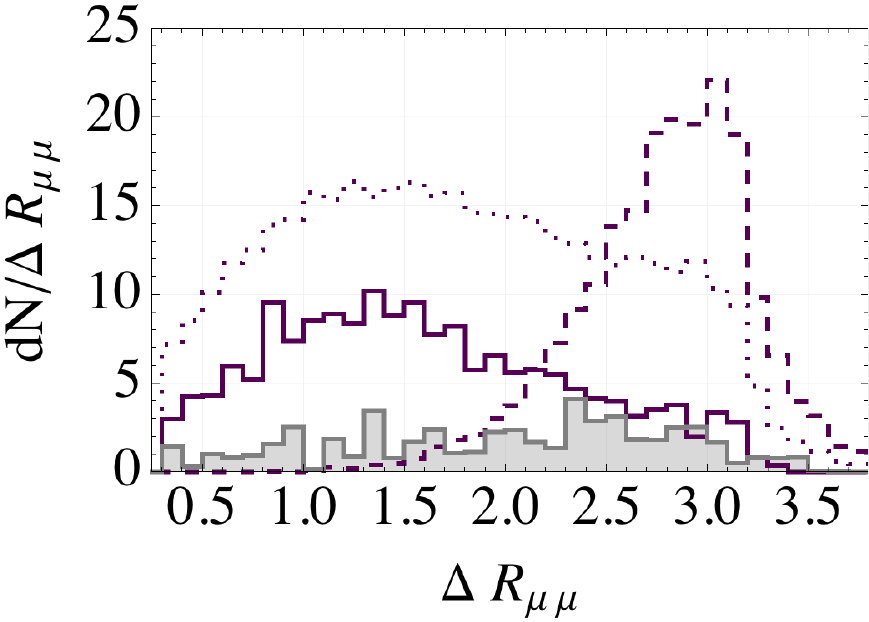} \hfill
  \includegraphics[width=.333\columnwidth]{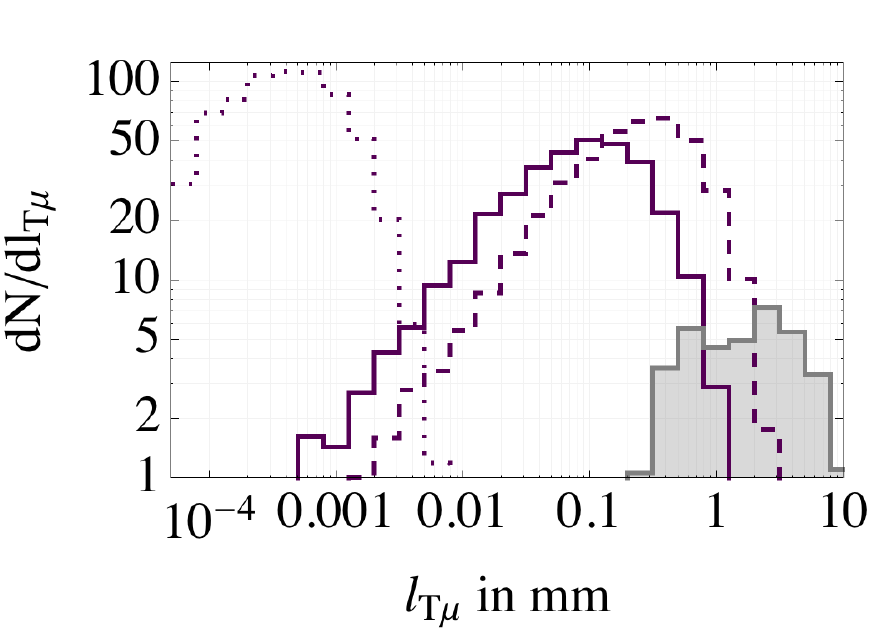} 
  \caption{Event distributions after basic cuts for three signal points $m_\Delta = (62, 142, 142) \text{ GeV}$ and $m_N = (25, 25, 65) \text{ GeV}$ in solid, dashed and dotted lines. Throughout, $M_{W_R} = 3 \text{ TeV}, s_\theta = 0.05$ and $\sqrt s = 13 \text{ TeV}, \mathcal L = 100 \text{ fb}^{-1}$. The gray shaded area corresponds to the sum of genuine backgrounds: $t \overline t X$, $VX$ and $WWjj$ as in Tab.~\ref{tabBckgDL2}. In the lepton displacement plot (bottom-right) its non-prompt component is evident, while the prompt component is outside the plot, at zero displacement.}
   \label{figKinDnn}
\end{figure}

We consider a number of SM backgrounds coming from top and multi-production of gauge bosons, see Tab.~\ref{tabBckgDL2}. The most relevant one turns out to be $t \overline t$ with one prompt lepton coming from $W$ and the other same-sign lepton from $b$ decay which is typically displaced. The additional $t \overline t X$, $X = W, Z, h$ are further suppressed. A similar sized contribution comes from the $WZ, ZZ$ and $Wh$ channels, while $Zh$ and $W^\pm W^\pm jj$ are again smaller. In all these backgrounds, one of the leptons is prompt and in some cases the second one is displaced, as seen in the lower right plot of Fig.~\ref{figKinDnn}.

%
Apart from these sources, additional contributions to the background come from jet fakes, charge mis-identification and secondary photo-production. Jet fakes are the most significant and we make an effort to estimate their rate, as described in the Appendix~\ref{SecJetFakes}. The main sources of fake leptons turn out to be $W+jjj$, $VV + \text{jets}$ and QCD jets. The latter has a prohibitively large cross-section to simulate and should be estimated from data. The behaviour of the background with imposed cuts is clear from Tab.~\ref{tabBckgDL2} for muons and electrons. Similar pattern of prompt plus displaced leptons is also evident in this background. 

%
Charge mis-identification may give a sub-dominant contribution in the electron channel~\cite{Khachatryan:2016olu}, while for muons the mis-id rate is negligible~\cite{Khachatryan:2015gha}. Moreover, the mis-id rate is smaller for low $p_T$ electrons, which is the case at hand. The underlying processes are similar to the ones we consider for jet fakes in the Appendix~\ref{SecJetFakes}. These are prompt (e.g. $Z$+jets) and lead to prompt same-sign lepton pairs, therefore we expect them to be eliminated by the $e$ displacement cut. While charge mis-id is not explicitly included in the detector simulation, the corresponding kinematical distributions are very similar to those coming from jet-fakes; therefore, we take into account this background by over-estimating the jet-fake rates in the electron case to fully reproduce the experimental data in Fig.~\ref{figJFValid}.

\paragraph{Further signal characteristics.} The $\Delta \to NN$ channel has a number of revealing features, similar to the $h \to NN$ case~\cite{Maiezza:2015lza}. Apart from the LNV character of the final state due to Majorana neutrinos, one may reconstruct $M_N$ and $m_\Delta$ by observing the invariant masses of leptons and jets, $m_{\ell\ell4j}$ and $m_{\ell2j}$ in different flavor channels. 

After detector simulation and jet clustering, these characteristics are somewhat degraded. Nevertheless, some of the features remain, for example peaks in the invariant mass distributions. As seen in Fig.~\ref{figKinDnn}, invariant masses of $\ell \ell$, $\ell j$ and $\ell \ell j j$ as well as the muon separation $\Delta R$ may be useful to further discriminate against the background.

\paragraph{Displacement and sensitivity.} Another prominent feature of heavy neutrinos with electroweak masses is their fairly long decay length, either due to the suppression of the large interaction scale or small Dirac masses. This is another example of displaced signatures at the LHC that have recently been receiving attention in the theoretical~\cite{llt} as well as experimental community~\cite{lle}. 

In the regions of interest for this study, the $m_N$ is somewhat below the LR scale and, similar to the muon in the SM, its lifetime increases significantly. The macroscopic length then leads to the displacement of $\ell j j$ vertices coming from decays of $N$s. In the $N$ eigenframe, we have
\begin{equation} \label{eqTauN}
  c \tau_N^0 \simeq 0.1 \text{ mm} \left(\frac{40 \text{ GeV}}{m_N}\right)^5 \left(\frac{M_{W_R}}{5 \text{ TeV}}\right)^4,
\end{equation}
which is visibly displaced from the interaction point.\footnote{The decay rate is taken to be dominated by $W_R^*$. For large LR scales  $M_{W_R} \gtrsim 70 \text{ TeV} \left(m_N/60 \text{ GeV}\right)^{1/4}$, the Dirac mixing amplitude may become comparable.} The transverse displacement is plotted on the lower right panel of Fig.~\ref{figKinDnn}, where the impact of the boost factor becomes evident.

A recent experimental search for pair-produced neutralinos and gluinos~\cite{CMS:2016kmw} exploits a similar feature of the MSSM with RPV to reduce the background and look for a pair of displaced $tbs$ vertices. We adopt the quoted vertexing efficiency of 50\% and require both leptons to be transversally displaced $l_{T\ell} > 0.1 \text{ mm}$ for signals with lifetimes longer than $1 \text{ mm}$, which covers most of the parameter space, as seen in Fig.~\ref{figSensComb}. This ``$l_{T\ell}$'' cut significantly reduces the backgrounds, as seen in the last rows of Table~\ref{tabBckgDL2}, enhancing the overall sensitivity.
 
The estimate of the final sensitivity after displacement cuts demonstrates that the $\Delta \to N N$ channel may probe fairly high scales of LR breaking in the $\mathcal{O}(\text{TeV})$ range, with better prospects for the muonic final state. These results will be shown in Fig.~\ref{figSensComb}, together with the $h \to \Delta \Delta \to 4 N$ channel, which is the subject of the following section.

Clearly, the above strategy is tailored for sensitivity to higher $W_R$ masses, where displacement is large. For $W_R$ masses below $5 \text{ TeV}$ the signal is rather prompt and one should not use the displaced lepton cut $l_{T\ell} > 0.1 \text{ mm}$, but rather veto the displaced leptons. However, the signal rate is high even without displacement cuts and the sensitivity estimate covers essentially the same $m_N$ region but does not extend to high $W_R$ masses. For simplicity, we only show the sensitivity to displaced signatures in Fig.~\ref{figSensComb}.

%
\subsection{$pp \to h \to \Delta \Delta \to 4 N$} \label{subsechto4N}

Pair production of $\Delta$ is an intriguing possibility since one ends up producing four $N$s. Due to the Majorana nature of $N$, there are $2^4$ possibilities for the charge of the final state lepton, as seen in Fig.~\ref{figFeynDiag}. These can be characterized by the amount of lepton number violation $\Delta L = 4, 2, 0$. The expected ratio of such events is then
\begin{equation} \label{eqRatDL}
  \Delta L_4 : \Delta L_2 : \Delta L_0 = 1 : 4 : 3 ,
\end{equation}
thus half of the $4N$ events will break $L$ by two and $1/8$ of them by four units. We concentrate on these final states because of low backgrounds. Again, the $\Delta L = 0$ is plagued by the SM and might be re-considered in case large displacements may eliminate the backgrounds. 

The types of signals in this channel can be separated into four non-overlapping regions of interest called $\mathcal R^{\# \ell}_{\Delta L}$, which are defined by the required number of leptons $\# \ell$ and the apparent violation of lepton number $\Delta L$. The di-lepton channel $\mathcal R^2_2$ with $\Delta L = 2$ is selected as in the $\Delta \to NN$ case above. The same-sign trilepton signal $\mathcal R^3_3$ with apparent $L$ violation by three units $\Delta L \simeq 3$ appears when one final state charged lepton goes missing, either due to kinematical, geometrical or isolation cuts. Three same-sign leptons then come either from the $\Delta L = 4$ case where one same-sign lepton is missing or from the $\Delta L = 2$ channel where the opposite-sign lepton disappears. Finally, when four charge leptons are required, they can combine into $\ell^\pm \ell^\pm \ell^\pm \ell^\mp$ with $\Delta L = 2$, which is the $\mathcal R^4_2$ or $\ell^\pm \ell^\pm \ell^\pm \ell^\pm$ with $\Delta L = 4$, the $\mathcal R^4_4$.

The $\mathcal R^4_4$ signal is conceptually most interesting since it would allow for observation of the breaking of lepton number by four units. This puts the LHC in quite a unique position since, to our knowledge, there is no competing low energy alternative. This is in contrast to $\Delta L = 2$ case, probed by the KS process, and where $0\nu2\beta$ experiments play a complementary role~\cite{Mohapatra:1980yp, Tello:2010am}. Even though the first search for the quadruple neutrino-less beta decay has been performed only recently~\cite{0nu4bNemo3:2016}, the rate from the $\Delta^2 h^2$ vertex and $4N$ exchange is hopelessly small~\cite{Heeck:2013rpa}.

\begin{figure} \centering  
  \includegraphics[width=.244\columnwidth]{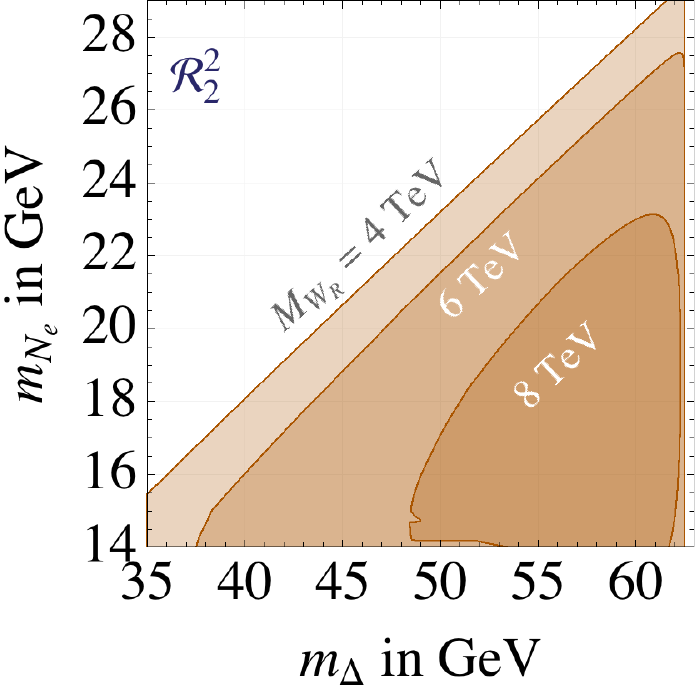} 
  \includegraphics[width=.244\columnwidth]{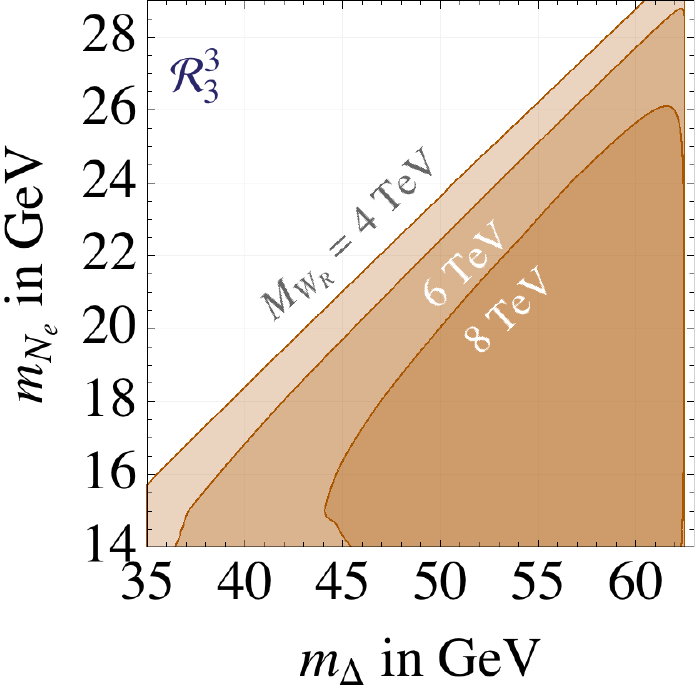}
  \includegraphics[width=.244\columnwidth]{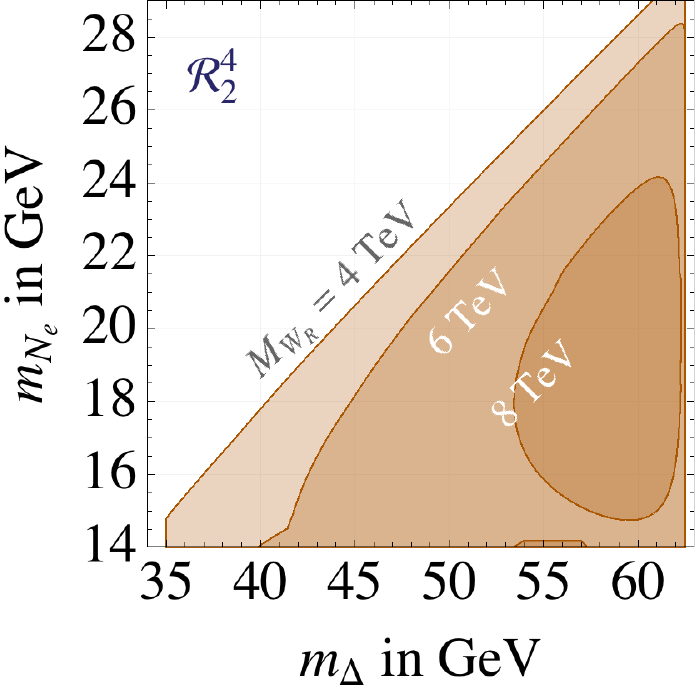} 
  \includegraphics[width=.244\columnwidth]{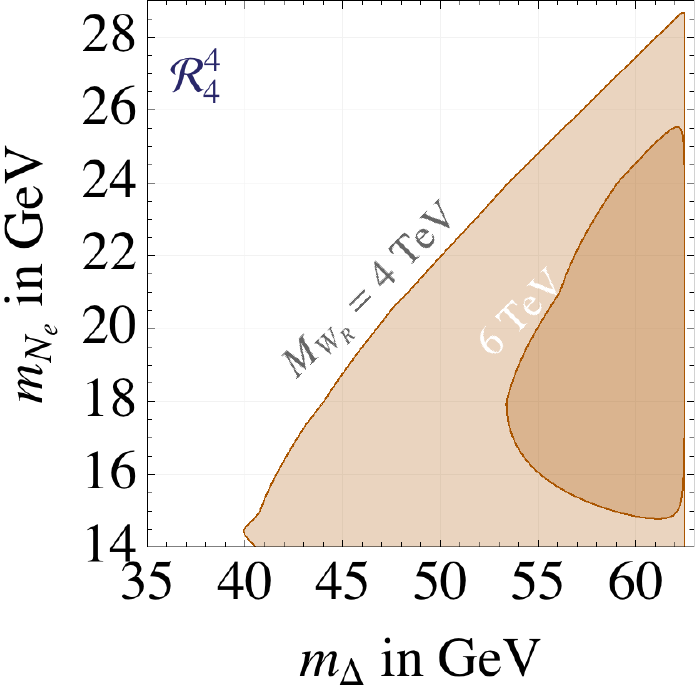}

  \includegraphics[width=.244\columnwidth]{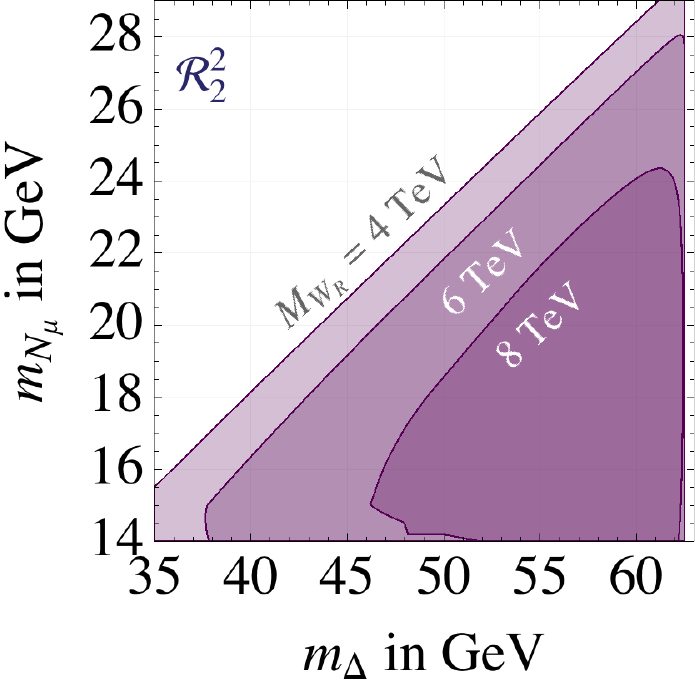} 
  \includegraphics[width=.244\columnwidth]{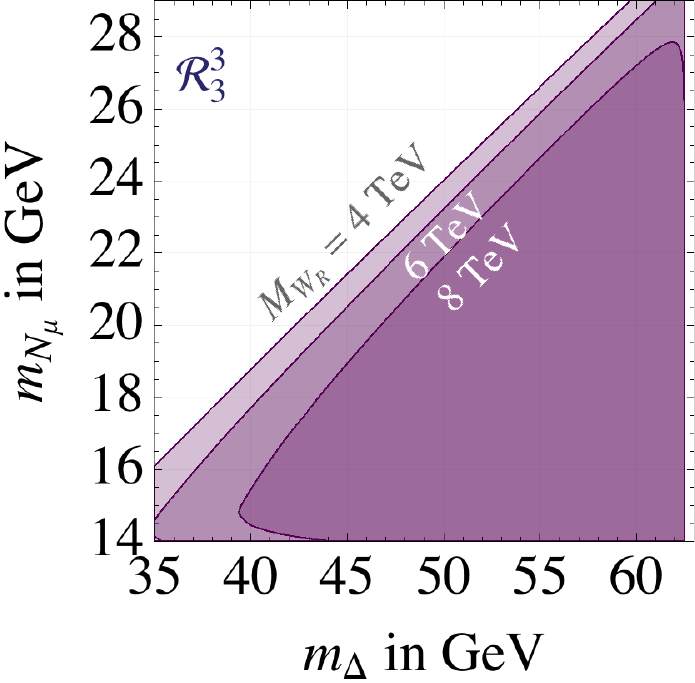}
  \includegraphics[width=.244\columnwidth]{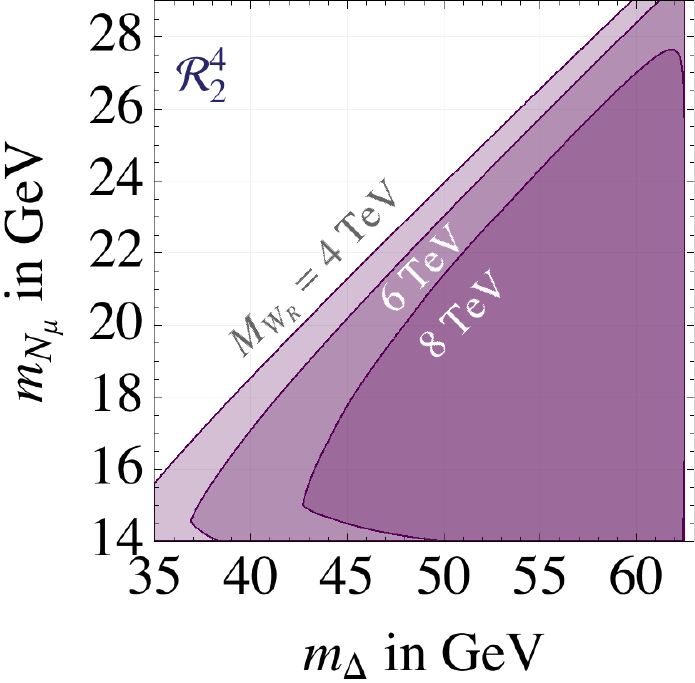} 
  \includegraphics[width=.244\columnwidth]{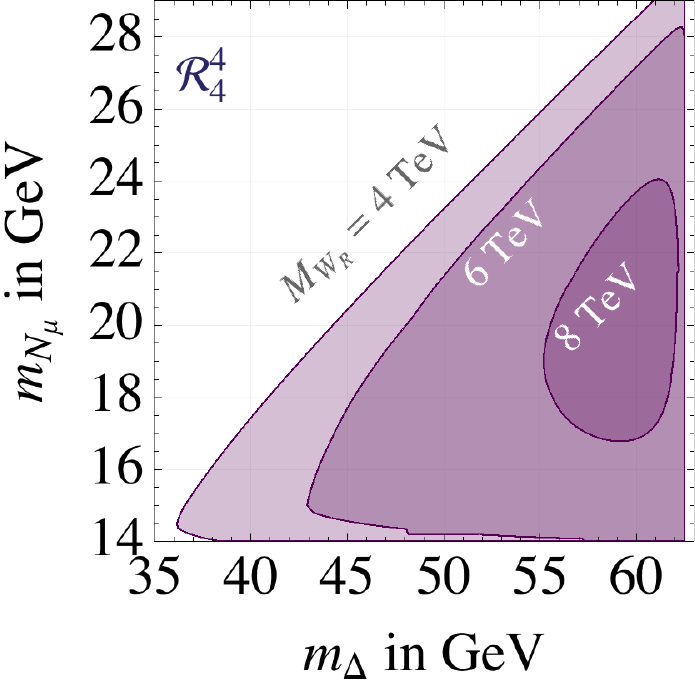}  
  \caption{Contours of estimated sensitivity ($S/\sqrt{S+B}  = 5$) for the $h \to \Delta \Delta \to 4N$ process in the electron and muon channels. The plots from left to right refer to the signal regions $\mathcal R^{\# \ell}_{\Delta L}$ described in the text. The contours refer to different $M_{W_R}$, while $\sqrt s = 13 \text{ TeV}, s_\theta = 0.1$ and the luminosity is $100 \text{ fb}^{-1}$. The electron channel yields similar plots with a smaller sensitivity.}
  \label{figSens_hDD4N_NoB}
\end{figure}

\paragraph{Efficiency, backgrounds and sensitivity.}

As in \S \ref{subsecDto2N} above, the final state particles are somewhat soft. We find that the di-muon trigger with $p_{T \mu_2} > 10 \text{ GeV}$ keeps the most number of events in $\mathcal R^2_2$, while in the other three regions the tri-muon trigger with $p_{T \mu_2} > 6 \text{ GeV}$ is most efficient. In the parameter space of interest defined by Figs.~\ref{figBrhDD} and \ref{figSigBrhgg3TeV}, the selection efficiency for $\mathcal R^2_2$ ranges from $3-5\%$, in the $\mathcal R^3_3$ from $1-4\%$, while for the four-leptons it goes from $6 \permil - 3\%$ in $\mathcal R^4_2$ and $1-9\permil$ in $\mathcal R^4_4$, mainly due to~\eqref{eqRatDL}.

Let us consider the backgrounds for this channel. In the di-lepton case, the selection is the same as in \S \ref{subsecDto2N} and the backgrounds are those listed in Table~\ref{tabBckgDL2}. As for the tri- and quad-lepton final state, we simulate the prompt multi-production of $V$s and tops using the same procedure as for the signal. After the selection of charged leptons, less than one event survives, even without additional cuts. This comes mainly from $t \overline t Z$, $t \overline t h$ and $WZZ$; the four $V$, four top and and $VVt\overline t$ leptonic rates are negligible. After the missing energy cut $\slashed E_T < 30 \text{ GeV}$ is imposed, background rates for $\mathcal R^{3,4}_i$ practically vanish.

The final estimate of the LHC's sensitivity in the four regions is shown on Fig.~\ref{figSens_hDD4N_NoB}. The most sensitive ones are $\mathcal R^3_3$ and $\mathcal R^4_2$ with three and four leptons. In $\mathcal R^2_2$ the signal is larger, but the backgrounds are significant as well, while in the $\mathcal R^4_4$ case, the rate is suppressed by the combinatorics of Eq.~\eqref{eqRatDL}. 

%
\subsection{Summary and $0\nu2\beta$ connection}

\begin{figure} \centering
  \includegraphics[height=.34 \columnwidth]{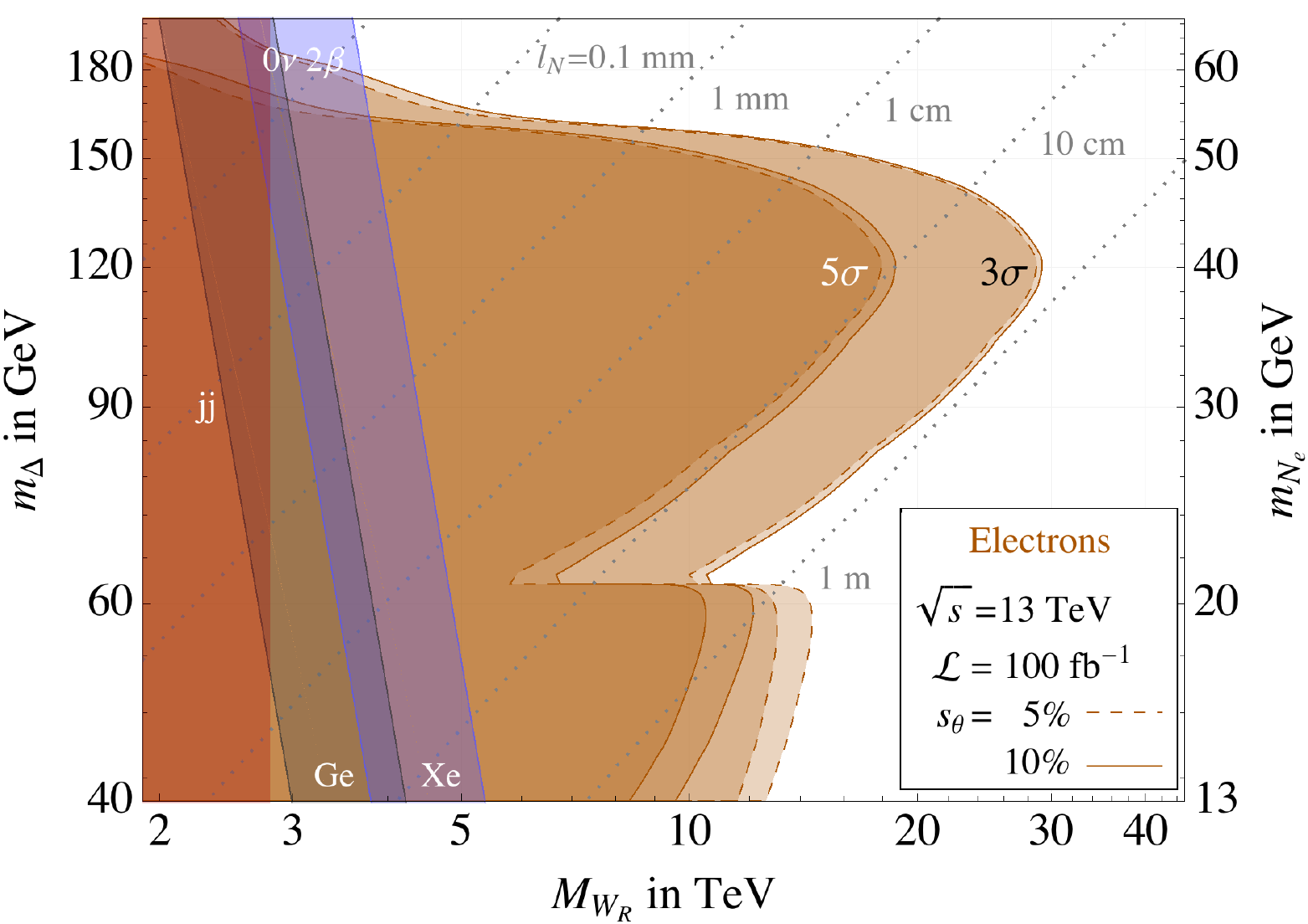} \hfill
  \includegraphics[height=.34 \columnwidth]{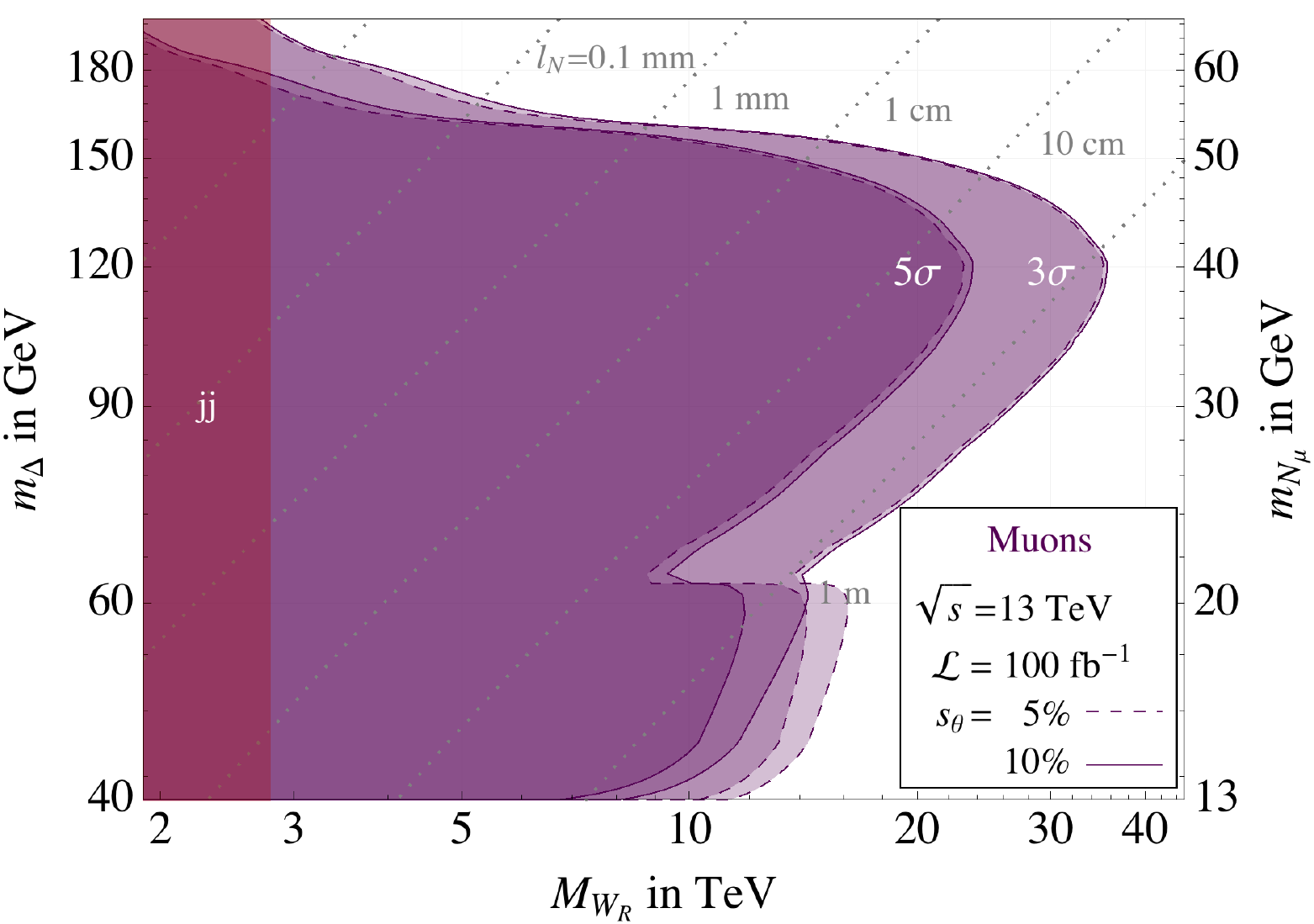} 
  \caption{Contours of estimated combined sensitivities of the $h \to NN, \Delta \to NN$ and $\Delta \Delta \to 4 N$ channels at 3 and 5 $\sigma$ with solid(dashed) contours corresponding to $s_\theta = 0.05(0.1)$. The left panel shows the sensitivity with loose electrons in orange and the right one for muons in purple. The dotted lines show the lab-frame $N$ decay length in the $\Delta \to NN$ channel, with $m_N = m_\Delta/3$.  The red region marks the current limit on $M_{W_R}$ from dijet searches~\cite{LHC:dijets2016}. The grey and light blue shaded regions show the current limits from $0\nu2\beta$ searches on $^{76}\text{Ge}$ and $^{136}\text{Xe}$ from Gerda~\cite{Agostini:2013mzu} and KamLAND-Zen~\cite{KamLAND-Zen:2016pfg}, respectively.}
   \label{figSensComb}
\end{figure}

The proposed signals involving single and pair production of $\Delta$ are combined and the sensitivity to the LR scale is estimated in Fig.~\ref{figSensComb}. This estimate is indeed quite encouraging with a $5 \sigma$ prospect well in the $10 \text{ TeV}$ range for muons. Because of lower efficiencies and higher backgrounds, the reach in the electron case is lower. The sensitivity peaks at 60 and 120 GeV for the $4N$ and $2N$ final states, respectively. Perhaps counter-intuitively, the reach is practically independent of the Higgs mixing in the range under consideration $s_\theta=0.05$--0.1. This is due to the opposite effect of reducing $s_\theta$: the production of $\Delta$ is reduced, but the $\Delta\to NN$ branching ratio increases, as evident in fig.~\ref{figBrD}.

At the same time, this portion of parameter space is allowed by existing searches. In particular, the search in the KS process becomes increasingly ineffective below about $100 \text{ GeV}$~\cite{Ferrari:2000sp, Nemevsek:2011hz, Mitra:2016kov}. For $N$ with a lifetime beyond the size of the detectors, the $W' \to \ell + \slashed{E}_T$ searches take over~\cite{LHC:ellnu2016}, however this happens for $m_N \lesssim 15$ GeV, therefore for much lower masses than those considered here. Dijet searches were also updated recently~\cite{LHC:dijets2016}; they are nearly independent of $m_N$ and stay below the $3 \text{ TeV}$ limit. As a result, the channel proposed here is complementary to the other searches. 

Arguably the most significant connection between $\Delta$ processes at the LHC and low energies is the search for $0\nu2\beta$ decay. Experimental searches have improved the existing limits on $^{76}\text{Ge}$~\cite{Agostini:2013mzu} and $^{136}\text{Xe}$~\cite{KamLAND-Zen:2016pfg} by an order of magnitude. The heavy neutrino exchange~\cite{Mohapatra:1980yp, Tello:2010am} can be conveniently expressed by the effective Majorana mass
\begin{equation}
  m_{ee}^N = \frac{p^2 V_{eN}^2}{m_N} \left( \frac{M_W}{M_{W_R}} \right)^4,
\end{equation} 
which is limited by experimental searches with the use of nuclear matrix elements updated and collected in~\cite{Barea:2012zz}. These calculations are still plagued by rather large uncertainties, hence the resulting constraint on the $m_{\Delta}-M_{W_R}$ parameter space (in Fig.~\ref{figSensComb}, $m_N = m_\Delta/3$ is set for illustration) is spread out in shaded grey and blue regions in Fig.~\ref{figSensComb}. Despite the lack of accuracy and precision, it is clear that the two regions overlap throughout the parameter space, even in the $\Delta \Delta \to 4 N_e$ channel.

%
%
\section{$e^+ e^-$ colliders} \label{SecEpEm}

\begin{figure} \centering
  \includegraphics[height=.32\columnwidth]{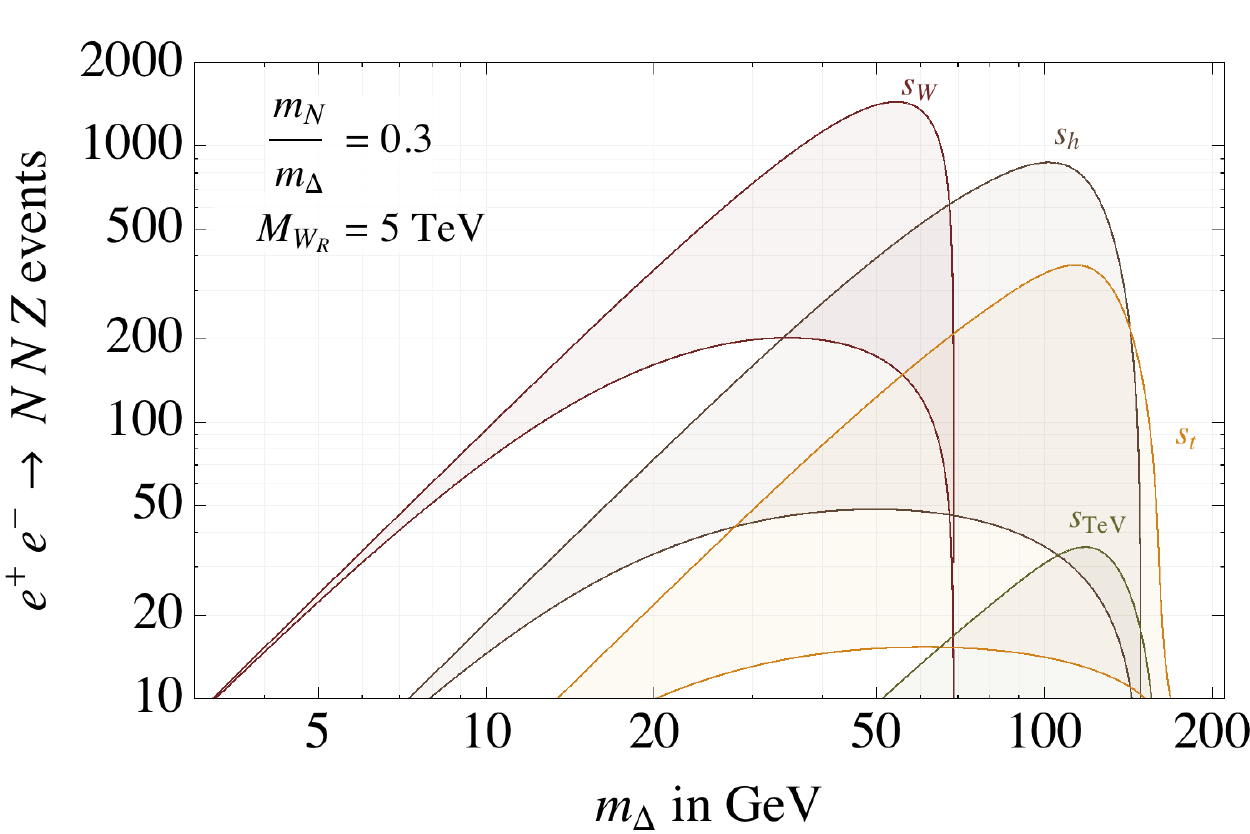} \hfill
  \includegraphics[height=.32\columnwidth]{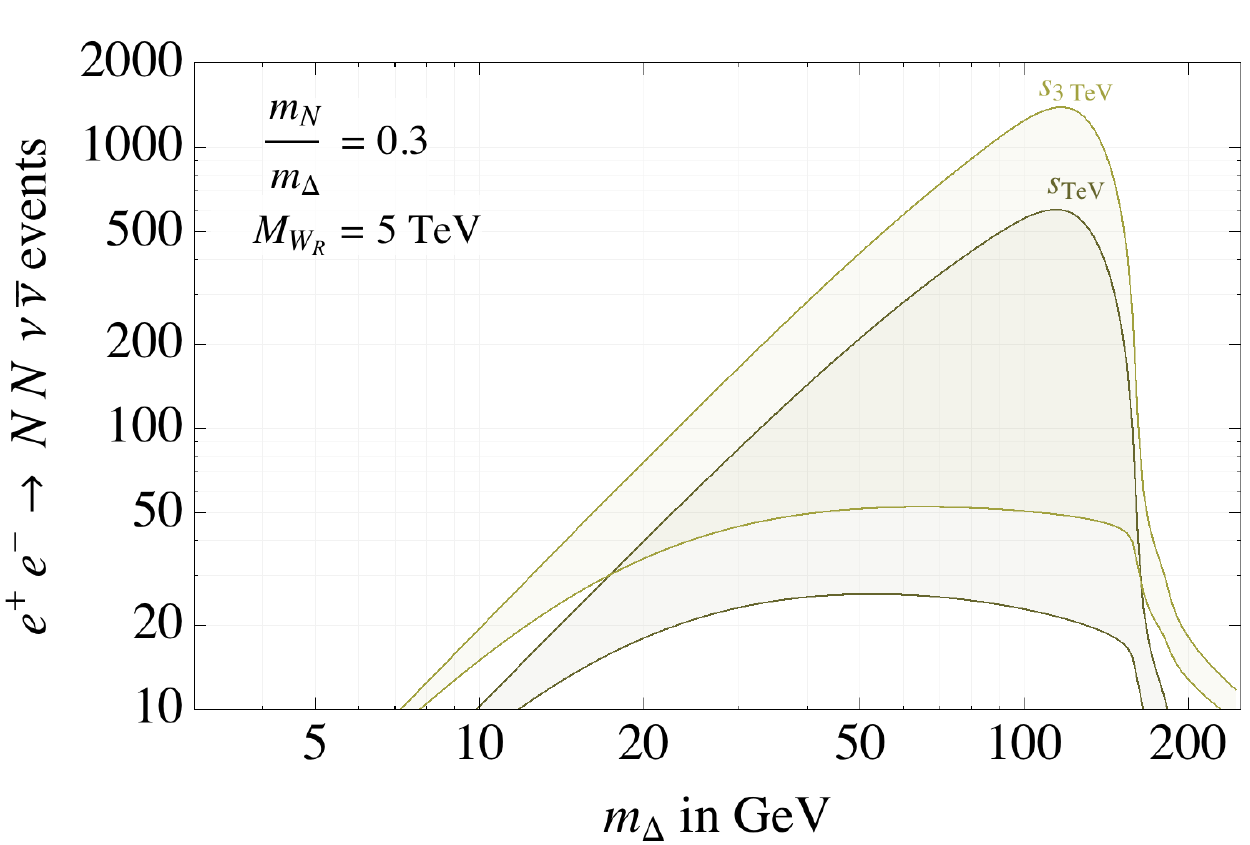}
  \caption{Signal event rates for $e^+ e^- \to N N Z$ and $e^+ e^- \to N N \nu \overline \nu$ productions at lepton colliders for different $\sqrt s$, universal luminosity of $\mathcal L = 1 \text{ ab}^{-1}$, $m_N = m_\Delta/3$. The shaded regions cover the range $s_\theta \in (0.01, 0.1)$.}
  \label{figNev_epemD_NoB}
\end{figure}

Hadronic colliders create a noisy environment with plenty of hadronic activity and issues related to triggering and jet fakes. On the other hand, lepton colliders such as $e^+ e^-$ machines are cleaner and with projected luminosity~\cite{Gomez-Ceballos:2013zzn} may provide a handle on soft and displaced signals, such as the production of heavy neutrinos~\cite{Blondel:2014bra, Antusch:2015mia}. 

The production of $N$ in the mLRSM can be sizeable at lepton colliders, even for relatively high $W_R$ mass and small Higgs mixing. The two main triplet Higgs production channels at $e^+ e^-$ colliders are the associated $\Delta Z$ production
\begin{equation}
  \sigma_{e^+e^- \to \Delta Z} = s_\theta^2 \frac{G_F^2 M_Z^4}{96 \pi s} \left( \hat v_e^2 + \hat a_e^2 \right) \sqrt \lambda \,
  \frac{\lambda + 12 \left(M_Z^2/s \right)}{1 - \left(M_Z^2/s \right)^2},
\end{equation}
where $\lambda = \left(1 - m_h^2/s - M_Z^2/s \right)^2 - 4 m_h^2 M_Z^2/s^2$, $\hat v_e = -1$, $\hat a_e = -1 + 4 s_w^2$, and the $\Delta \nu \nu$ one via $WW$ fusion, see e.g.~\cite{Djouadi:2005gi}. 

The total integrated luminosity at LEP was too small to find more than $\sim 2$ $NNZ$ events from the collected data. On the other hand, the future $e^+ e^-$ machines may have sufficient sensitivity to look for heavy neutrinos from $\Delta$ decays. Various production c.m. energies are currently under consideration~\cite{Gomez-Ceballos:2013zzn} from the $Z$ pole at 90 GeV all the way to a $3 \text{ TeV}$ machine: $\sqrt{s_{W,h,t,\text{TeV},\text{3TeV}}}=\{0.16, 0.24, 0.35, 1, 3 \} \text{ TeV}$.

For $\sqrt s < \mathcal O(100) \text{ GeV}$, the dominant production of $\Delta$ occurs in the associated $\Delta Z$ channel, leading to the $N N Z$ final state with up to four leptons and no missing energy when $Z$ decays leptonically. The backgrounds depend on the c.m. energy and are particularly low below the $t \overline t$ threshold. Moreover, they can be reduced with cuts to a small level even above this energy. Conversely, for TeV machines, the $W$ VBF channel takes over and the $NN \nu \overline \nu$ final state dominates, as seen on the right panel of Fig.~\ref{figNev_epemD_NoB}. The exact capabilities of the detectors are presently unknown, therefore we only show the signal event counts for different $\sqrt s$ cases in the left panel of Fig.~\ref{figNev_epemD_NoB}.

%
%
\section{Conclusions} \label{SecConclude}

Many proposals were made to experimentally establish the Majorana nature of the neutrino and search for the associated breaking of lepton number. Perhaps the most exciting prospect, apart from the $0\nu2\beta$, would be a direct observation of $N$. Were its mass protected by gauge symmetry, a ``Majorana'' Higgs boson should exist. The purpose of this work has been to study its collider phenomenology in the energy range accessible to the LHC and to provide a roadmap for experimental searches. The study has been carried out in the context of the Left-Right symmetric model, which serves as a complete model of neutrino mass origin, linked to the spontaneous breaking of parity. 

We find that a substantial number of LNV (and possibly LFV) events can originate from the ``Majorana'' Higgs field that is responsible for the breaking of $B-L$. The presence of such new resonance can significantly enhance the production of $N$s at the LHC via the Higgs portal mixing. This happens when its mass is approximately below 160 GeV, beyond which the standard searches for resonances apply, for instance the $VV$ channels. Current Higgs data allows for a fairly large mixing through which gluon fusion production occurs. At the same time, the Higgs width is small enough to provide a large branching ratio of the SM Higgs and new resonances to exotic states, even when these are weakly coupled. Particularly interesting and clean channels are the production of two and four $N$s, leading to $\Delta L \simeq 2, 3, 4$ final states. While the former case is partially covered by $0\nu2\beta$ in a similar portion of parameter space (see Fig.~\ref{figSensComb}), the breaking of lepton number by four units has no viable counterpart at low energies; here, the LHC is in a privileged position.

The sensitivity estimates show that in the appropriate RH neutrino mass range of 10--65 GeV these channels turn the LHC into a sensitive probe of new interactions at high energy scales. Because of such high scale suppression, the $N$ decays at a macroscopic distance, which allows for additional background suppression on top of the LNV selection.  One of the main results of our work is shown on Fig.~\ref{figSensComb}, where the combination of channels demonstrates the sensitivity to the LR scale well into the 10 TeV range, even beyond the capability of direct searches for $W_R$. At the same time, this portion of parameter space is in direct correspondence to $0\nu2\beta$ searches at low energies and nicely fills the parameter space between the golden KS search, which applies at higher $N$ masses, and missing energy $W' \to \ell \nu$ searches which are relevant at smaller ones.  A potentially interesting outlook for these channels would be to determine the polarization of $N$, similar to the KS case~\cite{Han:2012vk}, and establish the feasibility of measuring Majorana CP phases~\cite{Barger:2011tb}.

Despite our effort to accurately estimate the sensitivities, in order to precisely assess the background, the input from data will clearly be indispensable, in particular what concerns the fake rates. Another way to further increase the signal yield is to include the $\Delta L = 0$ displaced signals. Moreover, the prospects for $\tau$ final states were not investigated here due to the experimental intricacies and backgrounds. The observational potential of future leptonic colliders, however, seems to be quite optimistic, with a possibility of producing a significant number of $N$ events.

Some of the conclusions in this work are valid also for related theoretical frameworks where a Higgs portal resonance that couples to heavy RH neutrinos is present. These may include a simple model with spontaneous breaking of $U(1)_{B-L}$  or a more generic singlet Higgs coupling to sterile neutrinos, without gauge symmetries~\cite{Shoemaker:2010fg}. The main difference is the decay of $N$ which proceeds through the mixing with light neutrinos via the Dirac mass, which leads to additional subdominant final states and somewhat longer lifetimes.

In summary, the Higgs portal provides a new frontier for lepton number violation searches at colliders that may be sensitive to high energy scales well into the TeV domain.

%
%
\acknowledgments We thank Goran Senjanovi\'c for very useful comments on the manuscript and Stefano Bertolini and Alessio Maiezza for useful discussions. MN is partially supported by the Slovenian Research Agency funding program P1-0035. JV would like to thank the Jo\v zef Stefan Institute for the hospitality given while this work has been completed. JV is especially grateful to Goran Senjanovi\'c, who followed this work from its beginning and for enlightening discussions during all its stages. FN is partially supported by the H2020 CSA Twinning project No.~692194 ``RBI-T-WINNING'' and by the Croatian Science Foundation (HRZZ) project PhySMaB.  JV was funded by Conicyt PIA/Basal FB0821.

%
%
\appendix

\section{Tri-linear Higgs couplings} \label{SecAppTripleH}

%
%
Let us discuss in some detail the Higgs trilinear couplings within the LRSM, from tree-level to rather substantial loop corrections. In both cases we follow the phenomenological approach and start with physical masses and mixings as inputs, solve the linear system for the quartic couplings in the potential and thus derive the mixed field tri-linears.

{\em Tree level.} First of all, the $h$ and $\Delta$ mixing with $\Delta_L$ is suppressed by small $\langle \Delta_L \rangle$, and for simplicity we set to zero the CP violating phases, such that the mass matrix of interest reduces to three coupled fields: $h, \Delta, H$. Even in this case, the tri-linear vertices in the LRSM are non-trivial functions of quartic parameters and $t_{\beta/2} \equiv v_2/v_1$. To derive them, we define the mass eigenstates as
\begin{align}
  h &= \phantom{-} \left(c_\theta - \phi s_\eta s_\theta \right) h_0 + c_\eta  s_\theta \Delta_0 - \left( s_\eta s_\theta + \phi c_\theta \right) H_0
   &\simeq   \phantom{-} c_\theta h_0  +c_\eta s_\theta \Delta _0 - s_\eta s_\theta H_0 \,,
  \\[1ex]
  \Delta &= - \left(s_\theta + \phi s_\eta c_\theta \right) h_0 + c_\eta c_\theta \Delta_0 - \left( s_\eta c_\theta - \phi s_\theta \right) H_0
    & \simeq - s_\theta h_0 + c_\eta c_\theta \Delta _0 + s_\eta c_\theta H_0 \,,
  \\[1ex]
  H &= \phantom{-} \phi  c_\eta h_0+s_\eta \Delta _0 + c_\eta H_0
    &\simeq \phantom{-}\phi\, h_0\,\ \ + s_\eta \Delta _0\ \ \ + c_\eta H_0 \,,
 \end{align}
and give the expressions of mixing angles in terms of potential parameters explicitly below in~\eqref{gentheta}. The $H$-$h_0$ angle $\phi$ can be treated linearly because it is very small, $\phi \sim \epsilon^2 = v^2/v_R^2$, and plays no significant role. The $H_0$ components in $\Delta$ and $h$ are naturally of order $\eta \sim \epsilon$, but they are further constrained by flavor for light $m_{h, \Delta} $~\cite{Maiezza:2016bzp}: $s_\eta \ll m_\Delta/m_H < 0.01$ and thus also quite small. On the other hand, because we consider the case of light $\Delta$, the mixing $\theta$ can be of order one.
 
\begin{figure} \centering
  \includegraphics[height=.32\columnwidth]{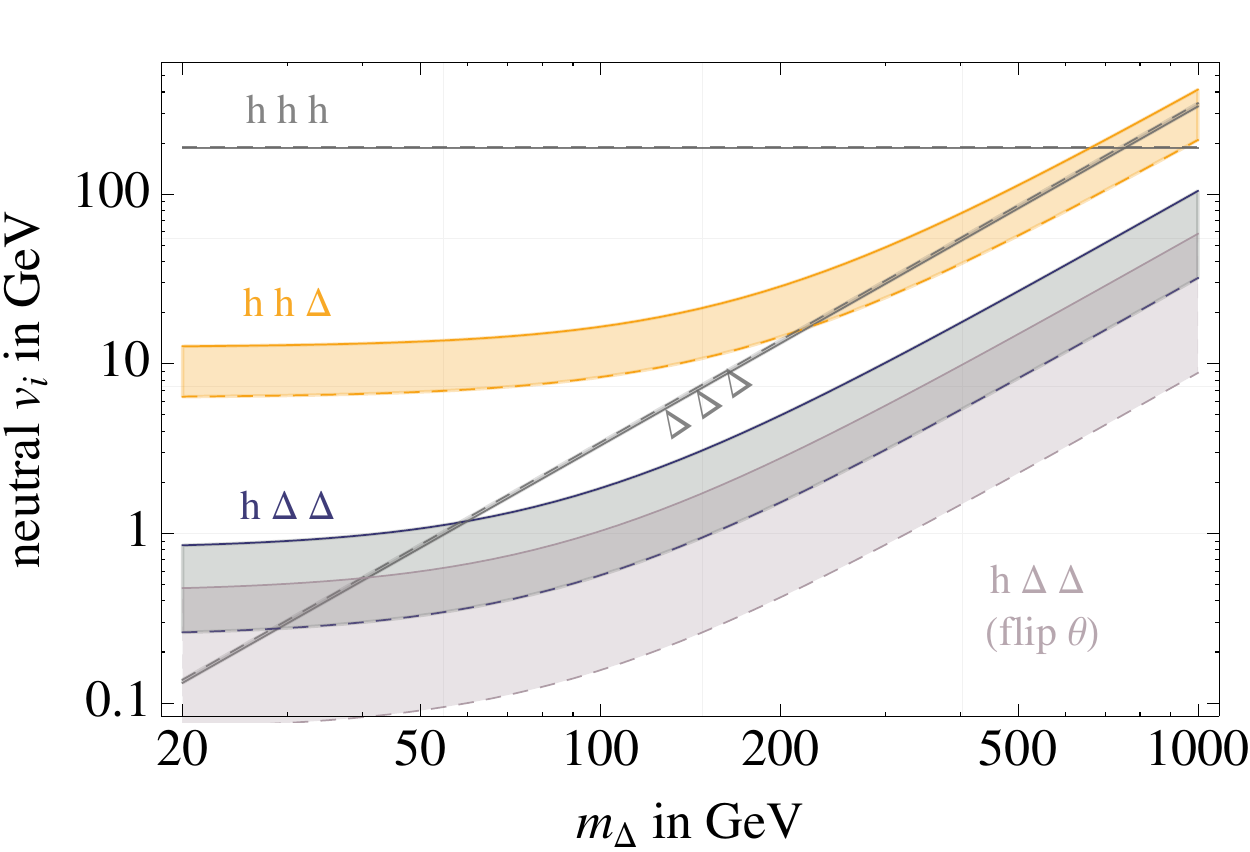}~~~
  \includegraphics[height=.32\columnwidth]{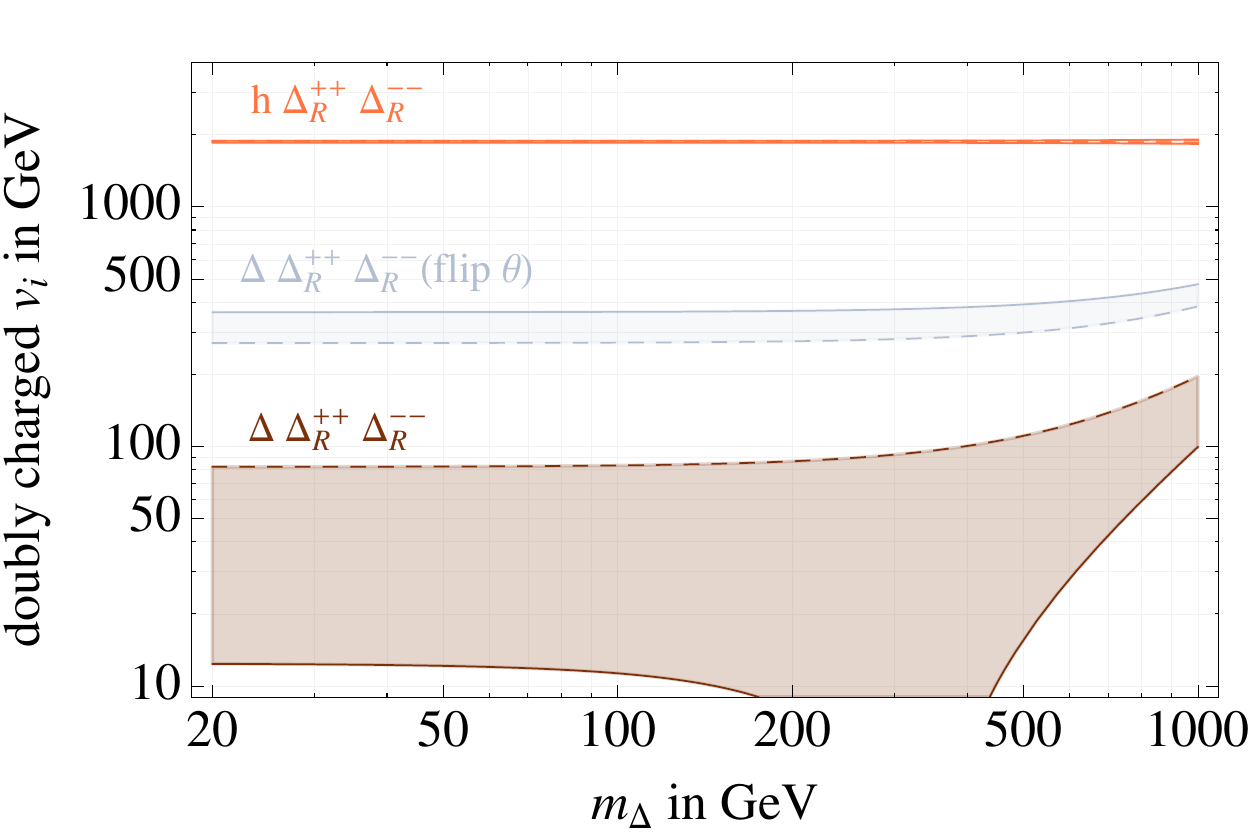}
  \caption{Left: Tree-level neutral tri-linear Higgs couplings from Eqs.~\eqref{eqVhhh}-\eqref{eqVDDD} with regions covering the $s_\theta \in (0.05,0.1)$ interval and $M_{W_R} = 4 \text{ TeV}$. Right: Tri-linear vertices with doubly charged scalars. The pink region shows the $v_{\Delta \Delta_R^{++} \Delta_R^{--}}$ vertex that enters the $\gamma \gamma$ and $Z \gamma$ rates and $m_{\Delta_R^{++}} = 1 \text{ TeV}$, $m_H = 17 \text{ TeV}$ was taken in agreement with $B \overline B$ and $h\to \gamma \gamma$ bound, discussed in~\cite{Maiezza:2016bzp}, that comes from the $v_{h \Delta_R^{++} \Delta_R^{--}}$ coupling.} 
  \label{figvHn}
\end{figure}

By solving for the quartic couplings in terms of physical masses and mixings, one can obtain the triple Higgs vertices, expanded here in small $\eta$,
\begin{align} \label{eqVhhh}
    v_{hhh} \simeq & \frac{3 g}{2 M_{W_L}} \Big[ m_h^2 \left(c_\theta^3 + \epsilon s_\theta^3 \right) 
    - 2\, \eta^2 m_H^2\, s_\theta^2 \left(c_\theta - \epsilon s_\theta \right)\Big]\,,
    \\ 
    \label{eqVhhD}
    v_{hh\Delta} \simeq & \frac{g}{4 M_{W_L}} \Big[\left(2 m_h^2+m_\Delta^2 \right) \left(\epsilon s_\theta - c_\theta  \right) s_{2 \theta} - 
    2\, \eta^2 m_H^2\, s_\theta \left(3 c_{2 \theta} - 3 \epsilon s_{2 \theta} + 1 \right) \Big]\,,
    \\ 
    \label{eqVhDD}
    v_{h\Delta\Delta} \simeq &\frac{g}{4 M_{W_L}} \Big[ \left(m_h^2 + 2 m_\Delta^2 \right) \left( \epsilon c_\theta + s_\theta \right) s_{2 \theta}
    - 2\, \eta ^2 m_H^2\, c_\theta \left(3 c_{2 \theta} - 3 \epsilon s_{2 \theta} - 1 \right) \Big]\, ,
    \\ 
    \label{eqVDDD}
    v_{\Delta \Delta \Delta} \simeq & \frac{3 g}{2 M_{W_L}} \Big[ m_\Delta^2 \left( \epsilon c_\theta^3 - s_\theta^3 \right)
    + 2\, \eta ^2 m_H^2\, c_\theta ^2 \left(\epsilon c_\theta + s_\theta \right) \Big]\,,
\end{align}
which is valid up to $\mathcal O (\epsilon^3)$. Notably, these expressions do not depend on $t_\beta$, which appears at $\mathcal O (\epsilon^4)$, while $\phi$ appears at $\mathcal O (\epsilon^3)$. Because of the flavour constraint $\eta^2 m_H^2 \ll m_\Delta^2$, the corrections $\eta^2$ due to the mixing with the heavy $H$ are sub-leading. It is thus safe to consider the limit in which $\eta$ is neglected. This is achieved formally if the couplings are such that $\alpha_2/\alpha_3+t_{\beta}$ is small, effectively reducing the mass matrix to a $2 \times 2$ system. In such case, the first terms in Eqs.~(\ref{eqVhhh})--(\ref{eqVDDD}) are exact to all orders in $v$, just as a function of the $h$, $\Delta$ masses and their mixing angle. In particular, the dependence on $m_H$, $t_\beta$ and other parameters such as $\alpha_2$ and $\lambda_i$ disappears. The numerical value of these vertices is plotted on the left panel of Fig.~\ref{figvHn}.

Considering that phenomenologically also $s_\theta$ has to be somewhat small, i.e. $s_\theta<0.1$--0.2, it is useful to show the behaviour in this limit:
\begin{align} 
  v_{hhh} &\simeq\frac{3 g \left(2-3\theta ^2\right) m_h^2}{4 M_{W_L}}\,, &
  v_{hh\Delta} &\simeq\frac{g\, \theta  (\epsilon\theta-1)   \left(2 m_h^2+m_{\Delta}^2\right)}{2 M_{W_L}}\,,
  \\
  \label{vhDDlin}
  v_{h\Delta\Delta} &\simeq\frac{g\,\theta (\theta+\epsilon ) \left(m_h^2+2 m_{\Delta }^2\right)}{2 M_{W_L}}\,, &
  v_{\Delta\Delta\Delta} &\simeq\frac{3 g\, \epsilon  \left(2-3 \theta^2\right) m_{\Delta }^2}{4 M_{W_L}}\, .
\end{align}
For $\theta =0$ the SM rule for the $hhh$ vertex is reproduced and the mixed $\Delta h$ vertices disappear. In \eqref{vhDDlin} the combination $(\theta+\epsilon)$ adds constructively or destructively depending on the sign of $\theta$. The destructive case is also shown in Fig.~\ref{figvHn} as ``flip $\theta$''.

Similarly one can derive the doubly charged triplet component couplings, which depend explicitly on $t_\beta$ and $m_H$ (see Fig.~\ref{figvHn}, right):
\begin{align} \label{eqVhDRpp} 
  v_{h \Delta_R^{++}\Delta_R^{--}}& \simeq \frac{\sqrt{2} g }{ M_{W_R}} \left[
    \left(m_{\Delta_R^{++}}^2 + \frac{1}{2} m_h^2 + c_\theta^2 \left(m_\Delta^2 - m_h^2 \right) \right) 
    s_\theta + \epsilon \, m_H^2 \left(c_\theta - \epsilon s_\theta \right) c_\beta^2 \right],
  \\
  \label{eqVDDRpp} 
  v_{\Delta \Delta_R^{++}\Delta_R^{--}} &\simeq \frac{\sqrt{2} g }{ M_{W_R}}\left[
    \left(m_{\Delta_R^{++}}^2 + \frac{1}{2} m_\Delta^2 - s_\theta^2 \left(m_\Delta^2 - m_h^2 \right) \right)
    c_\theta - \epsilon\, m_H^2 \left( \epsilon c_\theta + s_\theta \right) c_\beta^2 \right]
\end{align}
and equivalent expressions with $\Delta_R^{++} \to \Delta_L^{++}$.

\bigskip

\pagebreak[3]

The above expressions are all that is needed for the phenomenological analysis. For completeness, we
report also the expressions for the mass eigenvalues in terms of the potential parameters, in the
limit of small mixings:
\begin{align}
  m_h^2 &\simeq 4 \left( \bar\lambda_1 + \theta^2 \left( s_2 + \bar \lambda_1 - \bar \rho_1 \right) \right) v^2 \, ,
   \\[1ex]
  m_\Delta^2 &\simeq  4 \left( \bar \rho_1 - \theta^2 \left( s_2 + \bar \lambda_1 - \bar \rho_1 \right) - s_2 \right) v^2 \,,
   \\[1ex]
  m_H^2 &\simeq  \bar\alpha_3 v_R^2 + 4 s_2 v^2 \,,
\end{align}
where
\begin{align}
  \bar \rho_1 &= \frac{\rho_1}{\epsilon ^2} ,
  & \bar \alpha_3 &= \frac{\alpha_3}{c_\beta} , 
  & \bar \lambda_1&= \lambda_1 + 2 \lambda_4 s_\beta + \left( 2 \lambda_2 + \lambda_3 \right) s^2_\beta ,
  & s_2 &= \frac{\bar \alpha_3}{4} \, \left(\frac{\eta}{\epsilon}\right)^2.
\end{align}
In the first expression one must have $\rho_1 \sim \epsilon^2$ as explained in the text, since $m_\Delta$ is near the electroweak scale. Likewise, the expressions for the mixing angles are:
\begin{align} \label{gentheta}
  \theta & \simeq \frac{- 2 \alpha_1 + \bar \alpha_3 \left(1 - c_\beta + (\eta/\epsilon) t_\beta \right)}{
  4 \epsilon \left( \bar \lambda_1 - \bar \rho_1 + s_2 \right)},
  \\[1ex]
  \eta   & \simeq \epsilon \frac{-c_\beta \left(4 \alpha_2 + \bar \alpha_3 s_\beta \right)}{4\,\bar\alpha_3 } , 
  \\[1ex]
  \phi   & \simeq \epsilon^2 \frac{4 c_\beta \left(s_\beta  (2 \lambda_2+\lambda_3) + \lambda_4 \right) }{\bar\alpha_3},
\end{align}
where it becomes clear that the $h$-$H$ mixing $\phi$ is of order $\epsilon^2$ and thus negligible, while the $\Delta$-$H$ mixing $\eta$ is of order $\epsilon$. On the other hand, the $h$-$\Delta$ mixing $\theta$ can be of order one. For this to happen, in equation (\ref{gentheta})\ the numerator has to be suppressed like $\epsilon$ with respect to the denominator.

When $m_\Delta$ is above the electroweak scale and closer to $m_h$, the $\eta$ mixing can become sizeable without significantly affecting the flavor constraints. In this case, the $H$ also becomes a ``Majorana'' Higgs and couples to both $N$ and $\Delta$. With enough c.m.\ energy, the $H$ resonance may become visible and a heavier $\Delta$ could also be produced through the $H$ portal. Prospective channels $H \to NN$ and $H \to \Delta \Delta$ open up and would allow to probe higher $N$ and $\Delta$ masses. They are set by the Yukawa of $N$ and the triple vertex 
\begin{equation}
  v_{H \Delta \Delta} \simeq - g \frac{m_H^2}{M_{W_L}} \eta\left(\theta+\frac32 \epsilon\right).
\end{equation}
Although in the next decade the sensitivity to the heavy Higgs mass $m_H \gtrsim 20 \text{ TeV}$ will be considerably strengthened~\cite{Bertolini:2014sua}, such signals could become relevant for a future $\sqrt s \gg 14 \text{ TeV}$ collider and would require a separate study beyond the scope of the present work.

%
%
{\em Loop corrections.} In the phenomenologically interesting regime where $\Delta$ is fairly light, loop corrections to the $h-\Delta$ trilinears become important and may dominate the tree-level coupling. To derive the one-loop correction, we employ the scheme where the mass matrices, i.e. the masses and mixing angles after renormalization, remain the same as the tree-level ones. We then solve the same linear equations as above to derive the tri-linears. In the absence of mixing and $m_{h,\Delta} \to 0$, we have
\begin{align}
  v_{hhh}^{(1)} 				& \simeq c^{(1)} \left( 1 + \frac{17}{3}\frac{1}{r_{++}} \right) \varepsilon^3 v_R,
  &
  v_{hh\Delta}^{(1)} 			& \simeq c^{(1)} \, 11 \, \varepsilon^2 v_R ,
  \\
  v_{h\Delta\Delta}^{(1)}		& \simeq c^{(1)} \left( 4 + 10 \, r_{++} \right) \varepsilon v_R,
  &
  v_{\Delta\Delta\Delta}^{(1)}	& \simeq c^{(1)} \left( 8 + 16 \, r_{++}^2 \right) v_R,
\end{align}
where $c^{(1)} = 1/\sqrt{2}(4 \pi)^2 \, (m_H/v_R)^4$ and $r_{++} = (m_{\Delta^{++},\Delta_L^{0,+,++}}/m_H)^2$. These expressions become more involved when the mixings are turned on and are not particularly illuminating. In short, when $s_\theta \neq 0$, the $v_{\Delta\Delta\Delta}^{(1)}$ is reduced by $c_\theta^3$ while the other three receive a significant addition $\Delta v_{hhh, hh\Delta, h\Delta\Delta}^{(1)} \propto s_\theta^{1,2,3} v_{\Delta\Delta\Delta}^{(1)}(\theta=0)$.

\begin{figure}
   \includegraphics[height=.32\columnwidth]{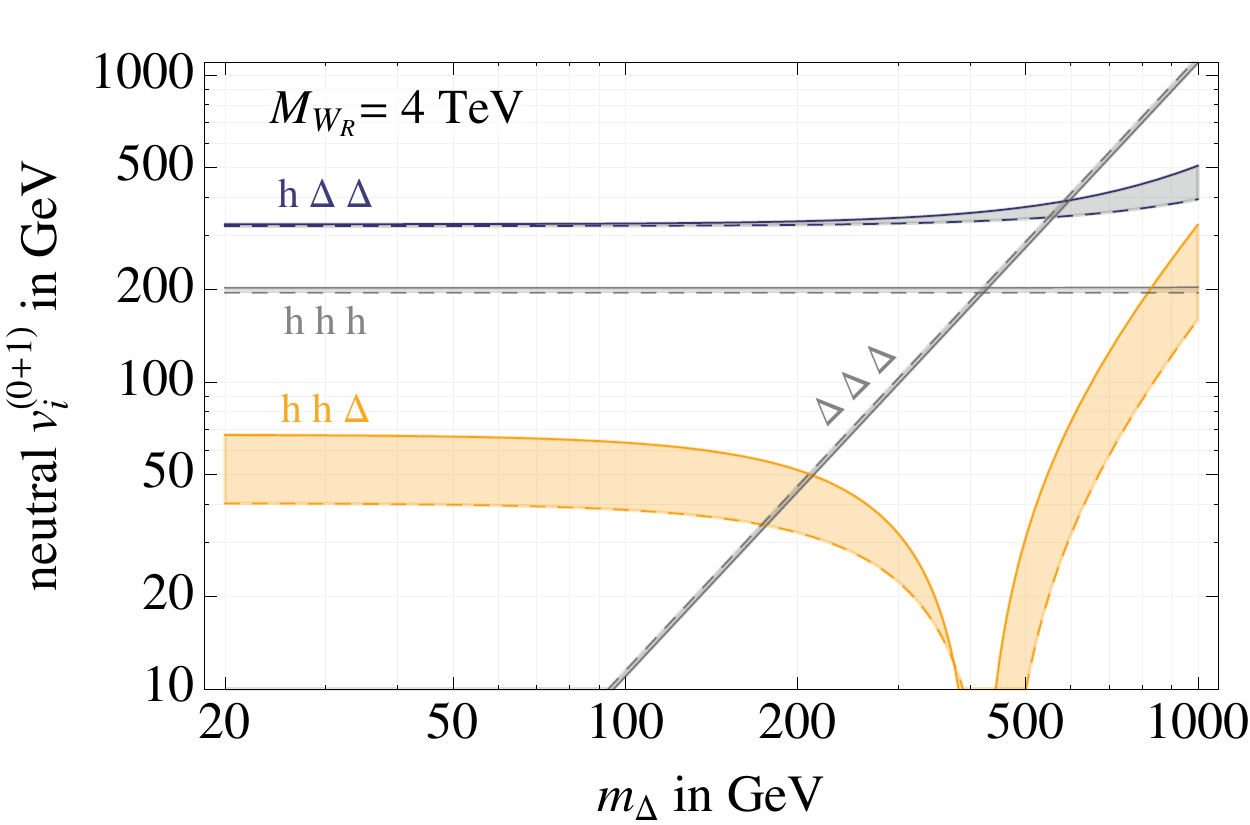}
   \includegraphics[height=.32\columnwidth]{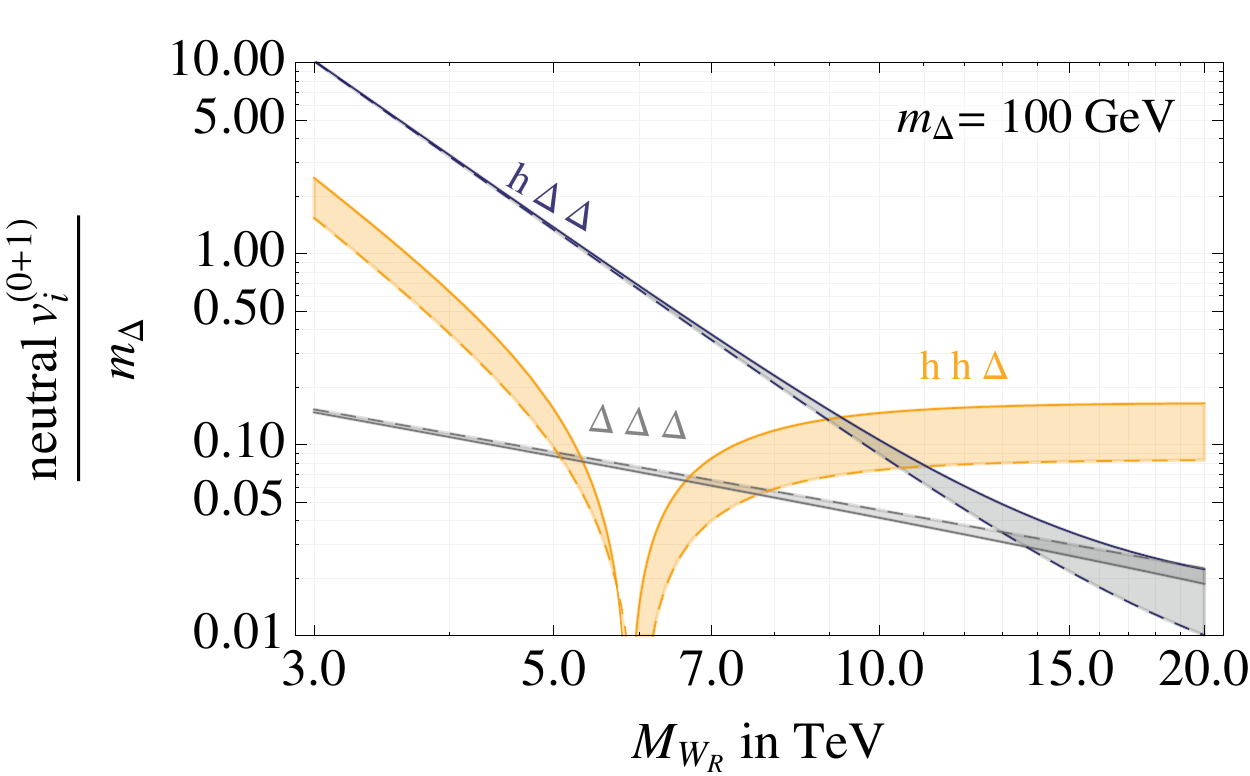}
   \caption{Left: Neutral Higgs trilinear couplings at one loop, obeying the vacuum stability condition in~\eqref{eqCondStab}, $s_\theta \in (0.05,0.1)$, $M_{W_R} = 4 \text{ TeV}, m_H = 17 \text{ TeV}$ and $r_{++} = 0.3$ (see text). The blue and yellow regions are the $v_{h \Delta \Delta}$ and $v_{h h \Delta}$ vertices that set the pair-production and associated production rates in Fig.~\ref{figDProd}. Right: Neutral tri-linear dependence on $M_{W_R}$ with fixed $m_\Delta = 100 \text{ GeV}$, all values as on the left but normalized to $m_\Delta$.}
   \label{figv3HnStable}
\end{figure}

For the Higgs masses of interest here, loops may dominate the tree-level values derived above~\cite{Coleman:1973jx} and cannot be predicted solely from $m_{h,\Delta}$ and $s_\theta$. Instead, their magnitudes are set by the heavier scalars $H, A, H^+, \Delta_L, \Delta^{++}$ and decouple only with increasing $v_R$.\footnote{This happens in the SM, where the Higgs trilinear would be dominated by the heavy top if $m_h < 40 \text{ GeV}$.}

However, such light Higgs masses jeopardize the spontaneous breaking; in the absence of fermions the quantum corrections render the unbroken phase energetically more favorable~\cite{StableQuantumVacuum}. To demonstrate the point, the effective potential with $\alpha_3$ and the heavy neutrino Yukawa coupling $Y_{\Delta}$ can be expanded in small field values $\Delta_0\ll v_R$ up to $\mathcal{O}(\Delta_0^5/v_R)$
\begin{equation}
 V_{eff}= \left(\text{C}-\rho_1 \right) v_R^4 + 
   4 \rho_1 v_R^2 \Delta_0^2 + \left( 4 \rho_1 + \frac{16}{3} \text{C} \right) v_R\Delta_0^3 + 
   \left( \rho_1 + \frac{16}{3}\text{C} \right) \Delta_0^4,
\end{equation}
where
\begin{equation} \label{eqa3vsYD}
 \text{C}= \frac{1}{32\pi^2}\left(\alpha_3^2 - 8 \, Y_\Delta^4 \right) .
\end{equation}
Requiring absolute stability of $V$ thus leads to an upper bound on the loop trilinear correction
\begin{equation} \label{eqCondStab}
  V_{eff}(v_R) \leq V_{eff}(0) \, \Rightarrow \, v_{\Delta \Delta \Delta}^{(1)} \leq \left( \frac{7}{3} \right) v_{\Delta \Delta \Delta}^{\text{tree level}} ,
\end{equation}
which can be achieved in the LRSM if at least one $N$ is above $M_{W_R}$ (as top in the SM) such that $C$ in~\eqref{eqa3vsYD} is suppressed. For the sake of illustration we choose~\eqref{eqCondStab} as a benchmark and show the trilinears in Fig.~\ref{figv3HnStable}. Requiring meta-stability, with  vacuum lifetime bigger than the age of the universe may significantly relax this bound. In any case this does not significantly alter the our results because it will principally increase only the $v_{\Delta \Delta \Delta}$ coupling, which does not play a major role in the phenomenological considerations.

%
%
\section{Loop functions} \label{SecAppLoops}

%
\paragraph{Loop functions for $\gamma\gamma$ and $Z \gamma$.} The loop functions of fermions, scalars and gauge bosons are
\begin{align} \label{eqLoopFf}
  F_f    &= -\frac{\sqrt{2}}{v} 2 \beta_f \left[1 + \left(1 - \beta_f \right) f(\beta_f)\right],
  & F_S   &= \frac{\beta_S}{m_S^2} \left[1 - \beta_S f(\beta_S) \right],
  \\
  F_W &= \frac{\sqrt{2}}{v} \left[2 + 3 \beta_W \left(1 + (2 - \beta_W) f(\beta_W)) \right) \right],
  & F_{W_R} &= \frac{v}{v_R} F_W(\beta_{W_R}),
\end{align}
where  $v \simeq 174 \text{ GeV}$. Defining $\beta_i = (2 m_i/m_\Delta)^2$, the dimensionless $f$ is the usual
\begin{align}
  f (\beta) = \begin{cases}  
 \arcsin \left(1/\sqrt \beta \right)^2, & \beta \geq 1,
  \\
  -\frac{1}{4} \left( \log \left( \frac{1 + \sqrt{1 - \beta}}{1 - \sqrt{1 - \beta}} \right) - i \pi \right)^2, & \beta < 1.
\end{cases}
\end{align}
The $Z \gamma$ functions are a bit more involved
\begin{align}
  G_f  &= \frac{2 \sqrt 2 \beta \lambda}{v (\beta - \lambda)^2}  \left[\beta - \lambda + ((\beta - 1) \lambda + \beta) (f(\beta) - f(\lambda)) - 2 \beta (g(\lambda) - g(\beta)) \right],
  \\
  G_S   &= \frac{\beta \lambda}{m_S^2 \left( \beta - \lambda \right)^2} \left[
  \lambda - \beta + \beta (\lambda \left( f(\lambda) - f(\beta) \right) + 2 ( g(\lambda) - g(\beta)))\right],
  \\
\begin{split}
  G_W &= \frac{\sqrt 2 c_w^2}{v (\beta - \lambda)^2} \biggl[
  (4 + 2 (\beta - \lambda) - 3 \beta \lambda) (\beta (1 + 2 (g(\beta) - g(\lambda))) - \lambda) -
  \\ \label{eqLoopGf}
   &\beta \left(f(\beta) - f(\lambda) \right) \left(\beta (3 \lambda (\lambda + 2) - 8) + 2 (2 - 3 \lambda) \lambda \right) \biggr], 
   G_{W_R} = \left( v /v_R \right) G_W(\beta_{W_R}),
\end{split}
\end{align}
and the $F$ functions above are recovered in the $M_Z \to 0$ limit with
\begin{align}
  g (\beta) = \begin{cases}  
    \frac{\sqrt{1-\beta}}{2} \left( \log \left( \frac{1 + \sqrt{1 - \beta}}{1 - \sqrt{1 - \beta}} \right) - i \pi \right), &
      \beta \geq 1, \\
    \sqrt{\beta - 1} \arcsin \left(1/\sqrt \beta \right), & \beta < 1.
\end{cases}
\end{align}

%
%
\paragraph{Loop functions for $\ell_i \ell_j$.} The loop function entering in the $\Delta \to \ell \ell'$ decays is
{\small
\begin{align} \label{eqFtLFV}
    \begin{split} \tilde f(\beta \leq 1) &= -\frac{\beta}{16} \biggl\{\pi^2 \beta + 4 i \pi \sqrt{1 - \beta} + \beta \biggl[-\log \left(1 - \beta + \sqrt{1-\beta} \right)^2 +
     4 i \pi \, \text{arctanh}\left(\sqrt{1 - \beta}\right) +
      \\ 
      &\log\left(\beta + \sqrt{1 - \beta} - 1 \right) \left(\log \left(\frac{1}{\sqrt{1 - \beta}} - 1\right) +  4 \, \text{arctanh}
      \left(\frac{\beta - 2 \sqrt{1 - \beta}}{\beta - 2}\right) \right) \biggr] +
      \\
      & 2 \sqrt{1 - \beta} \left(\log\left(1 - \beta \right) - 4 \, \text{arctanh}\left(\frac{\beta - 2 \sqrt{1 - \beta}}{\beta - 2}\right)\right) + 4 \biggr\},
   \end{split}
   \\
   \tilde f(\beta > 1) &=  -\frac{\beta}{4} \left[ \beta \, \text{arccsc}^2(\sqrt{\beta}) - 2 \sqrt{\beta - 1} \, \text{arccsc}(\sqrt \beta) + 1\right].
\end{align}}

%
%
\begin{figure}[t] \centering 
  \hfill
  \includegraphics[scale=0.5]{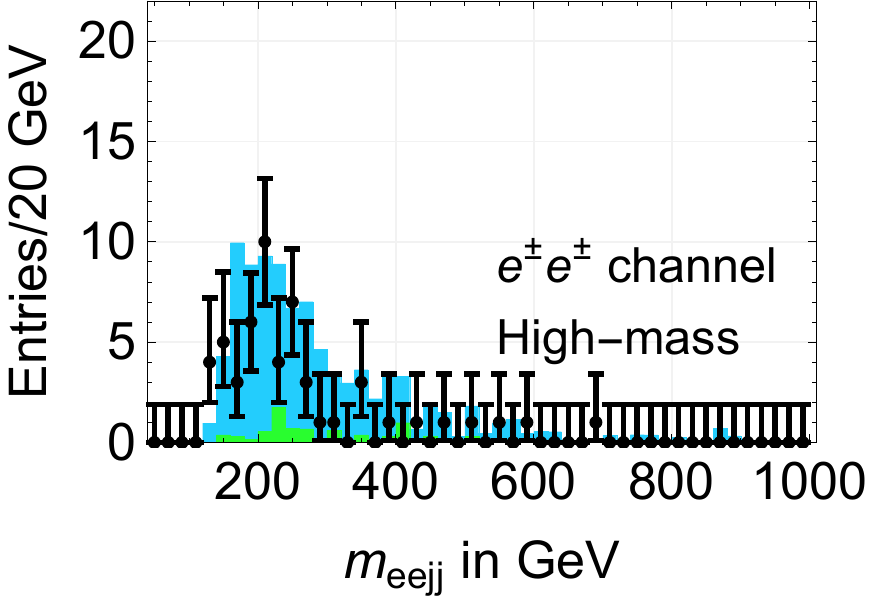} \hfill
  \includegraphics[scale=0.5]{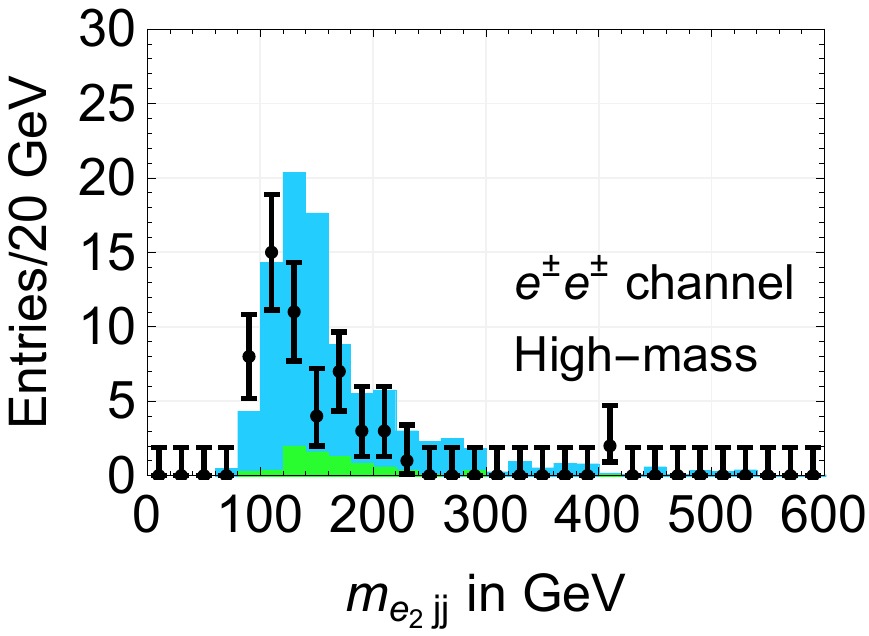} \hfill
  \includegraphics[scale=0.5]{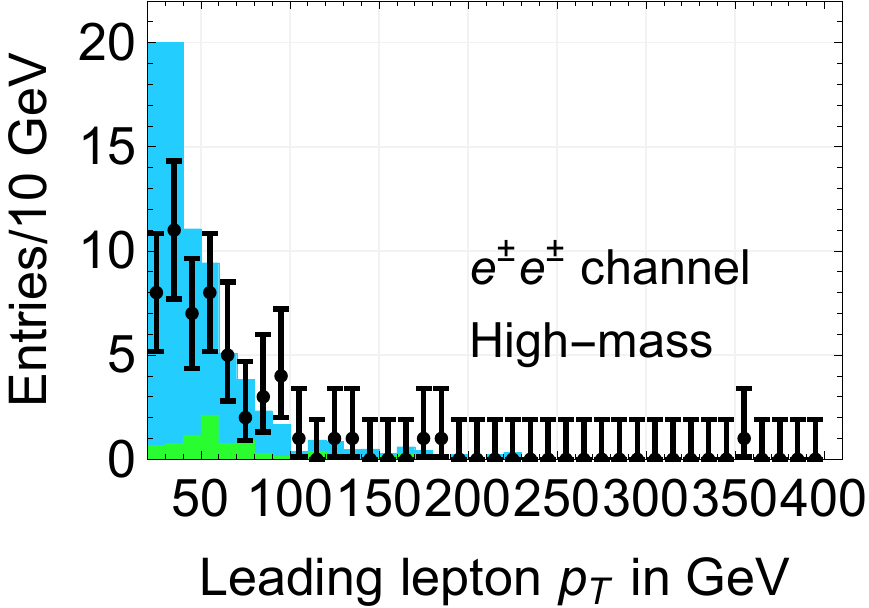} \hfill 

  \hfill
  \includegraphics[scale=0.5]{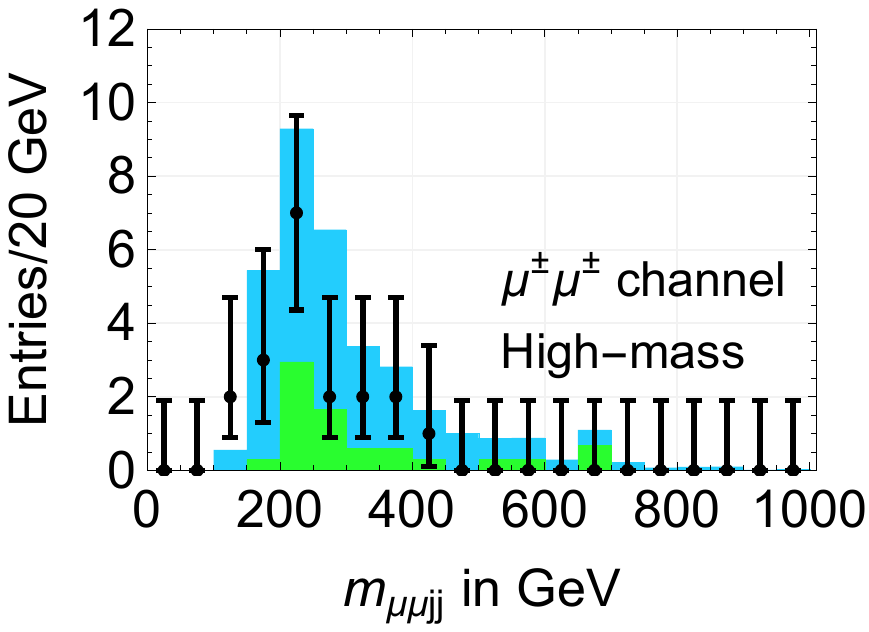} \hfill
  \includegraphics[scale=0.5]{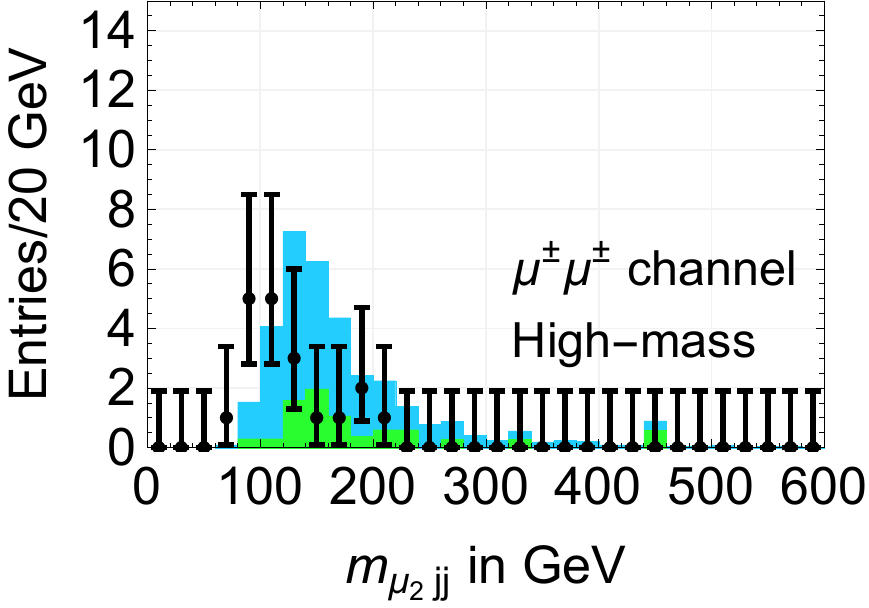} \hfill
  \includegraphics[scale=0.5]{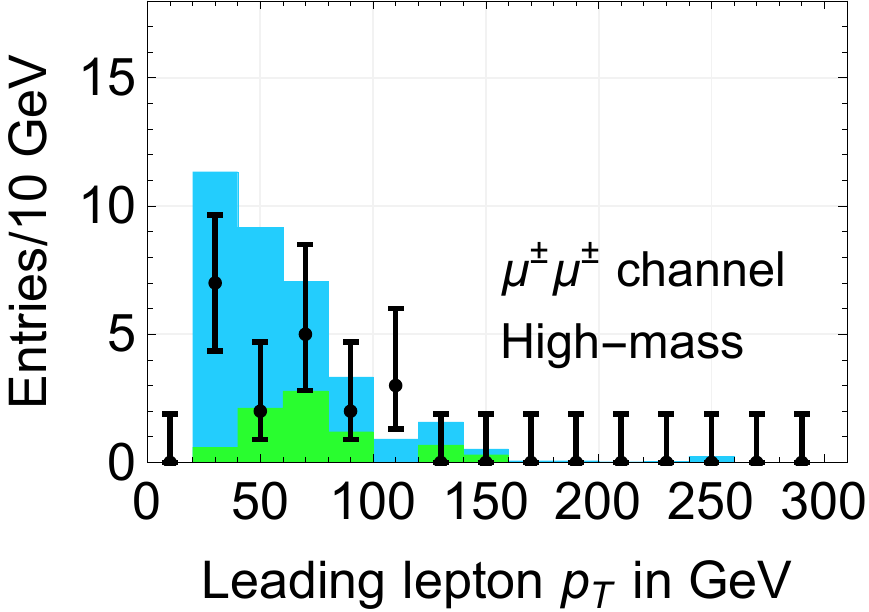} \hfill

  \hfill  
  \includegraphics[scale=0.5]{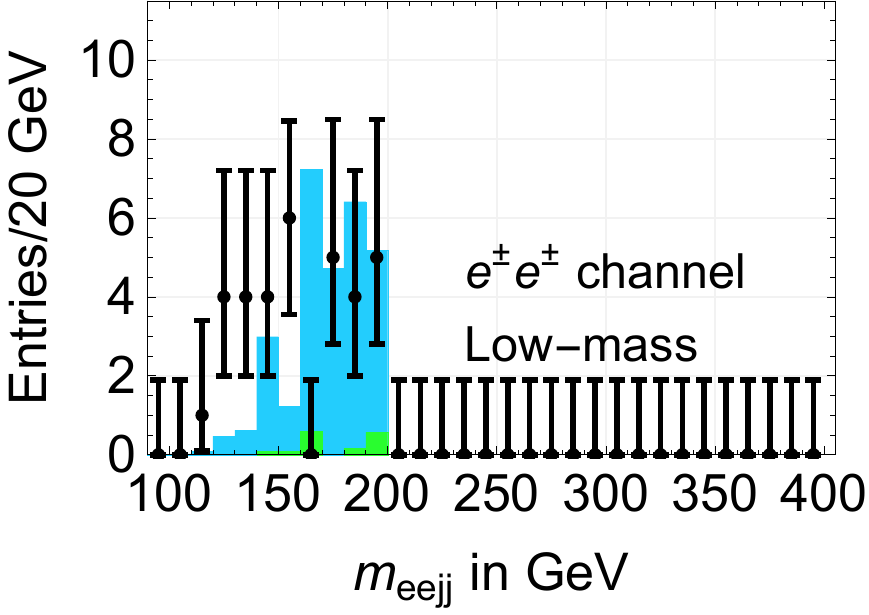} \hfill
  \includegraphics[scale=0.5]{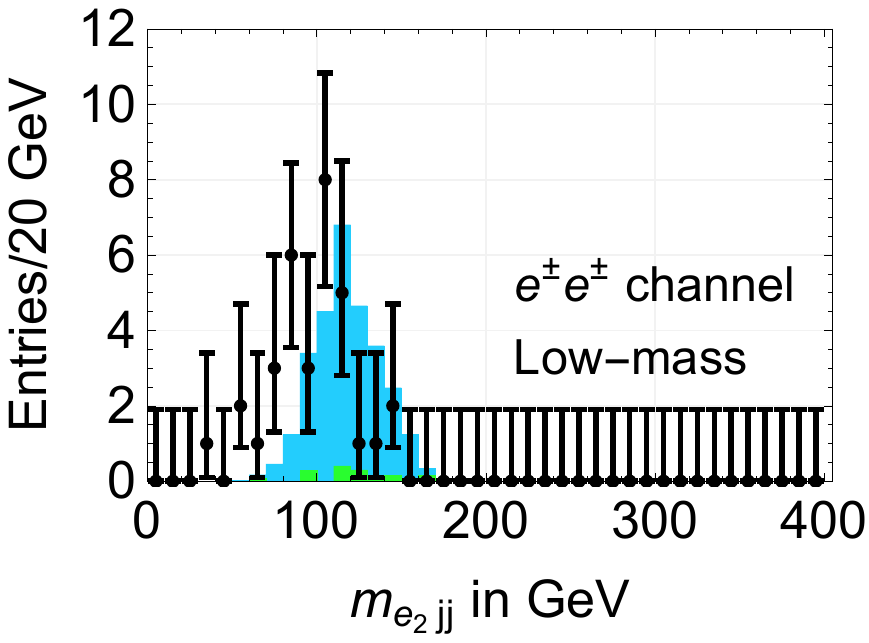} \hfill
  \includegraphics[scale=0.5]{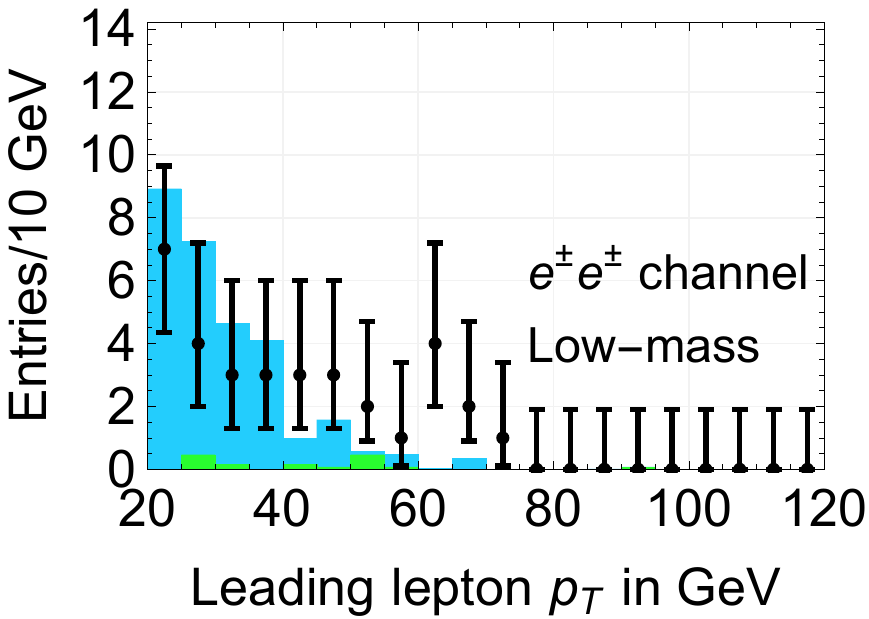} \hfill 

  \hfill
  \includegraphics[scale=0.5]{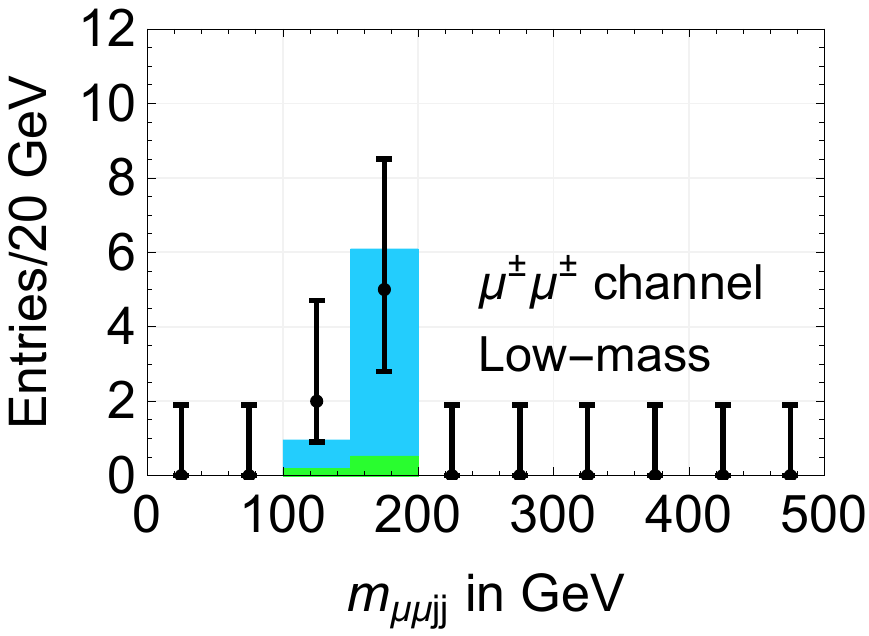} \hfill
  \includegraphics[scale=0.5]{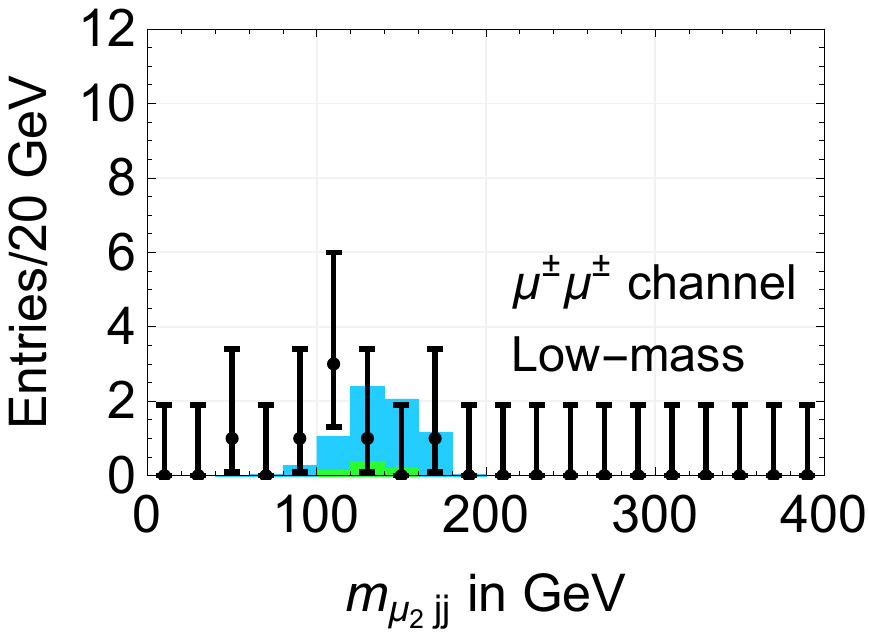} \hfill
  \includegraphics[scale=0.5]{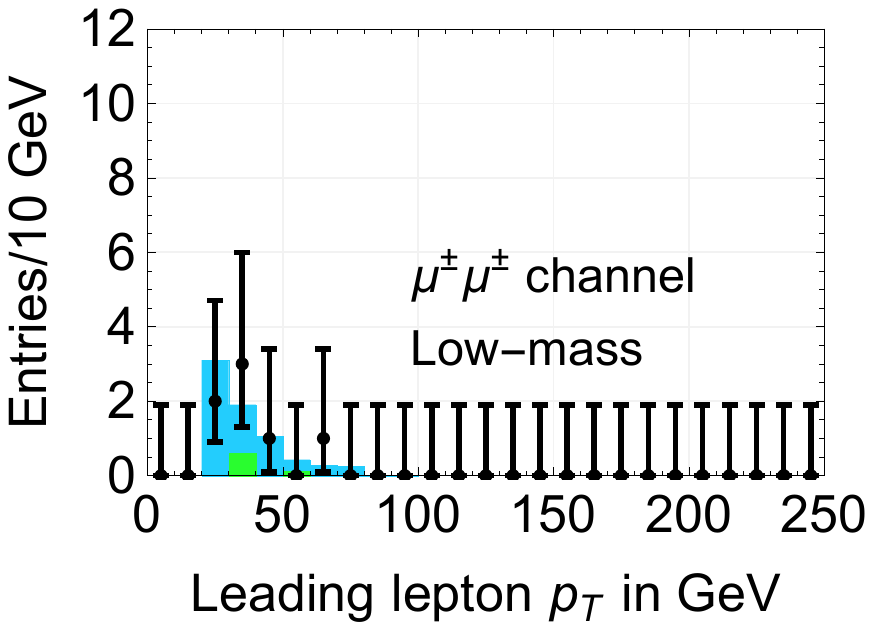} \hfill
  
  \caption{Validation of the fake lepton simulation with heavy Majorana searches at 8 TeV in the high and low mass regions. The  $W+3j$  and QCD jets are colored in blue.  The other subdominant processes with up to two matched jets are colored in green. Data points are taken from~\cite{Khachatryan:2015gha, Khachatryan:2016olu}, see text for details.}
  \label{figJFValid}
\end{figure}

\section{Jet fakes} \label{SecJetFakes}

One of the main backgrounds discussed above comes from the mis-identification of a jet for a charged lepton ($\ell = e, \mu$). To account for this source, we adopt a semi-empirical method advocated by~\cite{Curtin:2013zua} for the $t\bar{t}h$ channel and applied in \cite{Izaguirre:2015pga, Peng:2015haa, Dib:2016wge}.

The method postulates a mis-tag rate $\varepsilon_{j\rightarrow l}(p_T)$, which is the probability of a given jet to become mis-identified as a charged lepton that is here taken to be $p_T$ and $\eta$ independent. Additionally, the momentum of the fake lepton is softened
\begin{align}
  p_{T \ell} &= (1 - \alpha) p_{T\text{jet}}, & P(\alpha) &= \frac{1}{\mathcal{N}} \, e^{\frac{(\alpha-\mu)^2}{2\sigma^2}},
\end{align}
with a flavor and $p_T$ independent $\alpha$ obtained from a truncated normal distribution $P$ on the $[0,1]$ interval with the normalization constant $\mathcal{N}$. This procedure was implemented by augmenting the Delphes 3 \cite{deFavereau:2013fsa} \texttt{JetFakeParticle} class for fake lepton rate estimates.

The conversion probabilities $\varepsilon_{j\rightarrow l}$ and $\mu$ and $\sigma$ parameters are then determined from the 8 TeV heavy Majorana neutrino searches in the $\mu \mu$~\cite{Khachatryan:2015gha} and $ee/e\mu$ channels~\cite{Khachatryan:2016olu}. Fig.~\ref{figJFValid} shows how the data is reproduced with $\varepsilon_{j\rightarrow e(\mu)} = 5\times 10^{-4} (3 \times 10^{-4})$ and $\mu = 0.75$, $\sigma = 0.25$. The $e \mu$ data as well as other
pre-selection plots are also reproduced. The main source of the background comes from $W+jjj$ and from QCD jets in both low and high mass regions, while the prompt $VV = WZ, ZZ$ matched up to two jets are sub-dominant.


\def\arxiv#1[#2]{\href{http://arxiv.org/abs/#1}{[#2]}}
\def\Arxiv#1[#2]{\href{http://arxiv.org/abs/#1}{#2}}

%
%


\begin{thebibliography}{99}

\bibitem{Higgs:1964pj}
  P.~W.~Higgs,
  Phys.\ Rev.\ Lett.\  {\bf 13} (1964) 508;
  F.~Englert and R.~Brout,
  Phys.\ Rev.\ Lett.\  {\bf 13}, 321 (1964);
  G.~S.~Guralnik, C.~R.~Hagen and T.~W.~B.~Kibble,
  Phys.\ Rev.\ Lett.\  {\bf 13}, 585 (1964).

\bibitem{Weinberg:1967tq}
  S.~Weinberg,
  Phys.\ Rev.\ Lett.\  {\bf 19} (1967) 1264;

\bibitem{Khachatryan:2016vau}
  G.~Aad {\it et al.} [ATLAS and CMS Collaborations],
  \arxiv{1606.02266}[arXiv:1606.02266 [hep-ex]].

\bibitem{LROrigin}
  J.~C.~Pati and A.~Salam,
  Phys.\ Rev.\ D {\bf 10} (1974) 275
   Erratum: [Phys.\ Rev.\ D {\bf 11} (1975) 703];
  R.~N.~Mohapatra and J.~C.~Pati,
  Phys.\ Rev.\ D {\bf 11} (1975) 2558,
  Phys.\ Rev.\ D {\bf 11} (1975) 566.

\bibitem{LRSpont}
  G.~Senjanovi\'c and R.~N.~Mohapatra,
  Phys.\ Rev.\ D {\bf 12} (1975) 1502.
  G.~Senjanovi\'c,
  Nucl.\ Phys.\ B {\bf 153} (1979) 334.

\bibitem{MinkowskiMS79}
  P.~Minkowski,
  Phys.\ Lett.\ B {\bf 67} (1977) 421;
  R.~N.~Mohapatra and G.~Senjanovi\'c,
  Phys.\ Rev.\ Lett.\  {\bf 44} (1980) 912.

\bibitem{Majorana:1937vz}
  E.~Majorana,
 N. Cim.\  {\bf 14 } (1937)  171.

\bibitem{seesaw}
  T.~Yanagida, {\em Workshop on unified theories and baryon number in the universe}, ed. A. Sawada, A. Sugamoto (KEK, Tsukuba, 1979); S.~Glashow, {\em Quarks and leptons,  Carg\`ese 1979}, ed. M. L\'evy (Plenum, NY, 1980); M.~Gell-Mann, P.~Ramond, R.~Slansky, {\em Supergravity Stony Brook workshop, New York, 1979}, ed.\ P. Van Niewenhuizen, D. Freeman (North Holland, Amsterdam, 1980).

\bibitem{Schechter:1980gr}
  J.~Schechter and J.~W.~F.~Valle,
  Phys.\ Rev.\ D {\bf 22} (1980) 2227.

\bibitem{Keung:1983uu}
  W.~Y.~Keung and G.~Senjanovi\'c,
  Phys.\ Rev.\ Lett.\  {\bf 50} (1983) 1427.

\bibitem{Nemevsek:2011hz}
  M.~Nemev\v{s}ek, F.~Nesti, G.~Senjanovi\'c and Y.~Zhang,
  Phys.\ Rev.\ D {\bf 83} (2011) 115014
  \arxiv{1103.1627}[arXiv:1103.1627 [hep-ph]].

\bibitem{KSatLHC}
  V.~Khachatryan {\it et al.} [CMS Collaboration],
  Eur.\ Phys.\ J.\ C {\bf 74} (2014) no.11,  3149
  \arxiv{1407.3683}[arXiv:1407.3683 [hep-ex]],
  CMS-PAS-EXO-16-016, CMS-PAS-EXO-16-023
  S.~Chatrchyan {\it et al.} [CMS Collaboration],
  Phys.\ Rev.\ Lett.\  {\bf 109} (2012) 261802
  \arxiv{1210.2402}[arXiv:1210.2402 [hep-ex]],
  %
  ATL-PHYS-PUB-2009-072, ATL-COM-PHYS-2009-095, ATLAS-CONF-2011-115, ATLAS-CONF-2012-139,
  %
  G.~Aad {\it et al.} [ATLAS Collaboration],
  Eur.\ Phys.\ J.\ C {\bf 72} (2012) 2056
  \arxiv{1203.5420}[arXiv:1203.5420 [hep-ex]];
  JHEP {\bf 1507} (2015) 162
  \arxiv{1506.06020}[arXiv:1506.06020 [hep-ex]].

\bibitem{Mitra:2016kov}
  M.~Mitra, R.~Ruiz, D.~J.~Scott and M.~Spannowsky,
  \Arxiv{1607.03504}[arXiv:1607.03504 [hep-ph]].

\bibitem{Mohapatra:1980yp}
  R.~N.~Mohapatra and G.~Senjanovi\'c,
  Phys.\ Rev.\ D {\bf 23} (1981) 165.
   
\bibitem{Tello:2010am}
  V.~Tello, M.~Nemev\v{s}ek, F.~Nesti, G.~Senjanovi\'c and F.~Vissani,
  Phys.\ Rev.\ Lett.\  {\bf 106} (2011) 151801
  \arxiv{1011.3522}[arXiv:1011.3522 [hep-ph]].
  M.~Nemev\v{s}ek, F.~Nesti, G.~Senjanovi\'c and V.~Tello,
  \arxiv{1112.3061}[arXiv:1112.3061 [hep-ph]].
    
\bibitem{Das:2012ii} 
S.~P.~Das, F.~F.~Deppisch, O.~Kittel and J.~W.~F.~Valle,
Phys.\ Rev.\ D {\bf 86}, 055006 (2012)
\arxiv{1206.0256}[arXiv:1206.0256 [hep-ph]].
    
\bibitem{Vasquez:2014mxa}
  J.~C.~Vasquez,
  JHEP {\bf 1605} (2016) 176
  \arxiv{1411.5824}[arXiv:1411.5824 [hep-ph]].
    
\bibitem{Casas:2001sr}
  J.~A.~Casas and A.~Ibarra,
  Nucl.\ Phys.\ B {\bf 618} (2001) 171
  \arxiv{hep-ph/0103065}[hep-ph/0103065].

\bibitem{Nemevsek:2012iq}
  M.~Nemev\v{s}ek, G.~Senjanovi\'c and V.~Tello,
  Phys.\ Rev.\ Lett.\  {\bf 110} (2013) no.15,  151802
  \arxiv{1211.2837}[arXiv:1211.2837 [hep-ph]].

\bibitem{Datta:1993nm}
  A.~Datta, M.~Guchait and A.~Pilaftsis,
  Phys.\ Rev.\ D {\bf 50} (1994) 3195
  \arxiv{hep-ph/9311257}[hep-ph/9311257].
  C.~Y.~Chen, P.~S.~B.~Dev and R.~N.~Mohapatra,
  Phys.\ Rev.\ D {\bf 88} (2013) 033014
  \arxiv{1306.2342}[arXiv:1306.2342 [hep-ph]].

\bibitem{Izaguirre:2015pga}
  E.~Izaguirre and B.~Shuve,
  Phys.\ Rev.\ D {\bf 91} (2015) no.9,  093010
  \arxiv{1504.02470}[arXiv:1504.02470 [hep-ph]].
    
\bibitem{BhupalDev:2012zg}
  P.S.~Bhupal Dev, R.~Franceschini and R.N.~Mohapatra,
  Phys.\ Rev.\ D {\bf 86} (2012) 093010
  \arxiv{1207.2756}[arXiv:1207.2756 [hep-ph]];
  C.G.~Cely, A.~Ibarra, E.~Molinaro and S.T.~Petcov,
  Phys.\ Lett.\ B {\bf 718} (2013) 957
  \arxiv{1208.3654}[arXiv:1208.3654 [hep-ph]].

\bibitem{Nieves:1986uk}
 J.F.~Nieves, D.~Chang and P.B.~Pal,
 Phys.\ Rev.\ D {\bf 33} (1986) 3324.
    
\bibitem{0nu2bMix}
  J.~Barry and W.~Rodejohann,
  JHEP {\bf 1309} (2013) 153
  \arxiv{1303.6324}[arXiv:1303.6324 [hep-ph]].
  W.~C.~Huang and J.~Lopez-Pavon,
  Eur.\ Phys.\ J.\ C {\bf 74} (2014) 2853
  \arxiv{1310.0265}[arXiv:1310.0265 [hep-ph]].

\bibitem{Beall:1981ze}
  G.~Beall, M.~Bander and A.~Soni,
  Phys.\ Rev.\ Lett.\  {\bf 48} (1982) 848.

\bibitem{Mohapatra:1983ae}
  R.N.~Mohapatra, G.~Senjanovi\'c and M.D.~Tran,
  Phys.\ Rev.\ D {\bf 28} (1983) 546;
  K.~Kiers, J.~Kolb, J.~Lee, A.~Soni and G.-H.~Wu,
  Phys.\ Rev.\ D {\bf 66} (2002) 095002
  \arxiv{hep-ph/0205082}[hep-ph/0205082].

\bibitem{Ecker:1985vv}
  G.~Ecker and W.~Grimus,
  Nucl.\ Phys.\  B {\bf 258}, 328 (1985).

\bibitem{Zhang:2007fn}
  Y.~Zhang, H.~An, X.~Ji and R.N.~Mohapatra,
  Phys.\ Rev.\ D {\bf 76} (2007) 091301
  \arxiv{0704.1662}[arXiv:0704.1662 [hep-ph]] and
  Nucl.\ Phys.\ B {\bf 802} (2008) 247
  \arxiv{0712.4218}[arXiv:0712.4218 [hep-ph]].

\bibitem{Maiezza:2010ic}
  A.~Maiezza, M.~Nemev\v{s}ek, F.~Nesti and G.~Senjanovi\'c,
  Phys.\ Rev.\ D {\bf 82}, 055022 (2010)
  \arxiv{1005.5160}[arXiv:1005.5160 [hep-ph]].
 
\bibitem{Bertolini:2012pu}
  S.~Bertolini, J.O.~Eeg, A.~Maiezza and F.~Nesti,
  Phys.\ Rev.\ D {\bf 86} (2012) 095013
  \arxiv{1206.0668}[arXiv:1206.0668 [hep-ph]];
  S.~Bertolini, A.~Maiezza and F.~Nesti,
  Phys.\ Rev.\ D {\bf 88} (2013) 3,  034014
  \arxiv{1305.5739}[arXiv:1305.5739 [hep-ph]].
  
\bibitem{Bertolini:2014sua}
  S.~Bertolini, A.~Maiezza and F.~Nesti,
  Phys.\ Rev.\ D {\bf 89} (2014) 9, 095028
  \arxiv{1403.7112}[arXiv:1403.7112 [hep-ph]].
  
\bibitem{Maiezza:2014ala}
  A.~Maiezza and M.~Nemev\v{s}ek,
  Phys.\ Rev.\ D {\bf 90} (2014) 9,  095002
  \arxiv{1407.3678}[arXiv:1407.3678 [hep-ph]].
  
\bibitem{Senjanovic:1979cta}
  G.~Senjanovi\'c and P.~Senjanovi\'c,
  Phys.\ Rev.\ D {\bf 21}, 3253 (1980).
  
\bibitem{Dev:2016dja}
  P.~S.~B.~Dev, R.~N.~Mohapatra and Y.~Zhang,
  JHEP {\bf 1605} (2016) 174
  \arxiv{1602.05947}[arXiv:1602.05947 [hep-ph]].

\bibitem{Maiezza:2016bzp}
  A.~Maiezza, M.~Nemev\v{s}ek and F.~Nesti,
  Phys.\ Rev.\ D {\bf 94} (2016) no.3,  035008
  \arxiv{1603.00360}[arXiv:1603.00360 [hep-ph]].
  
\bibitem{Gunion:1986im}
  J.F.~Gunion, B.~Kayser, R.N.~Mohapatra, N.G.~Deshpande, J.~Grifols, A.~Mendez, F.I.~Olness and P.B.~Pal,
  PRINT-86-1324 (UC,DAVIS);
  J.~F.~Gunion, H.~E.~Haber, G.~L.~Kane and S.~Dawson,
  Front.\ Phys.\  {\bf 80} (2000) 1.
  
\bibitem{Pilaftsis:1991ug}
  A.~Pilaftsis,
  Z.\ Phys.\ C {\bf 55} (1992) 275
  [hep-ph/9901206].
  
\bibitem{Graesser:2007yj}
  M.L.~Graesser,
  Phys.\ Rev.\ D {\bf 76} (2007) 075006
  \arxiv{0704.0438}[arXiv:0704.0438 [hep-ph]] and
  \Arxiv{0705.2190}[arXiv:0705.2190 [hep-ph]].

\bibitem{Maiezza:2015lza}
  A.~Maiezza, M.~Nemev\v{s}ek and F.~Nesti,
  Phys.\ Rev.\ Lett.\  {\bf 115} (2015) 081802
  \arxiv{1503.06834}[arXiv:1503.06834 [hep-ph]].
  
\bibitem{singletHiggs}
  A.~Falkowski, C.~Gross and O.~Lebedev,
  JHEP {\bf 1505} (2015) 057
  \arxiv{1502.01361}[arXiv:1502.01361 [hep-ph]];
  S.~I.~Godunov, A.~N.~Rozanov, M.~I.~Vysotsky and E.~V.~Zhemchugov,
  Eur.\ Phys.\ J.\ C {\bf 76} (2016) no.1,  1
  \arxiv{1503.01618}[arXiv:1503.01618 [hep-ph]].

\bibitem{Buttazzo:2015bka}
  D.~Buttazzo, F.~Sala and A.~Tesi,
  JHEP {\bf 1511} (2015) 158
  \arxiv{1505.05488}[arXiv:1505.05488 [hep-ph]].
    
\bibitem{Clarke:2013aya}
  J.~D.~Clarke, R.~Foot and R.~R.~Volkas,
  JHEP {\bf 1402} (2014) 123
  \arxiv{1310.8042}[arXiv:1310.8042 [hep-ph]];
  J.~D.~Clarke,
  JHEP {\bf 1510} (2015) 061
  \arxiv{1505.00063}[arXiv:1505.00063 [hep-ph]].
    
\bibitem{Cirigliano:2004mv}
  V.~Cirigliano, A.~Kurylov, M.J.~Ramsey-Musolf and P.~Vogel,
  Phys.\ Rev.\ D {\bf 70} (2004) 075007
  \arxiv{0404233}[hep-ph/0404233].

\bibitem{Tello:2012qda}
  V.~Tello, SISSA Phd Thesis Sep. 2012.

\bibitem{Banerjee:2015hoa}
  S.~Banerjee, M.~Mitra and M.~Spannowsky,
  Phys.\ Rev.\ D {\bf 92} (2015) no.5,  055013
  \arxiv{1506.06415}[arXiv:1506.06415 [hep-ph]].

\bibitem{Kang:2015uoc}
  Z.~Kang, P.~Ko and J.~Li,
  Phys.\ Rev.\ D {\bf 93} (2016) no.7,  075037
  doi:10.1103/PhysRevD.93.075037
  [arXiv:1512.08373 [hep-ph]].

\bibitem{deFlorian:2016spz}
  D.~de Florian {\it et al.} [LHC Higgs Cross Section Working Group Collaboration],
  arXiv:1610.07922 [hep-ph].
    
\bibitem{Anastasiou:2016hlm}
  C.~Anastasiou, C.~Duhr, F.~Dulat, E.~Furlan, T.~Gehrmann, F.~Herzog, A.~Lazopoulos and B.~Mistlberger,
  \Arxiv{1605.05761}[arXiv:1605.05761 [hep-ph]].
    
\bibitem{Hessler:2014ssa}
  A.~G.~Hessler, A.~Ibarra, E.~Molinaro and S.~Vogl,
  Phys.\ Rev.\ D {\bf 91} (2015) 11,  115004
  \arxiv{1408.0983}[arXiv:1408.0983 [hep-ph]].
  
\bibitem{Maalampi:2002vx}
  J.~Maalampi and N.~Romanenko,
  Phys.\ Lett.\ B {\bf 532} (2002) 202
  \arxiv{hep-ph/0201196}[hep-ph/0201196].

\bibitem{LRMSMixModel} \href{https://sites.google.com/site/leftrighthep}{https://sites.google.com/site/leftrighthep}
  
\bibitem{Alloul:2013bka}
  A.~Alloul, N.D.~Christensen, C.~Degrande, C.~Duhr and B.~Fuks,
  Comput.\ Phys.\ Commun.\  {\bf 185} (2014) 2250
  \arxiv{1310.1921}[arXiv:1310.1921 [hep-ph]].

\bibitem{Roitgrund:2014zka}
  A.~Roitgrund, G.~Eilam and S.~Bar-Shalom,
  \Arxiv{1401.3345}[arXiv:1401.3345 [hep-ph]];
  Comput.\ Phys.\ Commun.\  {\bf 203} (2016) 18.
  
\bibitem{Alwall:2014hca}
  J.~Alwall, R.~Frederix, S.~Frixione, V.~Hirschi, F.~Maltoni, O.~Mattelaer, H.-S.~Shao and T.~Stelzer {\it et al.},
  JHEP {\bf 1407}, 079 (2014)
  \arxiv{1405.0301}[arXiv:1405.0301 [hep-ph]].

\bibitem{Sjostrand:2006za}
  T.~Sjostrand, S.~Mrenna and P.Z.~Skands,
  JHEP {\bf 0605} (2006) 026
  \arxiv{0603175}[hep-ph/0603175].

\bibitem{deFavereau:2013fsa}
  J.~de Favereau {\it et al.}  [DELPHES 3 Collaboration],
  JHEP {\bf 1402} (2014) 057
  \arxiv{1307.6346}[arXiv:1307.6346 [hep-ex]].

\bibitem{HiggsXSecWG}\href{https://twiki.cern.ch/twiki/bin/view/LHCPhysics/CERNYellowReportPageAt1314TeV2014}{https://twiki.cern.ch/twiki/bin/view/LHCPhysics/CERNYellowReportPageAt1314TeV2014}

\bibitem{TTbarXSec} \href{https://twiki.cern.ch/twiki/bin/view/LHCPhysics/TtbarNNLO}{https://twiki.cern.ch/twiki/bin/view/LHCPhysics/TtbarNNLO}

\bibitem{Melia:2011tj}
  T.~Melia, P.~Nason, R.~Rontsch and G.~Zanderighi,
  JHEP {\bf 1111} (2011) 078
  \arxiv{1107.5051}[arXiv:1107.5051 [hep-ph]].

\bibitem{ATLAS:2016iqc}
  The ATLAS collaboration [ATLAS Collaboration],
  ATLAS-CONF-2016-024.

\bibitem{Aad:2016jkr}
  G.~Aad {\it et al.} [ATLAS Collaboration],
  \Arxiv{1603.05598}[arXiv:1603.05598 [hep-ex]].
  
\bibitem{ATLASTrig} ATL-DAQ-PUB-2016-001, \href{http://cds.cern.ch/record/2136007}{http://cds.cern.ch/record/2136007}
  
\bibitem{Conte:2012fm}
  E.~Conte, B.~Fuks and G.~Serret,
  Comput.\ Phys.\ Commun.\  {\bf 184} (2013) 222
  \arxiv{1206.1599}[arXiv:1206.1599 [hep-ph]].

\bibitem{Senjanovic:2014pva}
  G.~Senjanovi\'c and V.~Tello,
  Phys.\ Rev.\ Lett.\  {\bf 114}, no. 7, 071801 (2015)
  \arxiv{1408.3835}[arXiv:1408.3835 [hep-ph]];
  G.~Senjanovi\'c and V.~Tello,
  \Arxiv{1502.05704}[arXiv:1502.05704 [hep-ph]].
    
\bibitem{llt}
  C.~Csaki, E.~Kuflik, S.~Lombardo, O.~Slone and T.~Volansky,
  JHEP {\bf 1508} (2015) 016
  doi:10.1007/JHEP08(2015)016
  [arXiv:1505.00784 [hep-ph]];
  C.~Csaki, E.~Kuflik, S.~Lombardo and O.~Slone,
  Phys.\ Rev.\ D {\bf 92} (2015) no.7,  073008
  doi:10.1103/PhysRevD.92.073008
  [arXiv:1508.01522 [hep-ph]];
  A.~Coccaro, D.~Curtin, H.~J.~Lubatti, H.~Russell and J.~Shelton,
  arXiv:1605.02742 [hep-ph];
  B.~C.~Allanach, M.~Badziak, G.~Cottin, N.~Desai, C.~Hugonie and R.~Ziegler,
  Eur.\ Phys.\ J.\ C {\bf 76} (2016) no.9,  482
  doi:10.1140/epjc/s10052-016-4330-3
  [arXiv:1606.03099 [hep-ph]].

\bibitem{lle}
  G.~Aad {\it et al.} [ATLAS Collaboration],
  Phys.\ Rev.\ D {\bf 92} (2015) no.7,  072004
  doi:10.1103/PhysRevD.92.072004
  [arXiv:1504.05162 [hep-ex]].
  G.~Aad {\it et al.} [ATLAS Collaboration],
  JINST {\bf 8} (2013) P07015
  doi:10.1088/1748-0221/8/07/P07015
  [arXiv:1305.2284 [hep-ex]].

\bibitem{CMS:2016kmw}
  CMS Collaboration [CMS Collaboration],
  CMS-PAS-SUS-14-020.

\bibitem{0nu4bNemo3:2016} NEMO3 talk at
  \href{http://neutrino2016.iopconfs.org/IOP/media/uploaded/EVIOP/event_948/10.25__5__waters.pdf}{Neutrino 2016.}

\bibitem{Heeck:2013rpa}
  J.~Heeck and W.~Rodejohann,
  Europhys.\ Lett.\  {\bf 103} (2013) 32001
  \arxiv{1306.0580}[arXiv:1306.0580 [hep-ph]].

\bibitem{Ferrari:2000sp}
 A.~Ferrari et al.
 Phys.\ Rev.\ D {\bf 62} (2000) 013001;
 S.N.~Gninenko et al. 
 Phys.\ Atom.\ Nucl.\  {\bf 70} (2007) 441.

\bibitem{LHC:ellnu2016}
  The ATLAS collaboration [ATLAS Collaboration],
  ATLAS-CONF-2016-061.
  CMS Collaboration [CMS Collaboration],
  CMS-PAS-EXO-15-006, CMS-PAS-EXO-16-006
  
\bibitem{LHC:dijets2016}
  G.~Aad {\it et al.} [ATLAS Collaboration],
  Phys.\ Lett.\ B {\bf 754} (2016) 302
  \arxiv{1512.01530}[arXiv:1512.01530 [hep-ex]],
  ATLAS-CONF-2016-069.
  and V.~Khachatryan {\it et al.} [CMS Collaboration],
  Phys.\ Rev.\ Lett.\  {\bf 116} (2016) no.7,  071801
  \arxiv{1512.01224}[arXiv:1512.01224 [hep-ex]],
  CMS-PAS-EXO-16-032.

\bibitem{Agostini:2013mzu}
  M.~Agostini {\it et al.}  [GERDA Collaboration],
  Phys.\ Rev.\ Lett.\  {\bf 111} (2013) 12,  122503
  \arxiv{1307.4720}[arXiv:1307.4720 [nucl-ex]]
  and talk at \href{https://www.mpi-hd.mpg.de/gerda/public/2016/t16_neutrino_gerda_ma.pdf}{Neutrino 2016}.

\bibitem{KamLAND-Zen:2016pfg}
  A.~Gando {\it et al.} [KamLAND-Zen Collaboration],
  Phys.\ Rev.\ Lett.\  {\bf 117} (2016) 082503
  \arxiv{1605.02889}[arXiv:1605.02889 [hep-ex]].
  
\bibitem{Barea:2012zz}
  J.~Barea, J.~Kotila and F.~Iachello,
  Phys.\ Rev.\ Lett.\  {\bf 109} (2012) 042501;
    J.~Barea, J.~Kotila and F.~Iachello,
  Phys.\ Rev.\ C {\bf 87} (2013) 1,  014315
  \arxiv{1301.4203}[arXiv:1301.4203 [nucl-th]];
   J.~Barea, J.~Kotila and F.~Iachello,
  Phys.\ Rev.\ C {\bf 91} (2015) 3,  034304.

\bibitem{Gomez-Ceballos:2013zzn}
  M.~Bicer {\it et al.} [TLEP Design Study Working Group Collaboration],
  JHEP {\bf 1401} (2014) 164
  \arxiv{1308.6176}[arXiv:1308.6176 [hep-ex]].
  
\bibitem{Blondel:2014bra}
  A.~Blondel {\it et al.} [FCC-ee study Team Collaboration],
  \Arxiv{1411.5230}[arXiv:1411.5230 [hep-ex]].

\bibitem{Antusch:2015mia}
  S.~Antusch and O.~Fischer,
  JHEP {\bf 1505} (2015) 053
  \arxiv{1502.05915}[arXiv:1502.05915 [hep-ph]],
  S.~Antusch, E.~Cazzato and O.~Fischer,
  \Arxiv{1604.02420}[arXiv:1604.02420 [hep-ph]].

\bibitem{Djouadi:2005gi}
  A.~Djouadi,
  Phys.\ Rept.\  {\bf 457} (2008) 1
  \arxiv{hep-ph/0503172}[hep-ph/0503172].
  
\bibitem{Curtin:2013zua}
  D.~Curtin, J.~Galloway and J.~G.~Wacker,
  Phys.\ Rev.\ D {\bf 88} (2013) no.9,  093006
  \arxiv{1306.5695}[arXiv:1306.5695 [hep-ph]].

\bibitem{Peng:2015haa}
  T.~Peng, M.~J.~Ramsey-Musolf and P.~Winslow,
  Phys.\ Rev.\ D {\bf 93} (2016) no.9,  093002
  \arxiv{1508.04444}[arXiv:1508.04444 [hep-ph]].
  
\bibitem{Dib:2016wge}
  C.~O.~Dib, C.~S.~Kim, K.~Wang and J.~Zhang,
  Phys.\ Rev.\ D {\bf 94} (2016) no.1,  013005
  doi:10.1103/PhysRevD.94.013005
  \arxiv{1605.01123}[arXiv:1605.01123 [hep-ph]].
    
\bibitem{Khachatryan:2015gha}
  V.~Khachatryan {\it et al.} [CMS Collaboration],
  Phys.\ Lett.\ B {\bf 748} (2015) 144
  \arxiv{1501.05566}[arXiv:1501.05566 [hep-ex]].

\bibitem{Khachatryan:2016olu}
  V.~Khachatryan {\it et al.} [CMS Collaboration],
  JHEP {\bf 1604} (2016) 169
  \arxiv{1603.02248}[arXiv:1603.02248 [hep-ex]].

\bibitem{Han:2012vk}
  T.~Han, I.~Lewis, R.~Ruiz and Z.~g.~Si,
  Phys.\ Rev.\ D {\bf 87} (2013) no.3,  035011
  Erratum: [Phys.\ Rev.\ D {\bf 87} (2013) no.3,  039906]
  \arxiv{1211.6447}[arXiv:1211.6447 [hep-ph]].
  
\bibitem{Barger:2011tb}
  V.~Barger, W.~Y.~Keung and B.~Yencho,
  Phys.\ Rev.\ D {\bf 84} (2011) 015001
  \arxiv{1105.1780}[arXiv:1105.1780 [hep-ph]].
    
\bibitem{Shoemaker:2010fg}
  I.M.~Shoemaker, K.~Petraki and A.~Kusenko,
  JHEP {\bf 1009} (2010) 060
  \arxiv{1006.5458}[arXiv:1006.5458 [hep-ph]];
  P.~Humbert, M.~Lindner and J.~Smirnov,
  JHEP {\bf 1506} (2015) 035
  \arxiv{1503.03066}[arXiv:1503.03066 [hep-ph]].
   O.~Fischer,
  \Arxiv{1607.00282}[arXiv:1607.00282 [hep-ph]].

\bibitem{Coleman:1973jx}
  S.~R.~Coleman and E.~J.~Weinberg,
  Phys.\ Rev.\ D {\bf 7} (1973) 1888.
  doi:10.1103/PhysRevD.7.1888
  
\bibitem{StableQuantumVacuum}
  S.~Weinberg,
  Phys.\ Rev.\ Lett.\  {\bf 36} (1976) 294;
  A.~D.~Linde,
  JETP Lett.\  {\bf 23} (1976) 64, [Pisma Zh.\ Eksp.\ Teor.\ Fiz.\  {\bf 23} (1976) 73];
  J.~Basecq and D.~Wyler,
  Phys.\ Rev.\ D {\bf 39} (1989) 870.
      

      
\end{thebibliography}
\end{document}